# Cut-offs and light-spin flips in surface plasmon resonance between a chiral medium and a metal

Hyoung-In Lee[1,*], Christopher Gaul[2]

[1]*Research Institute of Mathematics, Seoul National University, 599 Gwanak-Ro, Gwanak-Gu, Seoul, Republic of Korea 08826*
[2]*Fundación ICAMCYL, International Center for Advanced Materials and Raw Materials of Castilla y León, León, Spain*
*Corresponding author: hileesam@naver.com*

**We have derived a dispersion relation governing the surface plasmon resonance established along a planar interface between a metal and a chiral medium (chiral case). Resulting numerical solutions are compared with the well-known results obtained for a metal-dielectric interface (achiral case). Comparisons show that the chiral case exhibits smaller phase speeds than the achiral case. Such a longitudinal deceleration in the chiral case is caused by the energy expended on establishing the longitudinal spin density, which turns out anti-symmetric with respect to medium chirality. Moreover, the low-frequency range is disallowed in the chiral case. Relevant energy redistributions, spin-orbital couplings, and applications are discussed. © 2022 Optica Publishing Group**
https://doi.org/

Science of chirality and/or helicity is a huge area encompassing biology, chemistry, physics, and mathematics among others. Chirality plays great roles on various scales ranging from electronic systems [1,2] through molecules [3] to metamaterials [4,5], and others [6]. The structural chirality of a molecule is linked to the spin-orbit coupling (SOC) in the electrons and/or photons [1,7]. Our study would shed a light on the mechanisms lying under light-driven rotary molecular motors, where varying structural asymmetries are responsive to illuminating light [8].

Meanwhile, electromagnetic (EM) waves intrinsically harbor mutual vortices between electric and magnetic fields. In addition to these primary vortices [6], secondary vortices could arise if a medium becomes chiral [9,10]. In this regard, suppose that either air or a liquid is dispersed with a dilute ensemble of chiral molecules [2,4,5,10]. Sensing or enantiomer detection of the chiral content of an embedding medium is of interests to practical applications [4,5,10-12].

Surface plasmon waves can be sustained along metal-dielectric (M-D) interfaces [13]. If a single planar M-D interface separates two semi-infinite spaces, there is an analytic solution to the dispersion relation as presented in numerous publications [7]. Figure 1(a) shows an 'achiral case' with the upper space is filled with an ordinary dielectric medium, whereas Fig. 1(b) depicts a 'chiral case' with the upper space filled with a chiral medium [4]. We represent the different dynamics of the two cases by combinations of a big horizontal arrow with one and two circular arrows, respectively. The horizontal arrow stands for the size of the translational energy that is directed along the horizontal $x$-axis.

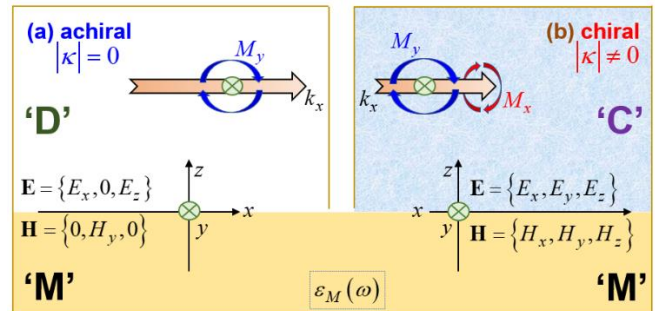

Fig. 1. The upper half-space is filled with (a) an achiral dielectric, and (b) a chiral medium. The lower half-space is filled with a metal. Surface plasmon waves propagate in the $x$-direction, being orthogonal to the transverse $y$-direction and the depth-wise $z$-direction.

In Fig. 1, the blue circular arrow marked by $M_y$ denotes the transverse (also being parallel to an interface) component of the spin density, which corresponds to the primary vortices. Meanwhile, Fig. 1(b) carries an additional red circular arrow for the longitudinal component $M_x$ of the spin density [1,14], which corresponds to the secondary vortices. In comparison to Fig. 1(a), Fig. 1(b) implies a redistribution among a translational energy and two types of rotary energies, where this redistribution is illustrated by differently sized arrows [2,4]. See Eq. (11) and Eq. (S4.13) of [15], where energy redistributions signify a SOC due to medium chirality.

We will derive a dispersion relation in the chiral case for the first time. From the associated numerical results, we found out two outstanding features: (i) frequency cut-offs, and (ii) decelerations in a translational motion due to an energy necessary to feed a longitudinal light spin. Our results could serve as a prototype for a metallic probe immersed in a chiral medium [8,15].

Electric and magnetic fields are denoted respectively by $\{\mathbf{E}, \mathbf{H}\}$, while $\mathrm{i}^2 \equiv -1$. All field variables follow $\exp(\mathrm{i}k_x x - \mathrm{i}\omega t)$ with $k_x > 0$ assumed for simplicity. The dimensionless Maxwell equations consist of $\nabla \times \mathbf{E} = \mathrm{i}\omega \mathbf{B}$ and $\nabla \times \mathbf{H} = -\mathrm{i}\omega \mathbf{D}$ along with $\nabla \cdot \mathbf{D} = \nabla \cdot \mathbf{B} = 0$. The subscripts $m \equiv M, D$ denotes respectively an achiral medium and a metal. Besides, $\{\varepsilon, \mu\}$ are relative electric permittivity and magnetic permeability. For a metal, $\mathbf{D} = \varepsilon_M \mathbf{E}$ and $\mathbf{B} = \mu_M \mathbf{H}$. For a chiral medium, $\mathbf{D} = \varepsilon_D \mathbf{E} + \mathrm{i}\kappa \mathbf{H}$ and $\mathbf{B} = \mu_D \mathbf{H} - \mathrm{i}\kappa \mathbf{E}$, whereas $\kappa \in \mathbb{R}$ is a medium chirality without loss [4-6,10-12]. See Section ('Sec.') S1 of Supplemental Document ('SD') and [15].

Consider the achiral case with $\kappa = 0$. In addition to $\Pi_m \equiv \exp(\mathrm{i}k_x x - \gamma_m |z|)$, transversal confinements are described by the decay rate $\gamma_m = \sqrt{k_x^2 - \omega^2 \varepsilon_m \mu_m}$ [1]. Figure 1(a) illustrates a transverse-magnetic (TM) wave, which is elliptically polarized on the longitudinal $zx$-plane. We thus obtain the dispersion relation $\varepsilon_D \gamma_M = |\varepsilon_M| \gamma_D$. In contrast, a transverse-electric (TE) wave does not admit a resonance [1,7]. A lossless metal is modeled by $\varepsilon_M = 1 - \omega_p^2 / \omega^2$ with $\omega_p$ being a plasma frequency, which is set at $\omega_p = 1$ [4,7]. See Sec. 2 of SD.

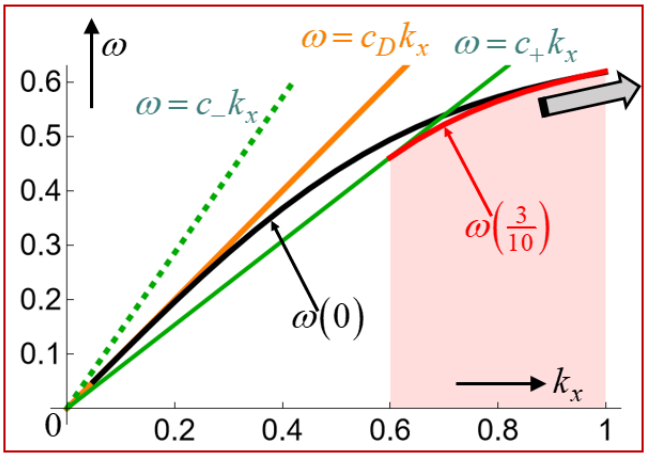

Fig. 2. Cutoffs on $\omega(k_x)$ as solutions to the dispersion relation over the interval $0 < k_x < 1$. It is chosen that $\varepsilon_D = \mu_D = 1$ and $\kappa = \tfrac{3}{10}$. Straight lines are given by $\omega = c_\pm k_x$ and $\omega = c_D k_x$.

On Fig. 2, the black solid curve pointed out by $\omega(0)$ displays the dispersion relation $\varepsilon_D \gamma_M = |\varepsilon_M| \gamma_D$ [7,13]. Certainly, $\omega$ on resonance increases monotonically with $k_x$. Additionally, we list below several key parameters.

$$\begin{cases} c_D^{-1} \equiv \sqrt{\varepsilon_D \mu_D} \\ c_\pm^{-1} \equiv \sqrt{\varepsilon_D \mu_D} \pm \kappa \end{cases}, \quad k_\pm \equiv \frac{\omega}{c_\pm}, \quad \gamma_\pm \equiv \sqrt{k_x^2 - k_\pm^2}. \quad (1)$$

Here, $\{c_D, c_\pm\}$ are the respective light speeds, where $c_D$ for the achiral case is recovered by setting $\kappa = 0$ in $c_\pm$. Besides, we assign $\varepsilon_D = \mu_M = \mu_D = 1$, whence $c_D = 1$. Furthermore, $\{k_\pm, \gamma_\pm\}$ are wave numbers and decay rates. A zoomed-in figure of Fig. 2 is offered as Fig. S2 of SD, which shows more clearly the reference lines $\omega_\pm \equiv c_\pm k_x$.

On the upper right corner of Fig. 2, a grey arrow indicates the usual asymptotic approach to $\{k_x, \omega\} = \{\infty, \tfrac{1}{\sqrt{2}}\}$. In brief, Fig. 2 displays approximate allowed ranges $0.6 < k_x < \infty$ and $0.461 < \omega < \tfrac{1}{\sqrt{2}} = 0.707$. Correspondingly, $-3.71 < \varepsilon_M < -1$.

For the chiral case, we introduce a pair $\mathbf{Q}_\pm$ of circular vectors [9,10], where the subscripts $\{+, -\}$ denote respectively 'left' and 'right'. Besides, we define $\Pi_\pm \equiv \exp(\mathrm{i}k_x x - \gamma_\pm z)$ with $\gamma_\pm > 0$ for a chiral medium [6,14].

With $\{\hat{\mathbf{x}}, \hat{\mathbf{y}}, \hat{\mathbf{z}}\}$ as the Cartesian unit vectors, we obtain $\sqrt{2}k_x \mathbf{P}_\pm = P_\pm (\gamma_\pm \hat{\mathbf{x}} \pm k_x \hat{\mathbf{y}} + \mathrm{i}k_x \hat{\mathbf{z}})\Pi_\pm$ with $P_\pm \in \mathbb{R}$ as solutions to both $\nabla \cdot \mathbf{P}_\pm = 0$ and $\nabla \times \mathbf{P}_\pm = \pm k_\pm \mathbf{P}_\pm$ [9,10]. Over $z \geq 0$, the EM fields are then constructed according to $\mathbf{E} = \mathbf{P}_+ - \mathbf{P}_-$ and $\mathbf{H} = -\mathrm{i}Z_D^{-1}(\mathbf{P}_+ + \mathbf{P}_-)$ as follows [14].

$$\sqrt{2}k_x \mathbf{E} = \sum_{\sigma=\pm} (\sigma \gamma_\sigma \hat{\mathbf{x}} + k_\sigma \hat{\mathbf{y}} + \sigma \mathrm{i}k_x \hat{\mathbf{z}}) P_\sigma \Pi_\sigma \\ \mathrm{i}Z_D \sqrt{2}k_x \mathbf{H} = \sum_{\sigma=\pm} (\gamma_\sigma \hat{\mathbf{x}} + \sigma k_\sigma \hat{\mathbf{y}} + \mathrm{i}k_x \hat{\mathbf{z}}) P_\sigma \Pi_\sigma \quad (2)$$

Here, $Z_D \equiv \sqrt{\mu_D / \varepsilon_D}$ is the impedance for the bulk dielectric medium. Here, both fields are elliptically polarized on all $xy$-, $yz$-, and $zx$-planes [14]. See Sec. S3 of SD.

In compliance to $\{\mathbf{E}, \mathbf{H}\}$ in Eq. (2) and on Fig. 1(b), the fields in a metal carry respectively three nonzero components over $z \leq 0$.

$$\sqrt{2}k_x \mathbf{E} = [G_{Ex}\hat{\mathbf{x}} + G_{Ey}\hat{\mathbf{y}} - \mathrm{i}(k_x/\gamma_M)G_{Ex}\hat{\mathbf{z}}]\Pi_M \\ \sqrt{2}k_x \mathbf{H} = [\mathrm{i}vG_{Ey}\hat{\mathbf{x}} + \mathrm{i}uG_{Ex}\hat{\mathbf{y}} + (k_x/\omega\mu_M)G_{Ey}\hat{\mathbf{z}}]\Pi_M \quad (3)$$

Here, medium-dependent parameters are $u \equiv \omega \varepsilon_M / \gamma_M$ and $v \equiv \gamma_M / \omega \mu_M$. We then apply the continuity relations: $(E_x)_{z=0+} = (E_x)_{z=0-}$ and the likes for $\{E_y, H_x, H_y\}$ [9,10,16]. We thus reach a solvability equation $\hat{\Upsilon}\boldsymbol{\Xi} = \mathbf{0}$, where $\hat{\Upsilon}$ is a 4-by-4 matrix $\hat{\Upsilon}$ and $\boldsymbol{\Xi}^T \equiv \{P_+, P_-, G_{Ex}, G_{Ey}\}$ is an eigenvector. By a vanishing determinant $|\hat{\Upsilon}| = 0$, we obtain the following dispersion relation $f(k_x, \omega, \kappa) = 0$ with $J_D^\kappa \equiv 1 + Z_D^2 uv$.

$$J_D^\kappa (k_+\gamma_- + k_-\gamma_+) + 2Z_D (u\gamma_+\gamma_- + vk_+k_-) = 0. \quad (4)$$

See Secs. S4 and S5 of SD respectively for its derivation and numerical solutions for a given set $\{\varepsilon_D, \mu_D, \varepsilon_M, \mu_M\}$ of data. We consider only the range $\omega < \omega_p = 1$ such that $\varepsilon_M < 0$. The evenness $f(k_x, \omega, -\kappa) = f(k_x, \omega, \kappa)$ is proved in Sec. S5 of SD.

Evanescent waves in Eq. (2) belong to structured light, thus being more practical than plane waves in sensing applications [10,15].

The chiral case displayed on Fig. 1(b) is reducible to the achiral case shown on Fig. 1(a). This amounts to saying that the chiral dispersion relation in Eq. (4) is reducible to the achiral dispersion relation $\varepsilon_D \gamma_M = |\varepsilon_M| \gamma_D$ [7,13]. See Sec. 4 of SD.

Figure 2 presents solutions to Eq. (4) on the $k_x \omega$-plane as the red solid curve $\omega\left(\frac{3}{10}\right)$, which means $\omega\left(\kappa = \frac{3}{10}\right)$. The frequencies exhibit monotonic increases with $k_x$. The orange solid line is the light line $\omega = c_D k_x$ in the bulk dielectric medium. In addition, there are two straight lines $\omega_{\pm}\left(\frac{3}{10}\right)$, where $\omega_{\pm}(\kappa) = c_{\pm} k_x$ as defined in Eq. (1). The superluminal curves $\omega_-(\kappa)$ are steeper than the subluminal curves $\omega_+(\kappa)$ [7].

Figure 2 shows that only the portion of the red solution curve below $\omega_+(\kappa)$ is allowed. This low-frequency cut-off or the low-wave-number cut-off phenomenon is one of key findings of this study [1,11]. The allowed frequency range is denoted, for instance, by the light-pink filling below $\omega\left(\frac{3}{10}\right)$.

The red solid curve $\omega\left(\frac{3}{10}\right)$ for the chiral case exhibits steeper gradient than the black solid curve $\omega(0)$ for the achiral case. Since $\omega/k_x$ is the phase speed in the longitudinal $x$-direction, the chiral case possesses smaller phase speeds than those of the achiral case. This result is what we call a 'chirality-induced deceleration (CID)' [17]. A given photon energy (a given frequency) is consumed for executing a rotational motion associated with a medium chirality so that the longitudinal phase speed becomes smaller.

Therefore, a low-frequency gap corresponds to a range of insufficient photon energy to excite a required rotational motion in a chiral medium [1,11,15]. This frequency cut-offs could be implemented for high-frequency pass filters or optical switches [3].

Figure S3 of SD presents a detailed account of the phase speed and the decay rates $\gamma_{\pm} \equiv \sqrt{k_x^2 - k_{\pm}^2}$ [1,14,15]. In practice, the medium chirality is mostly tiny on the order of $|\kappa| \approx \frac{1}{100}$ [5,6,10]. Notwithstanding, we take an exaggerated value of $|\kappa| = \frac{3}{10}$ in this study to show the effects of finite $\kappa$. In SD, we performed all numerical evaluations with two values of $\kappa = \frac{1}{10}, \frac{3}{10}$.

Based on the eigenvalue $k_x(\omega)$ or $\omega(k_x)$ presented on Fig. 2, the eigenvector $\Xi$ in $\widehat{\Upsilon}\Xi = \mathbf{0}$ is easily computed. We are particularly interested in the chiral medium occupying the spatial zone over $z > 0$. In this zone, the real value $P_+/P_-$ is the left-right mixture ratio, which is essentially the TM-TE mixture ratio for elliptical ('circular' included) polarizations [14]. Instead of the two $\{P_+, P_-\}$, we can thus introduce a single complex scalar $\Gamma \in \mathbb{C}$.

$$\beta_{\pm} \equiv k_{\mp} + \gamma_{\mp} \frac{Z_D \omega \varepsilon_M}{\gamma_M}, \quad P_{\pm} \equiv iZ_D \beta_{\pm} \Gamma. \tag{5}$$

See Secs. S5 of SD. We now evaluate the electric-field intensity $|\mathbf{E}|^2$ and electric spin density $\mathbf{M} \equiv \omega^{-1} \operatorname{Im}(\mathbf{E}^* \times \mathbf{E})$ as follows.

$$q|\mathbf{E}|^2 = \beta_+^2 \exp(-2\gamma_+ z) + \beta_-^2 \exp(-2\gamma_- z) - k_x^{-2} \beta_+ \beta_- \left(\gamma_+ \gamma_- - k_+ k_- + k_x^2\right) \exp\left[-(\gamma_+ + \gamma_-)z\right]. \tag{6}$$

$$\begin{cases} \Psi_{xx}^{even}(\kappa, z) \equiv k_+ \beta_+ \exp(-\gamma_+ z) + k_- \beta_- \exp(-\gamma_- z) \\ \Psi_{xy}^{odd}(\kappa, z) \equiv \beta_+ \exp(-\gamma_+ z) - \beta_- \exp(-\gamma_- z) \\ \Psi_{yy}^{odd}(\kappa, z) \equiv \gamma_+ \beta_+ \exp(-\gamma_+ z) - \gamma_- \beta_- \exp(-\gamma_- z) \end{cases} \tag{7}$$

$$\Rightarrow \begin{cases} qk_x \omega M_x = \Psi_{xx}^{even}(\kappa, z) \Psi_{xy}^{odd}(\kappa, z) \\ qk_x \omega M_y = -\Psi_{yy}^{odd}(\kappa, z) \Psi_{xy}^{odd}(\kappa, z) \end{cases}$$

Here, $q \equiv Z_D^{-2} |\Gamma|^{-2}$. See Secs. S6-S7 of SD. The depth-wise component vanishes, namely, $M_z = 0$. Meanwhile, we discovered the series of symmetry and complementarity properties: $k_{\pm}(-\kappa) = k_{\mp}(\kappa)$, $\gamma_{\pm}(-\kappa) = \gamma_{\mp}(\kappa)$, and $\beta_{\pm}(-\kappa) = \beta_{\mp}(\kappa)$ from pertinent constructions introduced so far for $\{k_{\pm}, \gamma_{\pm}, \beta_{\pm}\}$. Hence, Eqs. (6) and (7) provide a pair of symmetries: $|\mathbf{E}|^2(-\kappa) \equiv |\mathbf{E}|^2(\kappa)$ and $M_y(-\kappa) \equiv M_y(\kappa)$ [4].

More important from Eq. (7) is the anti-symmetry $M_x(-\kappa) \equiv -M_x(\kappa)$ in the longitudinal spin density, which we could legitimately call a 'chirality-induced (longitudinal) spin flip' [1,15,17]. A reversal in the chirality flux is also implied [5]. Notice additionally in Eq. (6) that $|\mathbf{E}|^2$ is not separable, whereas both $\{M_x, M_y\}$ in Eq. (7) are separable respectively into two factors. Spatial distributions of $\{|\mathbf{E}|^2, M_x, M_y\}$ in the $z$-direction are offered in See Secs. S6 and S7 of SD.

It is well-known that the achiral case on Fig. 1(a) gives rise to a single nonzero component $M_y$ as also proved in Sec. in S2 of SD. This is true since the nonzero components $\{E_z, E_x\}$ are out of phase by $\pm 90^o$. In our chiral case on Fig. 1(b), we encounter an additional nonzero longitudinal light-spin component $M_x$ due to medium chirality [14]. It turns out that $M_y(z) < 0$, whereas $M_x(z) > 0$ for a given $\kappa = \frac{3}{10} > 0$ [14]. The common component $M_y$ is induced by a depth-wise confinement [1].

Let $\int h(z) \equiv \int_{0+}^{\infty} h(z) dz$ denote a depth-wise integration over a chiral medium for $h(z) \in \mathbb{R}$ according to $\int_{0+}^{\infty} e^{-bz} dz = b^{-1}$. By this way, we formed both $\eta_x \equiv \omega \int M_x / \int |\mathbf{E}|^2$, being the longitudinal degree of circular polarization (DoCP), and $\eta_y \equiv \omega \int M_y / \int |\mathbf{E}|^2$, being the transverse DoCP [5]. It is obvious that $|\eta_x|, |\eta_y| \leq 1$ via the Cauchy-Schwarz inequality. Of course, $\eta_x^2 + \eta_y^2 \leq 1$ as well [5,14]. Besides, we have evaluated the ratio $\eta_y^x \equiv \int M_x / \int M_y$ that corresponds to the inverse pitch per turn in case with helical molecules [1,17].

Figure 3(a) displays three parameters $\{\eta_x, \eta_y, \eta_y^x\}$ against $k_x$. Since $\omega$ is uniformly increasing with $k_x$ [5], the vertical dotted line defined by $\eta_y^x = 1$ on Fig. 3(a) delineates the higher- and lower-frequency regions respectively to the right and left.

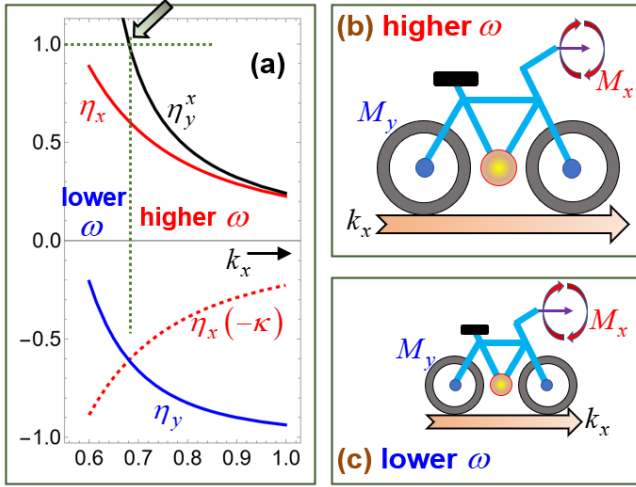

Fig. 3. (a) The ratios $\{\eta_x, \eta_y, \eta_y^x\}$ of the depth-wise integrals. (b) and (c) Cartoons respectively for the higher and lower frequencies. It is chosen that $\varepsilon_D = \mu_D = 1$ and $\kappa = \frac{3}{10}$.

On Fig. 3(a), we find contrasting trends between the magnitudes $\{|\eta_x|, |\eta_y|\}$. That is, the longitudinal $|\eta_x|$ decreases with increasing $\omega$, whereas $|\eta_y|$ increases with increasing $\omega$. The dotted red curve $\eta_x(-\kappa)$ is obtained with $\kappa = -\frac{3}{10}$, thus being anti-symmetric to the solid red curve $\eta_x$ obtained with $\kappa = +\frac{3}{10}$, as stated for Eq. (7). Additional evaluations with $\kappa = \frac{1}{10}$ are made on Fig. S9 of SD, which shows $\partial |\eta_x|/\partial |\kappa| > 0$ and $\partial |\eta_y|/\partial |\kappa| < 0$. We can thus confirm both energy redistributions and an enhanced light-spin polarization not with $|\eta_y|$ but with $|\eta_x|$ [17].

We illustrate the two differing situations on Fig. 3(a) by resorting to the caricatures on Figs. 3(b) and 3(c). There are bicycles, the larger on Fig. 3(b) and the smaller on Fig. 3(c). The size of each bicycle is proportional roughly to each of $\{k_x, \omega, M_y\}$, being respectively of fixed signs. The pinwheel on top of the driver's handle denotes $M_x$ with its sign undergoing flips depending on the environmental winds. The vertical dotted line in green on Fig. 3(a) for $\eta_y^x$ serves a sort of transition from the weaker to the stronger translational waves. We call the transition from Fig. 3(c) to Fig. 3(b) a 'baby-to-adult transition', where a slow and playful baby cyclist transits to a speedy and workaholic adult cyclist. See Fig. S9 and Sec. 7 of SD.

Due to lack of space, we will present elsewhere further accounts of the EM helicity $\omega^{-1}\mathrm{Im}(\boldsymbol{E} \cdot \boldsymbol{H}^*)$ and the Poynting vector $\omega^{-1}\mathrm{Re}(\boldsymbol{E} \times \boldsymbol{H}^*)$ along with other reactive parameters [6,7,9,13,15]. In practice, the imaginary parts of $\{\kappa, \varepsilon_M\}$ standing for respective losses are normally on the same order of magnitudes [5,6,10]. Those losses and/or dephasing could be handled in principle [17].

To conclude, we have proved the cut-offs of the lower frequencies on surface plasmon resonances via Fig. 2. This frequency filtering is associated with the reduced phase speed due to a nonzero medium chirality. Focusing on the chiral case, Fig. 3 corroborates the symmetry and anti-symmetry properties respectively of the transverse and longitudinal spin densities. The redistribution among one translational energy and two kinds of rotational energies are also illustrated. Our results would help us to better understand the plasmonic interactions between a metallic structure of more realistic geometry such as wires and spheres immersed in a surrounding medium containing chiral molecules [4].

**Funding.** National Research Foundation (NRF) of Republic of Korea under Grant NRF-2018R1D1A1B07045905.
C.G. acknowledges support by the European Union's Horizon 202 research and innovation programme under the Marie Sklodowska-Curie grant agreement No 883256.

**Disclosures.** The author declares no conflicts of interest.

**Data availability.** There is no underlying data set.

**Supplemental document**. See Supplement 1 for supporting content.

# Cut-offs and light-spin flips in surface plasmon resonance between a chiral medium and a metal: supplemental document

HYOUNG-IN LEE[1,*], CHRISTOPHER GAUL[2]

[1]Research Institute of Mathematics, Seoul National University, 599 Gwanak-Ro, Gwanak-Gu, Seoul, Republic of Korea 08826
[2]Fundación ICAMCYL, International Center for Advanced Materials and Raw Materials of Castilla y León, León, Spain
*Corresponding author: hileesam@naver.com
[2]Christopher Gaul: c.gaul@icamcyl.com

**Section S1. Dimensionless forms of governing equations**

We denote dimensionless parameters and variables by the overbar $\overline{\phantom{a}}$. Consider the Maxwell equations in dimensionless forms.

$$\begin{cases} \overline{\nabla} \times \overline{\mathbf{E}} = i\overline{\omega}\overline{\mathbf{B}} \\ \overline{\nabla} \times \overline{\mathbf{H}} = -i\overline{\omega}\overline{\mathbf{D}} \end{cases}, \quad \begin{cases} \overline{\nabla} \cdot \overline{\mathbf{D}} = 0 \\ \overline{\nabla} \cdot \overline{\mathbf{B}} = 0 \end{cases}, \quad \begin{cases} \overline{\mathbf{D}} = \overline{\varepsilon}\overline{\mathbf{E}} + i\overline{\kappa}\overline{\mathbf{H}} \\ \overline{\mathbf{B}} = \overline{\mu}\overline{\mathbf{H}} - i\overline{\kappa}\overline{\mathbf{E}} \end{cases}. \quad (S1.1)$$

Here, we have already employed the temporal oscillation factor $\exp(-i\overline{\omega}\overline{t})$ for all field variables. The pair $\{\overline{\varepsilon}, \overline{\mu}\}$ of dimensional electric permittivity and magnetic permeability is rewritten as $\{\overline{\varepsilon} \equiv \overline{\varepsilon}_0 \varepsilon, \overline{\mu} \equiv \overline{\mu}_0 \mu\}$. Here, $\{\overline{\varepsilon}_0, \overline{\mu}_0\}$ are the dimensional electric permittivity and magnetic permeability in vacuum. Hence, $\{\varepsilon, \mu\}$ are the dimensionless or relative electric permittivity and magnetic permeability. In addition, $\overline{\kappa}$ is the chirality parameter or medium chirality.

Let $\{\overline{\omega}_{ref}, \overline{t}_{ref} \equiv 1/\overline{\omega}_{ref}\}$ be the reference frequency and reference time, whence $\exp(-i\overline{\omega}\overline{t}) \equiv \exp(-i\omega t)$. Let us introduce the following set of reference parameters.

$$\overline{\omega}_{ref} \equiv \overline{\omega}_p, \quad \overline{t}_{ref} \equiv \frac{1}{\overline{\omega}_{ref}}, \quad \overline{c}_0 \equiv \frac{1}{\sqrt{\overline{\varepsilon}_0 \overline{\mu}_0}}, \quad \overline{Z}_0 \equiv \sqrt{\frac{\overline{\mu}_0}{\overline{\varepsilon}_0}}, \quad \overline{E}_{ref}, \quad \overline{H}_{ref} = \frac{\overline{E}_{ref}}{\overline{Z}_0}$$

$$\overline{L}_{ref} \equiv \overline{c}_0 \overline{t}_{ref} \equiv \frac{\overline{c}_0}{\overline{\omega}_{ref}} \equiv \frac{1}{\overline{k}_0}, \quad \overline{k}_0 \equiv \frac{\overline{\omega}_{ref}}{\overline{c}_0} \qquad (S1.2)$$

Let us further define the reference magnitude $\bar{E}_{ref}$ for the electric field. We stress that only $\{\bar{\omega}_{ref}, \bar{E}_{ref}\}$ are arbitrary reference parameters at our disposal. We normally choose $\bar{\omega}_{ref} = \bar{\omega}_p$ in dimensional form, namely, the plasma frequency is equal to the reference frequency. Therefore, $\omega_p = 1$ in dimensionless form.

Employing the above set of reference values, relevant dimensionless parameters and variables are defined below.

$$\omega \equiv \frac{\bar{\omega}}{\bar{\omega}_{ref}}, \quad t \equiv \frac{\bar{t}}{\bar{t}_{ref}}, \quad \nabla \equiv \frac{\bar{\nabla}}{\bar{k}_0}, \quad \varepsilon \equiv \frac{\bar{\varepsilon}}{\bar{\varepsilon}_0}, \quad \mu \equiv \frac{\bar{\mu}}{\bar{\mu}_0}, \quad , \quad \kappa \equiv \bar{c}_0 \bar{\kappa}$$
$$\mathbf{E} \equiv \frac{\bar{\mathbf{E}}}{\bar{E}_{ref}}, \quad \mathbf{H} \equiv \frac{\bar{\mathbf{H}}}{\bar{H}_{ref}}, \quad \mathbf{D} \equiv \frac{\bar{\mathbf{D}}}{\bar{\varepsilon}_0 \bar{E}_{ref}}, \quad \mathbf{B} \equiv \frac{\bar{c}_0 \bar{\mathbf{B}}}{\bar{E}_{ref}} \quad (S1.3)$$

Consequently, the Maxwell equations are cast into the dimensionless forms as follows.

$$\begin{cases} \nabla \times \mathbf{E} = i\omega \mathbf{B} \\ \nabla \times \mathbf{H} = -i\omega \mathbf{D} \end{cases}, \begin{cases} \nabla \cdot \mathbf{D} = 0 \\ \nabla \cdot \mathbf{B} = 0 \end{cases}, \begin{cases} \mathbf{D} = \varepsilon \mathbf{E} + i\kappa \mathbf{H} \\ \mathbf{B} = \mu \mathbf{H} - i\kappa \mathbf{E} \end{cases}. \quad (S1.4)$$

The fields and their respective curls are interrelated among themselves as follows.

$$\begin{cases} K = \begin{pmatrix} \kappa & i\mu \\ -i\varepsilon & \kappa \end{pmatrix}, \quad K^{-1} = \frac{1}{\kappa^2 - \varepsilon\mu} \begin{pmatrix} \kappa & -i\mu \\ i\varepsilon & \kappa \end{pmatrix} \\ \nabla \times \begin{Bmatrix} \mathbf{E} \\ \mathbf{H} \end{Bmatrix} = K \begin{Bmatrix} \mathbf{E} \\ \mathbf{H} \end{Bmatrix} \Rightarrow \begin{Bmatrix} \mathbf{E} \\ \mathbf{H} \end{Bmatrix} = K^{-1} \nabla \times \begin{Bmatrix} \mathbf{E} \\ \mathbf{H} \end{Bmatrix} \end{cases} \Rightarrow$$
$$\nabla \times \begin{Bmatrix} \mathbf{E} \\ \mathbf{H} \end{Bmatrix} = \begin{Bmatrix} \kappa \mathbf{E} + i\mu \mathbf{H} \\ -i\varepsilon \mathbf{E} + \kappa \mathbf{H} \end{Bmatrix} \quad \begin{Bmatrix} \mathbf{E} \\ \mathbf{H} \end{Bmatrix} = \frac{1}{\kappa^2 - \varepsilon\mu} \begin{Bmatrix} \kappa(\nabla \times \mathbf{E}) - i\mu(\nabla \times \mathbf{H}) \\ i\varepsilon(\nabla \times \mathbf{E}) + \kappa(\nabla \times \mathbf{H}) \end{Bmatrix} \quad (S1.4)$$

The last form indicates that the medium chirality is multiplying spatial gradients (curls) of fields so that it could serve as a curvature [S1].

**Section S2. Solutions to the achiral case**

This section provides a detailed derivation of the surface plasmon resonance across a planar metal-dielectric interface sitting between a loss-free dielectric and a loss-free metal both being in the semi-infinite space. Readers familiar with this textbook matter could skip the entirety of this section. Notwithstanding, this section is worth reading because of the sharp contrast with the results for the chiral case.

With $\{\hat{\mathbf{x}}, \hat{\mathbf{y}}, \hat{\mathbf{z}}\}$ as the Cartesian unit vectors, the $x$-ordinate is longitudinal, the $y$-ordinate is transverse, and the $z$-ordinate is depth-wise. In fact, the $z$-ordinate is also transverse, but it is called depth-wise to differentiate it from the $y$-ordinate.

Let us start with the following transverse-magnetic (TM) fields, while consulting Fig. 1(a) of Primary Document.

$$TM: \begin{cases} \begin{cases} \gamma_D > 0 \\ \gamma_M > 0 \end{cases}, \begin{cases} \Pi_D \equiv \exp(ik_x x - \gamma_D z) \\ \Pi_M \equiv \exp(ik_x x + \gamma_M z) \end{cases} \Rightarrow \begin{cases} \nabla \times \mathbf{E} = i\omega\mu\mathbf{H} \\ \nabla \times \mathbf{H} = -i\omega\varepsilon\mathbf{E} \end{cases} \Rightarrow \\ z > 0: \begin{cases} \mathbf{E}_D \equiv (E_{Dx}\hat{\mathbf{x}} + E_{Dz}\hat{\mathbf{z}})\exp(ik_x x - \gamma_D z) \\ \mathbf{H}_D \equiv H_{Dy}\hat{\mathbf{y}}\exp(ik_x x - \gamma_D z) \end{cases} \\ z < 0: \begin{cases} \mathbf{E}_M \equiv (E_{Mx}\hat{\mathbf{x}} + E_{Mz}\hat{\mathbf{z}})\exp(ik_x x + \gamma_M z) \\ \mathbf{H}_M \equiv H_{My}\hat{\mathbf{y}}\exp(ik_x x + \gamma_M z) \end{cases} \end{cases}. \quad (S2.1)$$

Here, the common temporal phase factor $\exp(-i\omega t)$ is assumed for all field variables. In this way, the depth-wise $z$-ordinate is associated with evanescent (or exponential) confinement.

Hence, both Faraday law $\nabla \times \mathbf{E} = i\omega\mu\mathbf{H}$ and the Ampère law $\nabla \times \mathbf{H} = -i\omega\varepsilon\mathbf{E}$ (both being alternatively called the 'curl equations') are written as follows.

$$\begin{cases} \nabla \times \mathbf{E}_D = \left(\frac{\partial E_{Dz}}{\partial y} - \frac{\partial E_{Dy}}{\partial z}\right)\hat{\mathbf{x}} + \left(\frac{\partial E_{Dx}}{\partial z} - \frac{\partial E_{Dz}}{\partial x}\right)\hat{\mathbf{y}} + \left(\frac{\partial E_{Dy}}{\partial x} - \frac{\partial E_{Dx}}{\partial y}\right)\hat{\mathbf{z}} \\ = \left(\frac{\partial E_{Dx}}{\partial z} - \frac{\partial E_{Dz}}{\partial x}\right)\hat{\mathbf{y}} = (-\gamma_D E_{Dx} - ik_x E_{Dz})\hat{\mathbf{y}} = i\omega\mu_D H_{Dy}\hat{\mathbf{y}} \\ \nabla \times \mathbf{E}_M = \left(\frac{\partial E_{Mx}}{\partial z} - \frac{\partial E_{Mz}}{\partial x}\right)\hat{\mathbf{y}} = (\gamma_M E_{Mx} - ik_x E_{Mz})\mathbf{y} = i\omega\mu_M H_{My}\hat{\mathbf{y}} \\ \nabla \times \mathbf{H}_D = \left(\frac{\partial H_{Dz}}{\partial y} - \frac{\partial H_{Dy}}{\partial z}\right)\hat{\mathbf{x}} + \left(\frac{\partial H_{Dx}}{\partial z} - \frac{\partial H_{Dz}}{\partial x}\right)\hat{\mathbf{y}} + \left(\frac{\partial H_{Dy}}{\partial x} - \frac{\partial H_{Dx}}{\partial y}\right)\hat{\mathbf{z}} \\ = -\frac{\partial H_{Dy}}{\partial z}\hat{\mathbf{x}} + \frac{\partial H_{Dy}}{\partial x}\hat{\mathbf{z}} = \gamma_D H_{Dy}\hat{\mathbf{x}} + ik_x H_{Dy}\mathbf{z} = -i\omega\varepsilon_D(E_{Dx}\mathbf{x} + E_{Dz}\mathbf{z}) \\ \nabla \times \mathbf{H}_M = -\frac{\partial H_{My}}{\partial z}\hat{\mathbf{x}} + \frac{\partial H_{My}}{\partial x}\hat{\mathbf{z}} = -\gamma_M H_{My}\hat{\mathbf{x}} + ik_x H_{My}\hat{\mathbf{z}} = i\omega|\varepsilon_M|(E_{Mx}\hat{\mathbf{x}} + E_{Mz}\hat{\mathbf{z}}) \end{cases}. \quad (S2.2)$$

Hereinafter, the subscripts $m \equiv M, D$ denotes respectively an achiral dielectric and a metal. For convenience, we set $\varepsilon_M \equiv -|\varepsilon_M|$ for a metal since we are taking the relative electric permittivity to be negative for a metal. Moreover, we made use of the fact that all field variables are independent of $y$-coordinate. This independence of the transverse $y$-direction will not be honored for the achiral case.

Collecting the final formulas, we find the following set of differential relations.

$$TM: \begin{cases} -\gamma_D E_{Dx} - ik_x E_{Dz} = i\omega\mu_D H_{Dy} \\ \gamma_M E_{Mx} - ik_x E_{Mz} = i\omega\mu_M H_{My} \end{cases}, \quad \begin{cases} \gamma_D H_{Dy} = -i\omega\varepsilon_D E_{Dx} \\ ik_x H_{Dy} = -i\omega\varepsilon_D E_{Dz} \\ -\gamma_M H_{My} = i\omega|\varepsilon_M| E_{Mx} \\ ik_x H_{My} = i\omega|\varepsilon_M| E_{Mz} \end{cases}$$

$$\Rightarrow TM: \begin{cases} z > 0: \begin{cases} \mathbf{E}_D \equiv \dfrac{H_{Dy}}{\omega\varepsilon_D}(i\gamma_D\hat{\mathbf{x}} - k_x\hat{\mathbf{z}})\Pi_D \\ \mathbf{H}_D \equiv H_{Dy}\hat{\mathbf{y}}\Pi_D \end{cases} \\ z < 0: \begin{cases} \mathbf{E}_M \equiv \dfrac{H_{My}}{\omega|\varepsilon_M|}(i\gamma_M\hat{\mathbf{x}} + k_x\hat{\mathbf{z}})\Pi_M \\ \mathbf{H}_M \equiv H_{My}\hat{\mathbf{y}}\Pi_M \end{cases} \end{cases}. \quad (S2.3)$$

Hence, all nonzero field components are expressed only in terms of the pair $\{H_{Dy}, H_{My}\}$, whence the name 'TM' arises. We have employed only the second set of four relations.

The two leading relations are processed to provide the solution to the Helmholtz equations for $\{H_{Dy}, H_{My}\}$ in the following fashion.

$$TM: \begin{cases} -\gamma_D\left(\dfrac{\gamma_D}{-i\omega\varepsilon_D}\right)H_{Dy} - ik_x\left(\dfrac{ik_x}{-i\omega\varepsilon_D}\right)H_{Dy} = i\omega\mu_D H_{Dy} \\ \gamma_M\left(\dfrac{-\gamma_M}{i\omega|\varepsilon_M|}\right)H_{My} - ik_x\left(\dfrac{ik_x}{i\omega|\varepsilon_M|}\right)H_{My} = i\omega\mu_M H_{My} \end{cases}$$

$$\Rightarrow \begin{cases} -\gamma_D^2 + k_x^2 = \omega^2\varepsilon_D\mu_D \\ -\gamma_M^2 + k_x^2 = -\omega^2|\varepsilon_M|\mu_M \end{cases} \Rightarrow \begin{cases} \gamma_D = \sqrt{k_x^2 - \omega^2\varepsilon_D\mu_D} \\ \gamma_M = \sqrt{k_x^2 + \omega^2|\varepsilon_M|\mu_M} \end{cases}. \quad (S2.4)$$

Therefore, we have determined the decay rates respectively in a dielectric and in a metal. Although the Helmholtz equations are derivable from both Faraday law $\nabla\times\mathbf{E} = i\omega\mu\mathbf{H}$ and the Ampère law $\nabla\times\mathbf{H} = -i\omega\varepsilon\mathbf{E}$, they are quite useful and convenient in deriving the decay rate.

Across the metal-dielectric interface, we then apply the continuity conditions $\{E_{Dx} = E_{Mx}, H_{Dy} = H_{My}\}$ for the four of the interfacial components.

$$\begin{cases} E_{Dx} = E_{Mx} \\ H_{Dy} = H_{My} \end{cases} \Rightarrow \hat{\mathbf{x}}: \dfrac{H_{Dy}}{\omega\varepsilon_D}i\gamma_D = \dfrac{H_{My}}{\omega|\varepsilon_M|}i\gamma_M \Rightarrow TM: \dfrac{\gamma_D}{\varepsilon_D} = \dfrac{\gamma_M}{|\varepsilon_M|}. \quad (S2.5)$$

It turns out that $H_{Dy} = H_{My}$ has already been incorporated in the above fields, whereas the remaining condition $E_{Dx} = E_{Mx}$ helps us to derive the desired dispersion relation.

Taking squares of both sides, we get the following expression for the wave number explicit in the frequency.

$$|\varepsilon_M|\gamma_D = \varepsilon_D \gamma_M \implies |\varepsilon_M|^2 \gamma_D^2 = \varepsilon_D^2 \gamma_M^2$$
$$\implies |\varepsilon_M|^2 \left(k_x^2 - \omega^2 \varepsilon_D \mu_D\right) = \varepsilon_D^2 \left(k_x^2 + \omega^2 |\varepsilon_M|\mu_M\right) \implies$$
$$\left(|\varepsilon_M|^2 - \varepsilon_D^2\right)k_x^2 = |\varepsilon_M|^2 \omega^2 \varepsilon_D \mu_D + \varepsilon_D^2 \omega^2 |\varepsilon_M|\mu_M \implies$$
$$k_x^2 = \omega^2 \frac{|\varepsilon_M|\varepsilon_D \left(|\varepsilon_M|\mu_D + \varepsilon_D \mu_M\right)}{|\varepsilon_M|^2 - \varepsilon_D^2}, \quad \begin{cases} k_x > 0 \\ \mu_D = \mu_M > 0 \end{cases} \quad . \quad \text{(S2.6)}$$
$$\implies k_x \equiv k_x(\omega) = \omega\sqrt{\varepsilon_D \mu_D}\sqrt{\frac{|\varepsilon_M|}{|\varepsilon_M| - \varepsilon_D}}$$
$$c_D \equiv \frac{1}{\sqrt{\varepsilon_D \mu_D}}, \quad k_D \equiv \frac{\omega}{c_D} = \omega\sqrt{\varepsilon_D \mu_D} \implies \frac{k_x}{k_D} = \sqrt{\frac{|\varepsilon_M|}{|\varepsilon_M| - \varepsilon_D}}$$

For concreteness, both frequency and wave number are assumed positive for concreteness, viz., $\omega, k_x > 0$. We introduced the additional assumption that $\mu_D = \mu_M$. By the way, we found it convenient to define the wave number squared $\omega^2 \varepsilon_D \mu_D \equiv k_D^2$ in the dielectric.

An analytic solution to $|\varepsilon_M|\gamma_D = \varepsilon_D \gamma_M$ can be found for a particular type of $\varepsilon_M$. To this, a loss-free Drude model is adopted with a simple choice of attendant parameters. We then proceed in the following way.

$$\begin{cases} k_x > 0 \\ \mu_D = \mu_M > 0 \end{cases}, \quad \varepsilon_M = \varepsilon_{fix}\left(1 - \frac{\omega_p^2}{\omega^2}\right), \quad \begin{cases} \varepsilon_{fix} = 1 \\ \omega_p = 1 \end{cases} \implies \varepsilon_M = 1 - \frac{1}{\omega^2} < 0$$
$$\omega > 0 \implies 0 < \omega < 1 \implies |\varepsilon_M| = \frac{1}{\omega^2} - 1 = \frac{1-\omega^2}{\omega^2} > 0 \implies \quad . \quad \text{(S2.7)}$$
$$\frac{k_x}{k_D} \equiv \frac{k_x}{\omega}\sqrt{\varepsilon_D \mu_D} = \sqrt{\frac{|\varepsilon_M|}{|\varepsilon_M| - \varepsilon_D}} = \sqrt{\frac{\frac{1-\omega^2}{\omega^2}}{\frac{1-\omega^2}{\omega^2} - \varepsilon_D}} = \sqrt{\frac{1-\omega^2}{1-(1+\varepsilon_D)\omega^2}}$$

The phase speed is thus always smaller than unity due the deceleration by plasmons in comparison to the phase speed vacuum. This deceleration in this achiral case will be further aggravated by an additional deceleration due to nonzero medium chirality.

We also found a couple of singularity conditions together with a boundedness condition.

$$\begin{cases} k_x = 0: \ \omega^2 = 1 \\ k_x = \infty: \ \omega^2 = \dfrac{1}{1+\varepsilon_D} \end{cases}$$

$$v_{ph} \equiv \frac{\omega}{k_x} = \frac{1}{\sqrt{\varepsilon_D \mu_D} \sqrt{\dfrac{1-\omega^2}{1-(1+\varepsilon_D)\omega^2}}} = \frac{1}{\sqrt{\varepsilon_D \mu_D}} \sqrt{\frac{1-(1+\varepsilon_D)\omega^2}{1-\omega^2}} \ . \tag{S2.8}$$

$$c_D \equiv \frac{1}{\sqrt{\varepsilon_D \mu_D}} \ \Rightarrow \ \frac{v_{ph}}{c_D} = \sqrt{\frac{1-(1+\varepsilon_D)\omega^2}{1-\omega^2}} < 1$$

Under the set $k_x > 0$, $\mu_D = \mu_M > 0$ of special conditions, let us consider a set of four Stokes parameters $\{I_{D0}, I_{D1}, I_{D2}, I_{D3}\}$. For convenience, we confine ourselves only to the dielectric.

$$k_D^2 \equiv \omega^2 \varepsilon_D \mu_D, \ k_x^2 = k_D^2 \frac{1-\omega^2}{1-(1+\varepsilon_D)\omega^2}$$

$$\mathbf{E}_D \equiv \frac{H_{Dy}}{\omega \varepsilon_D}(i\gamma_D \hat{\mathbf{x}} - k_x \hat{\mathbf{z}}) \Pi_D \equiv E_{Dx} \hat{\mathbf{x}} + E_{Dz} \hat{\mathbf{z}} \ \Rightarrow$$

$$I_{D0} \equiv |E_{Dz}|^2 + |E_{Dx}|^2 = \frac{k_x^2 + \gamma_D^2}{k_x^2} |E_{Dz}|^2 = \frac{2k_x^2 - k_D^2}{k_x^2} |E_{Dz}|^2 \tag{S2.9}$$

$$= \frac{2k_D^2 \dfrac{1-\omega^2}{1-(1+\varepsilon_D)\omega^2} - k_D^2}{k_D^2 \dfrac{1-\omega^2}{1-(1+\varepsilon_D)\omega^2}} |E_{Dz}|^2 = \frac{2(1-\omega^2) - [1-(1+\varepsilon_D)\omega^2]}{1-\omega^2} |E_{Dz}|^2$$

$$= \frac{2(1-\omega^2) - 1 + (1+\varepsilon_D)\omega^2}{1-\omega^2} |E_{Dz}|^2 = \frac{1+(\varepsilon_D - 1)\omega^2}{1-\omega^2} |E_{Dz}|^2 > 0$$

The zeroth Stokes parameter is hence always positive.

Furthermore, consider the first Stokes parameter, while recalling the defining relation of the decay rate $\gamma_D^2 = k_x^2 - k_D^2$ in an achiral dielectric.

$$S_{D1} \equiv |E_{Dz}|^2 - |E_{Dx}|^2 = \frac{k_x^2 - \gamma_D^2}{k_x^2}|E_{Dz}|^2 = \frac{k_D^2}{k_x^2}|E_{Dz}|^2 = \frac{k_D^2}{k_D^2 \dfrac{1-\omega^2}{1-(1+\varepsilon_D)\omega^2}}|E_{Dz}|^2$$

$$= \frac{k_D^2}{k_D^2 \dfrac{1-\omega^2}{1-(1+\varepsilon_D)\omega^2}}|E_{Dz}|^2 = \frac{1-(1+\varepsilon_D)\omega^2}{1-\omega^2}|E_{Dz}|^2 \Rightarrow \qquad \text{(S2.10)}$$

$$\begin{cases} S_{D1} > 0: & \omega^2 < (1+\varepsilon_D)^{-1} \\ S_{D1} = 0: & \omega^2 = (1+\varepsilon_D)^{-1} \\ S_{D1} \to -\infty: & \omega^2 \uparrow 1 \end{cases}$$

Therefore, the first Stokes parameter undergoes its sign flip at $\omega^2 = (1+\varepsilon_D)^{-1}$ from positive to negative as frequency is increased. Recall that we are working on the right-handed coordinates so that the cyclic orders $x \to y \to z \to x$ and $\hat{\mathbf{x}} \to \hat{\mathbf{y}} \to \hat{\mathbf{z}} \to \hat{\mathbf{x}}$ are employed.

Let us now form the second and third Stokes parameters.

$$\mathbf{E}_D \equiv \frac{H_{Dy}}{\omega \varepsilon_D}(i\gamma_D \hat{\mathbf{x}} - k_x \hat{\mathbf{z}})\Pi_D \equiv E_{Dx}\hat{\mathbf{x}} + E_{Dz}\hat{\mathbf{z}}, \quad k_x, \gamma_D > 0 \Rightarrow$$

$$E_{Dz}E_{Dx}^* = E_{Dz}\left(-\frac{i\gamma_D}{k_x}E_{Dz}\right)^* = \frac{i\gamma_D}{k_x}E_{Dz}E_{Dz}^* = \frac{i\gamma_D}{k_x}|E_{Dz}|^2$$

$$S_{D2} \equiv 2\operatorname{Re}(E_{Dz}E_{Dx}^*) = 2\operatorname{Re}\left(\frac{i\gamma_D}{k_x}|E_{Dz}|^2\right) = 0 \qquad \text{(S2.11)}$$

$$S_{D3} \equiv -2\operatorname{Im}(E_{Dz}E_{Dx}^*) = -2\operatorname{Im}\left(\frac{i\gamma_D}{k_x}|E_{Dz}|^2\right) = -\frac{2\gamma_D}{k_x}|E_{Dz}|^2 < 0$$

As a reference, we now from a circular (rotatory) basis to form the second and third Stokes parameters.

$$\begin{cases} \hat{\mathbf{l}} \equiv \tfrac{1}{\sqrt{2}}(\hat{\mathbf{z}}+i\hat{\mathbf{x}}) \\ \hat{\mathbf{r}} \equiv \tfrac{1}{\sqrt{2}}(\hat{\mathbf{z}}-i\hat{\mathbf{x}}) \end{cases} \Rightarrow \begin{cases} \hat{\mathbf{l}}+\hat{\mathbf{r}} = 2\tfrac{1}{\sqrt{2}}\hat{\mathbf{z}} = \sqrt{2}\hat{\mathbf{z}} \\ \hat{\mathbf{l}}-\hat{\mathbf{r}} = i2\tfrac{1}{\sqrt{2}}\hat{\mathbf{x}} = i\sqrt{2}\hat{\mathbf{x}} \end{cases} \Rightarrow \begin{cases} \hat{\mathbf{z}} = \tfrac{1}{\sqrt{2}}(\hat{\mathbf{l}}+\hat{\mathbf{r}}) \\ \hat{\mathbf{x}} = -i\tfrac{1}{\sqrt{2}}(\hat{\mathbf{l}}-\hat{\mathbf{r}}) \end{cases} \Rightarrow$$

$$\mathbf{E}_D \equiv \frac{H_{Dy}}{\omega \varepsilon_D}(i\gamma_D \hat{\mathbf{x}} - k_x \hat{\mathbf{z}})\Pi_D \equiv E_{Dx}\hat{\mathbf{x}} + E_{Dz}\hat{\mathbf{z}} \Rightarrow$$

$$\mathbf{E}_D \equiv E_{Dx}\hat{\mathbf{x}} + E_{Dz}\hat{\mathbf{z}} = -i\tfrac{1}{\sqrt{2}}(\hat{\mathbf{l}}-\hat{\mathbf{r}})E_{Dx} + E_{Dz}\tfrac{1}{\sqrt{2}}(\hat{\mathbf{l}}+\hat{\mathbf{r}}) \qquad \text{(S2.12)}$$

$$= \tfrac{1}{\sqrt{2}}(E_{Dz}-iE_{Dx})\hat{\mathbf{l}} + \tfrac{1}{\sqrt{2}}(E_{Dz}+iE_{Dx})\hat{\mathbf{r}} \equiv E_{Dl}\hat{\mathbf{l}} + E_{Dr}\hat{\mathbf{r}} \Rightarrow$$

$$\begin{cases} E_{Dl} = \tfrac{1}{\sqrt{2}}(E_{Dz}-iE_{Dx}) \\ E_{Dr} = \tfrac{1}{\sqrt{2}}(E_{Dz}+iE_{Dx}) \end{cases}$$

Let us check the third Stokes parameter as an exercise.

$$S_{D3} \equiv |E_{Dr}|^2 - |E_{Dl}|^2 = \left|\tfrac{1}{\sqrt{2}}(E_{Dz} + iE_{Dx})\right|^2 - \left|\tfrac{1}{\sqrt{2}}(E_{Dz} - iE_{Dx})\right|^2$$
$$= \tfrac{1}{2}(E_{Dz}^* - iE_{Dx}^*)(E_{Dz} + iE_{Dx}) - \tfrac{1}{2}(E_{Dz}^* + iE_{Dx}^*)(E_{Dz} - iE_{Dx})$$
$$= \tfrac{1}{2}\left[E_{Dz}^* iE_{Dx} - iE_{Dx}^* E_{Dz} - (iE_{Dx}^* E_{Dz} - E_{Dz}^* iE_{Dx})\right] = -iE_{Dx}^* E_{Dz} \quad \text{(S2.13)}$$
$$= -i\left(-\frac{i\gamma_D}{k_x}E_{Dz}\right)^* E_{Dz} = -i\frac{i\gamma_D}{k_x}E_{Dz}^* E_{Dz} = \frac{\gamma_D}{k_x}|E_{Dz}|^2 > 0$$

An opposite sign is obtained, while its magnitude is half the preceding value. Recall that the longitudinal direction is defined by the wave propagation direction. Because of the propagation factor $\Pi_D \equiv \exp(ik_x x - \gamma_D z)$, the $x$-direction is chosen to be the longitudinal direction. The longitudinal plane is then chosen to be the $zx$-plane since the evanescence is taking place in the $z$-direction, Correspondingly, the transverse plane is selected to be the $yz$-plane.

Overall, we ascertain the fact that the electric field $\mathbf{E}_D \equiv E_{Dx}\hat{\mathbf{x}} + E_{Dz}\hat{\mathbf{z}}$ is elliptically polarized on the longitudinal $zx$-plane. On the transverse $yz$-plane, we encounter a linearly polarized electric field since there is a sole nonzero component $E_{Dz}\hat{\mathbf{z}}$ on the $yz$-plane.

Consider the spin density $\operatorname{Im}(\mathbf{E}_D^* \times \mathbf{E}_D)$ in the dielectric.

$$\mathbf{E}_D \equiv \frac{H_{Dy}}{\omega\varepsilon_D}(i\gamma_D \hat{\mathbf{x}} - k_x \hat{\mathbf{z}})\Pi_D \equiv E_{Dx}\hat{\mathbf{x}} + E_{Dz}\hat{\mathbf{z}} \Rightarrow$$
$$\begin{cases}\operatorname{Im}(\mathbf{E}_D^* \times \mathbf{E}_D) = \operatorname{Im}(E_{Dz}E_{Dx}^*)\hat{\mathbf{y}} \\ S_{D3} \equiv -2\operatorname{Im}(E_{Dz}E_{Dx}^*)\end{cases} \Rightarrow \left[\operatorname{Im}(\mathbf{E}_D^* \times \mathbf{E}_D)\right]_y = -\tfrac{1}{2}S_{D3} \quad \text{(S2.14)}$$

Consequently, there exists only a transverse component of the spin density, which is related to the state of polarization. Unfortunately, we do not see a spin flip in the third Stokes parameter $S_{D3}$ in an achiral dielectric.

Meanwhile, let us take on the transverse-electric (TE) wave, which is not presented on Fig. 1(a) of Primary Document.

$$\begin{cases}\gamma_D > 0 \\ \gamma_M > 0\end{cases}, \begin{cases}\Pi_D \equiv \exp(ik_x x - \gamma_D z) \\ \Pi_M \equiv \exp(ik_x x + \gamma_M z)\end{cases} \Rightarrow \begin{cases}\nabla \times \mathbf{E} = i\omega\mu\mathbf{H} \\ \nabla \times \mathbf{H} = -i\omega\varepsilon\mathbf{E}\end{cases} \Rightarrow$$
$$TE : \begin{cases} z>0: \begin{cases}\mathbf{E}_D \equiv E_{Dy}\hat{\mathbf{y}}\exp(ik_x x - \gamma_{Dz}z) \\ \mathbf{H}_D \equiv (H_{Dx}\hat{\mathbf{x}} + H_{Dz}\hat{\mathbf{z}})\exp(ik_x x - \gamma_{Dz}z)\end{cases} \\ z<0: \begin{cases}\mathbf{E}_M \equiv E_{My}\hat{\mathbf{y}}\exp(ik_x x + \gamma_{Mz}z) \\ \mathbf{H}_M \equiv (H_{Mx}\hat{\mathbf{y}} + H_{Mz}\hat{\mathbf{z}})\exp(ik_x x + \gamma_{Mz}z)\end{cases}\end{cases} \quad \text{(S2.15)}$$

Here, the common temporal phase factor $\exp(-i\omega t)$ is assumed for all field variables. Hence, both Faraday law $\nabla \times \mathbf{E} = i\omega\mu\mathbf{H}$ and the Ampère law $\nabla \times \mathbf{H} = -i\omega\varepsilon\mathbf{E}$ (both being alternatively called the 'curl equations') are written as follows.

$$\begin{cases} \nabla \times \mathbf{E}_D = \left(\frac{\partial E_{Dz}}{\partial y} - \frac{\partial E_{Dy}}{\partial z}\right)\hat{\mathbf{x}} + \left(\frac{\partial E_{Dx}}{\partial z} - \frac{\partial E_{Dz}}{\partial x}\right)\hat{\mathbf{y}} + \left(\frac{\partial E_{Dy}}{\partial x} - \frac{\partial E_{Dx}}{\partial y}\right)\hat{\mathbf{z}} \\ = -\frac{\partial E_{Dy}}{\partial z}\hat{\mathbf{x}} + \frac{\partial E_{Dy}}{\partial x}\hat{\mathbf{z}} = \gamma_D E_{Dy}\hat{\mathbf{x}} + ik_x E_{Dy}\hat{\mathbf{z}} = i\omega\mu_D\left(H_{Dx}\hat{\mathbf{x}} + H_{Dz}\hat{\mathbf{z}}\right) \\ \nabla \times \mathbf{E}_M = -\frac{\partial E_{My}}{\partial z}\hat{\mathbf{x}} + \frac{\partial E_{My}}{\partial x}\hat{\mathbf{z}} = -\gamma_M E_{My}\hat{\mathbf{x}} + ik_x E_{My}\hat{\mathbf{z}} = i\omega\mu_M\left(H_{Mx}\hat{\mathbf{x}} + H_{Mz}\hat{\mathbf{z}}\right) \\ \nabla \times \mathbf{H}_D = \left(\frac{\partial H_{Dz}}{\partial y} - \frac{\partial H_{Dy}}{\partial z}\right)\hat{\mathbf{x}} + \left(\frac{\partial H_{Dx}}{\partial z} - \frac{\partial H_{Dz}}{\partial x}\right)\hat{\mathbf{y}} + \left(\frac{\partial H_{Dy}}{\partial x} - \frac{\partial H_{Dx}}{\partial y}\right)\hat{\mathbf{z}} \\ = \left(\frac{\partial H_{Dx}}{\partial z} - \frac{\partial H_{Dz}}{\partial x}\right)\hat{\mathbf{y}} = \left(-\gamma_D H_{Dx} - ik_x H_{Dz}\right)\hat{\mathbf{y}} = -i\omega\varepsilon_D E_{Dy}\hat{\mathbf{y}} \\ \nabla \times \mathbf{H}_M = \left(\frac{\partial H_{Mx}}{\partial z} - \frac{\partial H_{Mz}}{\partial x}\right)\hat{\mathbf{y}} = \left(\gamma_M H_{Mx} - ik_x H_{Mz}\right)\hat{\mathbf{y}} = i\omega|\varepsilon_M|E_{My}\hat{\mathbf{y}} \end{cases} \quad (S2.16)$$

Collecting the final formulas, we find the following.

$$TE: \begin{cases} \gamma_D E_{Dy} = i\omega\mu_D H_{Dx} \\ ik_x E_{Dy} = i\omega\mu_D H_{Dz} \\ -\gamma_M E_{My} = i\omega\mu_M H_{Mx} \\ ik_x E_{My} = i\omega\mu_M H_{Mz} \end{cases}, \quad \begin{cases} -\gamma_D H_{Dx} - ik_x H_{Dz} = -i\omega\varepsilon_D E_{Dy} \\ \gamma_M H_{Mx} - ik_x H_{Mz} = i\omega|\varepsilon_M|E_{My} \end{cases}$$

$$\Rightarrow TE: \begin{cases} z > 0: \begin{cases} \mathbf{E}_D \equiv E_{Dy}\hat{\mathbf{y}}\Pi_D \\ \mathbf{H}_D \equiv \dfrac{E_{Dy}}{\omega\mu_D}\left(-i\gamma_D\hat{\mathbf{x}} + k_x\hat{\mathbf{z}}\right)\Pi_D \end{cases} \\ z < 0: \begin{cases} \mathbf{E}_M \equiv E_{My}\hat{\mathbf{y}}\Pi_M \\ \mathbf{H}_M \equiv \dfrac{E_{My}}{\omega\mu_M}\left(i\gamma_M\hat{\mathbf{x}} + k_x\hat{\mathbf{z}}\right)\Pi_M \end{cases} \end{cases} \quad (S2.17)$$

Hence, all nonzero field components are expressed only in terms of the pair $\{E_{Dy}, E_{My}\}$, whence the name 'TE' arises. We have employed only the first set of four relations.

The two leading relations are processed to provide the solution to the Helmholtz equations for $\{E_{Dy}, E_{My}\}$ in the following fashion.

$$TE: \begin{cases} -\gamma_D \left(\dfrac{\gamma_D}{i\omega\mu_D}\right) E_{Dy} - ik_x \left(\dfrac{ik_x}{i\omega\mu_D}\right) E_{Dy} = -i\omega\varepsilon_D E_{Dy} \\ \gamma_M \left(\dfrac{-\gamma_M}{i\omega\mu_M}\right) E_{My} - ik_x \left(\dfrac{ik_x}{i\omega\mu_M}\right) E_{My} = i\omega|\varepsilon_M| E_{My} \end{cases} \quad (S2.18)$$

$$\Rightarrow \begin{cases} -\gamma_D^2 + k_x^2 = \omega^2 \varepsilon_D \mu_D \\ -\gamma_M^2 + k_x^2 = -\omega^2 |\varepsilon_M| \mu_M \end{cases} \Rightarrow \begin{cases} \gamma_D = \sqrt{k_x^2 - \omega^2 \varepsilon_D \mu_D} \\ \gamma_M = \sqrt{k_x^2 + \omega^2 |\varepsilon_M| \mu_M} \end{cases}$$

Therefore, we have determined the decay rates respectively in a dielectric and in a metal so that we obtained a pair of decay rates that is identical to that of the TM wave.

Across the metal-dielectric (M-D) interface, we then apply the continuity conditions $\{E_{Dx} = E_{Mx}, H_{Dy} = H_{My}\}$ for the four of the interfacial components.

$$\begin{cases} E_{Dx} = E_{Mx} \\ H_{Dy} = H_{My} \end{cases} \Rightarrow \hat{\mathbf{x}}: \dfrac{E_{Dy}}{\omega\mu_D}(-i\gamma_D) = \dfrac{E_{My}}{\omega\mu_M} i\gamma_M \Rightarrow$$

$$TE: \dfrac{\gamma_D}{\mu_D} + \dfrac{\gamma_M}{\mu_M} = 0, \quad 0 < \dfrac{\mu_D}{\mu_M} = -\dfrac{\gamma_D}{\gamma_M} < 0 \quad (S2.19)$$

It turns out that $E_{Dy} = E_{My}$ has already been incorporated in the above fields, whereas the remaining condition $H_{Dx} = H_{Mx}$ helps us to derive the desired dispersion relation. Consequently, there is no resonance for the TE wave. In other words, we learn two contrasting sign behaviors between the TM and TE waves.

**Section S3. Circular vectors and the electromagnetic field vectors in achiral medium**

In finding proper solutions to $\mathbf{Q}_\pm$ in a chiral medium, it is a direct way to assume the following form.

$$\begin{aligned} \Pi_\pm &\equiv \exp(ik_x x - \gamma_\pm z - i\omega t), \quad |f_{\pm x}|^2 + |f_{\pm y}|^2 + |f_{\pm z}|^2 = 1 \\ \mathbf{Q}_\pm &= Q_\pm \left(f_{\pm x}\hat{\mathbf{x}} + f_{\pm y}\hat{\mathbf{y}} + f_{\pm z}\hat{\mathbf{z}}\right) \Pi_\pm \end{aligned} \Rightarrow \quad (S3.1)$$

Here, the complex magnitude of the circular vectors $\mathbf{Q}_\pm$ are denoted by $Q_\pm$, which are unknown but will be determined later by applying appropriate interface conditions. The present pair of subscripts $\{+,-\}$ replaces the conventional pair of $\{L,R\}$, where a 'left' and a 'right' waves are respectively implied. This pair $\{+,-\}$ of notations turns out to greatly facilitate our ensuing formulas.

With a normalization condition $\left|f_{\pm x}\right|^2 + \left|f_{\pm y}\right|^2 + \left|f_{\pm z}\right|^2 = 1$, the polarization vectors are hence given by $f_{\pm x}\hat{\mathbf{x}} + f_{\pm y}\hat{\mathbf{y}} + f_{\pm z}\hat{\mathbf{z}}$. The common propagation factor $\Pi_{\pm} \equiv \exp(ik_x x - \gamma_{\pm} z - i\omega t)$ incorporates spatial dependences. Instead, $\{f_{\pm x}, f_{\pm y}, f_{\pm z}\}$ are assumed constant, thereby signifying spatially quasi-homogeneous waves. This is a direct (or inductive) approach that we have done. By this way, we have obtained the following solutions.

$$\gamma_{\pm} \equiv \sqrt{k_x^2 - \omega^2\left(\sqrt{\varepsilon\mu} \pm \kappa\right)^2} \equiv \sqrt{k_x^2 - k_{\pm}^2}, \quad k_x^2 \equiv \gamma_{\pm}^2 + k_{\pm}^2$$

$$\mathbf{Q}_{\pm} = Q_{\pm} \tfrac{1}{\sqrt{2}} k_x^{-1}\left(\gamma_{\pm}\hat{\mathbf{x}} \pm k_{\pm}\hat{\mathbf{y}} + ik_x\hat{\mathbf{z}}\right)\Pi_{\pm} = Q_{\pm} \tfrac{1}{\sqrt{2}}\left(\frac{\gamma_{\pm}}{k_x}\hat{\mathbf{x}} \pm \frac{k_{\pm}}{k_x}\hat{\mathbf{y}} + i\hat{\mathbf{z}}\right)\Pi_{\pm}$$

(S3.2)

The circular vectors $\mathbf{Q}_{\pm}$ refer to a pair of elliptically polarized fields on the transverse $yz$-plane due to the portion in consideration of the portion $(k_{\pm}\mathbf{y} \pm ik_x\mathbf{z})$ [S5]. In addition, both left and right waves are related to each other by various additional parameters given below.

$$\begin{cases} c_D \equiv \dfrac{1}{\sqrt{\varepsilon_D \mu_D}}, \quad c_{\pm} \equiv \dfrac{1}{\sqrt{\varepsilon_D \mu_D} \pm \kappa} \equiv \dfrac{c_D}{1 \pm c_D \kappa}, \quad \varepsilon\mu - \kappa^2 \equiv \dfrac{1}{c_+ c_-} \\ c_{avg}^{-+} \equiv \tfrac{1}{2}(c_- + c_+) = \dfrac{1}{c_D}\dfrac{1}{\varepsilon_D \mu_D - \kappa^2} = \dfrac{c_+ c_-}{c_D}, \quad \Delta_{-+} \equiv \tfrac{1}{2}(c_- - c_+) = \dfrac{\kappa}{\varepsilon_D \mu_D - \kappa^2} \end{cases}$$

(S3.3)

Here, $\{\varepsilon_D, \mu_D\}$ are the relative parameters for a bulk embedding dielectric (medium), where chiral molecules are thought to be uniformly dispersed. Furthermore, various wave numbers are defined below.

$$k_D \equiv \dfrac{\omega}{c_D} \equiv \omega\sqrt{\varepsilon_D \mu_D}, \quad k_{\pm} \equiv \dfrac{\omega}{c_{\pm}} \equiv \omega\left(\sqrt{\varepsilon_D \mu_D} \pm \kappa\right)$$

$$\begin{cases} c_{\pm} > 0 \\ k_{\pm} > 0 \end{cases} \Leftrightarrow \begin{cases} |c_D \kappa| < 1 \\ |\kappa| < \sqrt{\varepsilon_D \mu_D} \end{cases}$$

$$\begin{cases} k_+ + k_- = \omega\left(\sqrt{\varepsilon_D \mu_D} + \kappa\right) + \omega\left(\sqrt{\varepsilon_D \mu_D} - \kappa\right) = 2\omega\sqrt{\varepsilon_D \mu_D} = 2k_D \\ k_+ - k_- = \omega\left(\sqrt{\varepsilon_D \mu_D} + \kappa\right) - \omega\left(\sqrt{\varepsilon_D \mu_D} - \kappa\right) = 2\omega\kappa \\ k_+ k_- = \omega\left(\sqrt{\varepsilon_D \mu_D} + \kappa\right)\omega\left(\sqrt{\varepsilon_D \mu_D} - \kappa\right) = \omega^2\left(\varepsilon_D \mu_D - \kappa^2\right) \end{cases}$$

(S3.4)

We then take an inverse (or deductive) approach by verifying that this pair in Eq. (S3.2) of solutions satisfies the two conditions handled in the preceding section. Firstly,

$$\mathbf{U}_{\pm} \equiv \left(\gamma_{\pm}\hat{\mathbf{x}} \pm k_{\pm}\hat{\mathbf{y}} + ik_x\hat{\mathbf{z}}\right)\Pi_{\pm} \equiv \left(\gamma_{\pm}\hat{\mathbf{x}} \pm k_{\pm}\hat{\mathbf{y}} + ik_x\hat{\mathbf{z}}\right)\exp(ik_x x - \gamma_{\pm} z - i\omega t)$$

$$\nabla \cdot \mathbf{Q}_{\pm} \Rightarrow \nabla \cdot \left[\left(\gamma_{\pm}\hat{\mathbf{x}} \pm k_{\pm}\hat{\mathbf{y}} + ik_x\hat{\mathbf{z}}\right)\Pi_{\pm}\right] = ik_x \gamma_{\pm} - \gamma_{\pm}ik_x = 0$$

(S3.5)

The divergence-less property is hence trivially satisfied.

Secondly, consider the curl equations.

$$\nabla \times \mathbf{Q}_{\pm} \Rightarrow$$

$$\nabla \times \mathbf{U}_{\pm} = \left(\frac{\partial U_{\pm z}}{\partial y} - \frac{\partial U_{\pm y}}{\partial z}\right)\hat{\mathbf{x}} + \left(\frac{\partial U_{\pm x}}{\partial z} - \frac{\partial U_{\pm z}}{\partial x}\right)\hat{\mathbf{y}} + \left(\frac{\partial U_{\pm y}}{\partial x} - \frac{\partial U_{\pm x}}{\partial y}\right)\hat{\mathbf{z}}$$

$$= -\frac{\partial U_{\pm y}}{\partial z}\hat{\mathbf{x}} + \left(\frac{\partial U_{\pm x}}{\partial z} - \frac{\partial U_{\pm z}}{\partial x}\right)\hat{\mathbf{y}} + \frac{\partial U_{\pm y}}{\partial x}\hat{\mathbf{z}}$$

$$\frac{\nabla \times \mathbf{U}_{\pm}}{\Pi_{\pm}} = -(-\gamma_{\pm})(\pm k_{\pm})\hat{\mathbf{x}} + \left[(-\gamma_{\pm})\gamma_{\pm} - ik_x ik_x\right]\hat{\mathbf{y}} + ik_x(\pm k_{\pm})\hat{\mathbf{z}} \quad . \quad (S3.6)$$

$$= \pm\gamma_{\pm}k_{\pm}\hat{\mathbf{x}} + \left(-\gamma_{\pm}^2 + k_x^2\right)\hat{\mathbf{y}} \pm ik_{\pm}k_x\hat{\mathbf{z}} = \pm\gamma_{\pm}k_{\pm}\hat{\mathbf{x}} + k_{\pm}^2\hat{\mathbf{y}} \pm ik_{\pm}k_x\hat{\mathbf{z}}$$

$$= \pm k_{\pm}\left(\gamma_{\pm}\hat{\mathbf{x}} \pm k_{\pm}\hat{\mathbf{y}} + ik_x\hat{\mathbf{z}}\right) \equiv \pm k_{\pm}\frac{\mathbf{U}_{\pm}}{\Pi_{\pm}}$$

A specialty with $\nabla \times \mathbf{Q}_{\pm} = \pm k_{\pm}\mathbf{Q}_{\pm}$ lies in the sign reversal with the right wave. Of course, each component satisfies the scalar Helmholtz equation, thereby leading to the identical formula for the decay rate $k_x^2 \equiv \gamma_{\pm}^2 + k_{\pm}^2$. From physical perspectives, the translational wave number is greater than either of the decay rates, viz., $k_x^2 \geq \gamma_{\pm}^2$.

The EM fields in the chiral dielectric are thus found below.

$$\mathbf{E} = \mathbf{Q}_+ - iZ_D\mathbf{Q}_- = \tfrac{1}{\sqrt{2}}\frac{1}{k_x}\left[Q_+\left(\gamma_+\hat{\mathbf{x}} + k_+\hat{\mathbf{y}} + ik_x\hat{\mathbf{z}}\right)\Pi_+ - P_-\left(\gamma_-\hat{\mathbf{x}} - k_-\hat{\mathbf{y}} + ik_x\hat{\mathbf{z}}\right)\Pi_-\right]$$

$$\mathbf{H} = -\frac{i}{Z_D}\mathbf{Q}_+ + \mathbf{Q}_- = \left(-iZ_D^{-1}\right)(\mathbf{Q}_+ + \mathbf{P}_-) \quad . \quad (S3.7)$$

$$= \tfrac{1}{\sqrt{2}}\left(-iZ_D^{-1}\right)\frac{1}{k_x}\left[Q_+\left(\gamma_+\hat{\mathbf{x}} + k_+\hat{\mathbf{y}} + ik_x\mathbf{z}\right)\Pi_+ + P_-\left(\gamma_-\hat{\mathbf{x}} - k_-\hat{\mathbf{y}} + ik_x\mathbf{z}\right)\hat{\mathbf{z}}\Pi_-\right]$$

Here, we found it convenient to introduce $\{P_+ \equiv Q_+, P_- \equiv iZQ_-\}$ instead of $\{Q_+, Q_-\}$ for convenience, thus signifying a phase difference of $\pm 90°$ between the pair $\{Q_+, Q_-\}$ of two circular vectors. Correspondingly, we have $\{\mathbf{P}_+ \equiv \mathbf{Q}_+, \mathbf{P}_- \equiv iZ\mathbf{Q}_-\}$.

The preceding fields can be alternatively cast into the following concise forms.

$$\begin{cases} P_+ \equiv Q_+ \\ P_- \equiv iZQ_- \end{cases} \Rightarrow \begin{cases} \mathbf{P}_+ \equiv \mathbf{Q}_+ \\ \mathbf{P}_- \equiv iZ\mathbf{Q}_- \end{cases}, \begin{cases} \mathbf{E} = \mathbf{P}_+ - \mathbf{P}_- \\ \mathbf{H} = -iZ_D^{-1}(\mathbf{P}_+ + \mathbf{P}_-) \end{cases}$$

$$\begin{cases} \tilde{E}_x^{+-} \equiv \gamma_+Q_+\Pi_+ - \gamma_-P_-\Pi_- \\ \tilde{H}_x^{+-} \equiv \gamma_+Q_+\Pi_+ + \gamma_-P_-\Pi_- \end{cases}, \begin{cases} \tilde{E}_y^{+-} \equiv k_+Q_+\Pi_+ + k_-P_-\Pi_- \\ \tilde{H}_y^{+-} \equiv k_+Q_+\Pi_+ - k_-P_-\Pi_- \end{cases} \quad . \quad (S3.8)$$

$$\begin{cases} \tilde{E}_z^{+-} \equiv ik_x(Q_+\Pi_+ - P_-\Pi_-) \\ \tilde{H}_z^{+-} \equiv ik_x(Q_+\Pi_+ + P_-\Pi_-) \end{cases} \Rightarrow \begin{cases} \sqrt{2}k_x\mathbf{E} = \tilde{E}_x^{+-}\hat{\mathbf{x}} + \tilde{E}_y^{+-}\hat{\mathbf{y}} + \tilde{E}_z^{+-}\hat{\mathbf{z}} \\ \sqrt{2}k_x iZ_D\mathbf{H} = \tilde{H}_x^{+-}\hat{\mathbf{x}} + \tilde{H}_y^{+-}\hat{\mathbf{y}} + \tilde{H}_z^{+-}\hat{\mathbf{z}} \end{cases}$$

More concise summation forms are found below.

$$\begin{cases} P_+ \equiv Q_+ \\ P_- \equiv iZQ_- \end{cases}, \quad \begin{cases} \sqrt{2}k_x\mathbf{E} = \sum_{\sigma=\pm}\left(\sigma\gamma_\sigma\hat{\mathbf{x}} + k_\sigma\hat{\mathbf{y}} + \sigma ik_x\hat{\mathbf{z}}\right)P_\sigma\Pi_\sigma \\ iZ_D\sqrt{2}k_x\mathbf{H} = \sum_{\sigma=\pm}\left(\gamma_\sigma\hat{\mathbf{x}} + \sigma k_\sigma\hat{\mathbf{y}} + ik_x\hat{\mathbf{z}}\right)P_\sigma\Pi_\sigma \end{cases}. \quad (S3.9)$$

Here, the summation index $\sigma$ is reminiscent of signed photons of TM and TE natures. The existence of all three field components in $\mathbf{Q}_\pm$ indicates the fact that both TM and TE waves are interwoven. Obtaining the above solutions is a tedious, yet straightforward process. In retrospect, this step is a crucial one among all the steps taken in this study. Besides, $\sigma = \pm \equiv \pm 1$ smacks of fermions, whereas the forced assignment $\sigma = 0$ smells of bosons. We can take two hypothetical limits as $|\sigma| \to 0, \infty$ instead of $\sigma = \pm 1$. In other words, forcefully setting either of $\sigma = 0, \infty$ corresponds to the separation of a TM wave from a TE wave.

$$\sigma \to 0 \quad \Rightarrow \quad \begin{cases} \sqrt{2}k_x\mathbf{E} = (k_\sigma\hat{\mathbf{y}})Q_\sigma\Pi_\sigma \\ \sqrt{2}k_x iZ_D\mathbf{H} = (\gamma_\sigma\hat{\mathbf{x}} + ik_x\hat{\mathbf{z}})Q_\sigma\Pi_\sigma \end{cases} : \quad TE-like$$

$$\sigma \to \pm\infty \quad \Rightarrow \quad \begin{cases} \sqrt{2}k_x\mathbf{E} = (\gamma_\sigma\hat{\mathbf{x}} + ik_x\hat{\mathbf{z}})Q_\sigma\Pi_\sigma \\ \sqrt{2}k_x iZ_D\mathbf{H} = (k_\sigma\hat{\mathbf{y}})Q_\sigma\Pi_\sigma \end{cases} : \quad TM-like$$

(S3.10)

Only under such imaginary circumstances, we recover the TM and TE waves, respectively.

What we call the circular vectors $\mathbf{Q}_\pm$ satisfy the two equations $\nabla\cdot\mathbf{Q}_\pm = 0$ and $\nabla\times\mathbf{Q}_\pm = \pm k_\pm\mathbf{Q}_\pm$. The familiar Helmholtz equation $\nabla^2\mathbf{Q}_\pm + k_\pm^2\mathbf{Q}_\pm = \mathbf{0}$ is just derivable from these fundamental pair in the following manner.

$$\begin{cases} \nabla\cdot\mathbf{Q}_\pm = 0 \\ \nabla\times\mathbf{Q}_\pm = \pm k_\pm\mathbf{Q}_\pm \end{cases}, \quad \begin{cases} \mathbf{P}_+ \equiv \mathbf{Q}_+ \\ \mathbf{P}_- \equiv iZ\mathbf{Q}_- \end{cases} \Rightarrow \begin{cases} \nabla\cdot\mathbf{P}_\pm = 0 \\ \nabla\times\mathbf{P}_\pm = \pm k_\pm\mathbf{P}_\pm \end{cases} \Rightarrow$$

$$\begin{cases} \nabla\times(\nabla\times\mathbf{Q}_\pm) = \pm k_\pm(\nabla\times\mathbf{Q}_\pm) = \pm k_\pm(\pm k_\pm\mathbf{Q}_\pm) = k_\pm^2\mathbf{Q}_\pm \\ \nabla\times(\nabla\times\mathbf{Q}_\pm) = \nabla(\nabla\cdot\mathbf{Q}_\pm) - \nabla^2\mathbf{Q}_\pm = -\nabla^2\mathbf{Q}_\pm \end{cases} \Rightarrow \quad . \quad (S3.11)$$

$$\nabla^2\mathbf{Q}_\pm + k_\pm^2\mathbf{Q}_\pm = \mathbf{0}$$

The pair $\{k_+, k_-\}$ of wave numbers are defined previously in Eq. (S3.4). A true nature of the circular vector $\{\mathbf{Q}_+, \mathbf{Q}_-\}$ is the separability property or diagonalized nature. Notwithstanding, $\{\mathbf{Q}_+, \mathbf{Q}_-\}$ are neither parallel nor perpendicular to each other. In other words, $\{\mathbf{Q}_+, \mathbf{Q}_-\}$ are obliquely directed in general.

Let us show deductively further consequences of the above two properties. To this goal, let $\mathbf{Q}_\pm$ be related to the electromagnetic fields $\{\mathbf{E}, \mathbf{H}\}$ in the following way.

$$A \equiv \begin{pmatrix} 1 & -iZ_D \\ -iZ_D^{-1} & 1 \end{pmatrix}, \quad A^{-1} \equiv \tfrac{1}{2}\begin{pmatrix} 1 & iZ_D \\ iZ_D^{-1} & 1 \end{pmatrix} \Rightarrow$$

$$\begin{Bmatrix} \mathbf{E} \\ \mathbf{H} \end{Bmatrix} = A \begin{Bmatrix} \mathbf{Q}_+ \\ \mathbf{Q}_- \end{Bmatrix} \Rightarrow \begin{cases} \mathbf{E} = \mathbf{Q}_+ - iZ_D \mathbf{Q}_- \\ \mathbf{H} = -iZ_D^{-1} \mathbf{Q}_+ + \mathbf{Q}_- \end{cases}, \quad \begin{Bmatrix} \mathbf{Q}_+ \\ \mathbf{Q}_- \end{Bmatrix} = \tfrac{1}{2}\begin{pmatrix} 1 & iZ_D \\ iZ_D^{-1} & 1 \end{pmatrix}\begin{Bmatrix} \mathbf{E} \\ \mathbf{H} \end{Bmatrix}. \quad (S3.12)$$

Let us show deductively above two properties. Firstly,

$$\begin{cases} \mathbf{E} = \mathbf{Q}_+ - iZ_D \mathbf{Q}_- \\ \mathbf{H} = -iZ_D^{-1} \mathbf{Q}_+ + \mathbf{Q}_- \end{cases} \Rightarrow \nabla \cdot \begin{Bmatrix} \mathbf{E} \\ \mathbf{H} \end{Bmatrix} = \begin{Bmatrix} 0 \\ 0 \end{Bmatrix}. \quad (S3.13)$$

Secondly, we list several key properties related to the circular vectors.

$$\begin{cases} k_\pm \equiv \omega\left(\sqrt{\varepsilon_D \mu_D} \pm \kappa\right) \\ k_D \equiv \omega\sqrt{\varepsilon_D \mu_D} \equiv \dfrac{\omega}{c_D} \end{cases} \Rightarrow \begin{cases} k_+ + k_- = \omega\left(\sqrt{\varepsilon_D \mu_D} + \kappa + \sqrt{\varepsilon_D \mu_D} - \kappa\right) = 2k_D \\ k_+ - k_- = \omega\left(\sqrt{\varepsilon_D \mu_D} + \kappa - \sqrt{\varepsilon_D \mu_D} + \kappa\right) = 2\omega\kappa \end{cases}$$

$$\nabla \times \begin{Bmatrix} \mathbf{E} \\ \mathbf{H} \end{Bmatrix} = A \nabla \times \begin{Bmatrix} \mathbf{Q}_+ \\ \mathbf{Q}_- \end{Bmatrix} = \begin{pmatrix} 1 & -iZ_D \\ -iZ_D^{-1} & 1 \end{pmatrix}\begin{Bmatrix} k_+ \mathbf{Q}_+ \\ -k_- \mathbf{Q}_- \end{Bmatrix} = \begin{pmatrix} k_+ & iZ_D k_- \\ -iZ_D^{-1} k_+ & -k_- \end{pmatrix}\begin{Bmatrix} \mathbf{Q}_+ \\ \mathbf{Q}_- \end{Bmatrix}. \quad (S3.14)$$

$$= \begin{pmatrix} k_+ & iZ_D k_- \\ -iZ_D^{-1} k_+ & -k_- \end{pmatrix} \tfrac{1}{2} \begin{pmatrix} 1 & iZ_D \\ iZ_D^{-1} & 1 \end{pmatrix}\begin{Bmatrix} \mathbf{E} \\ \mathbf{H} \end{Bmatrix}$$

$$= \tfrac{1}{2}\begin{pmatrix} k_+ - k_- & iZ_D k_+ + iZ_D k_+ \\ -iZ_D^{-1} k_+ - iZ_D^{-1} k_- & k_+ - k_- \end{pmatrix}\begin{Bmatrix} \mathbf{E} \\ \mathbf{H} \end{Bmatrix} = \begin{pmatrix} \omega\kappa & iZ_D k_D \\ -iZ_D^{-1} k_D & \omega\kappa \end{pmatrix}\begin{Bmatrix} \mathbf{E} \\ \mathbf{H} \end{Bmatrix}$$

Here, we see that $\{k_+ + k_-, k_+ - k_-\}$ are respectively even and odd respect to the medium chirality $\kappa$ when everything else is held fixed. Therefore, we end up with the following pair.

$$\nabla \times \mathbf{E} = \omega\kappa \mathbf{E} + iZ_D k_D \mathbf{H} = \omega\kappa \mathbf{E} + i\sqrt{\dfrac{\mu_D}{\varepsilon_D}} \omega\sqrt{\varepsilon_D \mu_D} \mathbf{H} = i\omega\left(\mu_D \mathbf{H} - i\kappa \mathbf{E}\right) = i\omega \mathbf{B}$$

$$\nabla \times \mathbf{H} = -iZ_D^{-1} k_D \mathbf{E} + \omega\kappa \mathbf{H} = -i\sqrt{\dfrac{\varepsilon_D}{\mu_D}} \omega\sqrt{\varepsilon_D \mu_D} \mathbf{E} + \omega\kappa \mathbf{H} \quad . \quad (S3.15)$$

$$= -i\omega\left(\varepsilon_D \mathbf{E} + i\kappa \mathbf{H}\right) = -i\omega \mathbf{D} \Rightarrow \begin{cases} \mathbf{D} = \varepsilon_D \mathbf{E} + i\kappa \mathbf{H} \\ \mathbf{B} = \mu_D \mathbf{H} - i\kappa \mathbf{E} \end{cases}, \begin{cases} \nabla \times \mathbf{E} = i\omega \mathbf{B} \\ \nabla \times \mathbf{H} = -i\omega \mathbf{D} \end{cases}$$

Therefore, the Maxwell equations are successfully recovered as desired. Consider the Helmholtz equations in $\{\mathbf{E}, \mathbf{H}\}$, which take obviously the same forms as the circular vectors $\mathbf{Q}_\pm$.

$$\nabla^2 \mathbf{Q}_\pm + k_\pm^2 \mathbf{Q}_\pm = \mathbf{0} \Rightarrow \nabla^2 \begin{Bmatrix} \mathbf{Q}_+ \\ \mathbf{Q}_- \end{Bmatrix} + k_\pm^2 \begin{Bmatrix} \mathbf{Q}_+ \\ \mathbf{Q}_- \end{Bmatrix} = \mathbf{0} \Rightarrow$$

$$\nabla^2 A \begin{Bmatrix} \mathbf{Q}_+ \\ \mathbf{Q}_- \end{Bmatrix} + k_\pm^2 A \begin{Bmatrix} \mathbf{Q}_+ \\ \mathbf{Q}_- \end{Bmatrix} = \mathbf{0} \quad . \quad (S3.16)$$

$$\begin{Bmatrix} \mathbf{E} \\ \mathbf{H} \end{Bmatrix} = A \begin{Bmatrix} \mathbf{Q}_+ \\ \mathbf{Q}_- \end{Bmatrix} \Rightarrow \nabla^2 \begin{Bmatrix} \mathbf{E} \\ \mathbf{H} \end{Bmatrix} + k_\pm^2 \begin{Bmatrix} \mathbf{E} \\ \mathbf{H} \end{Bmatrix} = \mathbf{0} \Rightarrow \nabla^2 \begin{Bmatrix} \mathbf{E}_\pm \\ \mathbf{H}_\pm \end{Bmatrix} + k_\pm^2 \begin{Bmatrix} \mathbf{E}_\pm \\ \mathbf{H}_\pm \end{Bmatrix} = \mathbf{0}$$

This last pair should be understood to mean one pair $\nabla^2 \mathbf{E}_\pm + k_\pm^2 \mathbf{E}_\pm = \mathbf{0}$ and another pair $\nabla^2 \mathbf{H}_\pm + k_\pm^2 \mathbf{H}_\pm = \mathbf{0}$.

It is instructive to find the Helmholtz equations satisfied by $\{\mathbf{E}, \mathbf{H}\}$ without the appearance of $k_\pm$ [S2]. To this end, we proceed as follows.

$$Z_D \equiv \sqrt{\frac{\mu}{\varepsilon}}, \quad \begin{cases} \mathbf{E} = \mathbf{Q}_+ - iZ_D\mathbf{Q}_- \\ \mathbf{H} = -iZ_D^{-1}\mathbf{Q}_+ + \mathbf{Q}_- \end{cases} \Rightarrow \begin{cases} \mathbf{E} = \mathbf{Q}_+ - iZ_D\mathbf{Q}_- \\ -iZ_D\mathbf{H} = \mathbf{Q}_+ + iZ_D\mathbf{Q}_- \end{cases}.$$

$$\begin{cases} P_+ \equiv Q_+ \\ P_- \equiv iZQ_- \end{cases} \Rightarrow \begin{cases} \mathbf{P}_+ \equiv \mathbf{Q}_+ \\ \mathbf{P}_- \equiv iZ\mathbf{Q}_- \end{cases} \Rightarrow \begin{cases} \mathbf{E} = \mathbf{P}_+ - \mathbf{P}_- \\ iZ_D\mathbf{H} = \mathbf{P}_+ + \mathbf{P}_- \end{cases} \quad \text{(S3.17)}$$

We then work on the electric field for convenience.

$$\nabla \times \mathbf{E} = \nabla \times \mathbf{Q}_+ - iZ_D\nabla \times \mathbf{Q}_- = k_+\mathbf{Q}_+ + iZ_Dk_-\mathbf{Q}_-$$
$$\nabla \times (\nabla \times \mathbf{E}) = \nabla \times (k_+\mathbf{Q}_+) - iZ_D\nabla \times (-k_-\mathbf{Q}_-) = k_+\nabla \times \mathbf{Q}_+ + iZ_Dk_-\nabla \times \mathbf{Q}_-$$
$$\Rightarrow (sum) \begin{cases} k_-\nabla \times \mathbf{E} = k_-\nabla \times \mathbf{Q}_+ - iZ_Dk_-\nabla \times \mathbf{Q}_- \\ \nabla \times (\nabla \times \mathbf{E}) = k_+\nabla \times \mathbf{Q}_+ + iZ_Dk_-\nabla \times \mathbf{Q}_- \end{cases} \Rightarrow \quad \text{(S3.18)}$$
$$\nabla \times (\nabla \times \mathbf{E}) + k_-\nabla \times \mathbf{E} = (k_+ + k_-)\nabla \times \mathbf{Q}_+ = k_+(k_+ + k_-)\mathbf{Q}_+$$

In an analogous way, we form the following.

$$\begin{cases} \nabla \times \mathbf{E} = k_+\mathbf{Q}_+ + iZ_Dk_-\mathbf{Q}_- \\ k_-\mathbf{E} = k_-\mathbf{Q}_+ - iZ_Dk_-\mathbf{Q}_- \end{cases} \Rightarrow (sum) \quad \nabla \times \mathbf{E} + k_-\mathbf{E} = (k_+ + k_-)\mathbf{Q}_+$$
$$(difference) \begin{cases} \nabla \times (\nabla \times \mathbf{E}) + k_-\nabla \times \mathbf{E} = k_+(k_+ + k_-)\mathbf{Q}_+ \\ k_+\nabla \times \mathbf{E} + k_+k_-\mathbf{E} = k_+(k_+ + k_-)\mathbf{Q}_+ \end{cases} \quad \text{(S3.19)}$$
$$\Rightarrow \nabla \times (\nabla \times \mathbf{E}) + k_-\nabla \times \mathbf{E} = k_+\nabla \times \mathbf{E} + k_+k_-\mathbf{E} \Rightarrow$$
$$\nabla \times (\nabla \times \mathbf{E}) + (k_- - k_+)\nabla \times \mathbf{E} - k_+k_-\mathbf{E} = \mathbf{0}$$

When $\{k_- - k_+, k_+k_-\}$ are expressed in terms of $\{\varepsilon\mu, \omega, \kappa\}$ with the help of Eq. (S3.3), we get to the following.

$$\begin{cases} k_- - k_+ = \omega(\sqrt{\varepsilon\mu} - \kappa) - \omega(\sqrt{\varepsilon\mu} + \kappa) = -2\omega\kappa \\ k_+k_- = \omega(\sqrt{\varepsilon\mu} + \kappa)\omega(\sqrt{\varepsilon\mu} - \kappa) = \omega^2(\varepsilon\mu - \kappa^2) \end{cases} \Rightarrow \quad \text{(S3.20)}$$
$$\nabla \times (\nabla \times \mathbf{E}) - 2\omega\kappa\nabla \times \mathbf{E} - \omega^2(\varepsilon\mu - \kappa^2)\mathbf{E} = \mathbf{0}$$

The corresponding equation for the magnetic field can be obtained just by observation. Nonetheless, it is further instructive to proceed in an analogous way to see how the signs are reversed during development.

$$-iZ_D\mathbf{H} = \mathbf{Q}_+ + iZ_D\mathbf{Q}_- \Rightarrow$$
$$\nabla\times(-iZ_D\mathbf{H}) = \nabla\times\mathbf{Q}_+ + iZ_D\nabla\times\mathbf{Q}_- = k_+\mathbf{Q}_+ - iZ_Dk_-\mathbf{Q}_-$$
$$\nabla\times[\nabla\times(-iZ_D\mathbf{H})] = \nabla\times(k_+\mathbf{Q}_+) + iZ_D\nabla\times(-k_-\mathbf{Q}_-) = k_+\nabla\times\mathbf{Q}_+ - iZ_Dk_-\nabla\times\mathbf{Q}_- . \quad (S3.21)$$
$$\Rightarrow (sum) \begin{cases} k_-\nabla\times(-iZ_D\mathbf{H}) = k_-\nabla\times\mathbf{Q}_+ + iZ_Dk_-\nabla\times\mathbf{Q}_- \\ \nabla\times[\nabla\times(-iZ_D\mathbf{H})] = k_+\nabla\times\mathbf{Q}_+ - iZ_Dk_-\nabla\times\mathbf{Q}_- \end{cases} \Rightarrow$$
$$\nabla\times[\nabla\times(-iZ_D\mathbf{H})] + k_-\nabla\times(-iZ_D\mathbf{H}) = (k_+ + k_-)\nabla\times\mathbf{Q}_+ = k_+(k_+ + k_-)\mathbf{Q}_+$$

In an analogous way, we form the following Helmholtz-like equation.

$$\begin{cases} \nabla\times(-iZ_D\mathbf{H}) = k_+\mathbf{Q}_+ - iZ_Dk_-\mathbf{Q}_- \\ k_-(-iZ_D\mathbf{H}) = k_-\mathbf{Q}_+ + iZ_Dk_-\mathbf{Q}_- \end{cases} \Rightarrow$$
$$(sum)\ \nabla\times(-iZ_D\mathbf{H}) + k_-(-iZ_D\mathbf{H}) = (k_+ + k_-)\mathbf{Q}_+$$
$$(difference) \begin{cases} \nabla\times[\nabla\times(-iZ_D\mathbf{H})] + k_-\nabla\times(-iZ_D\mathbf{H}) = k_+(k_+ + k_-)\mathbf{Q}_+ \\ k_+\nabla\times(-iZ_D\mathbf{H}) + k_+k_-(-iZ_D\mathbf{H}) = k_+(k_+ + k_-)\mathbf{Q}_+ \end{cases} \Rightarrow \quad . \quad (S3.22)$$
$$\nabla\times[\nabla\times(-iZ_D\mathbf{H})] + k_-\nabla\times(-iZ_D\mathbf{H}) = k_+\nabla\times(-iZ_D\mathbf{H}) + k_+k_-(-iZ_D\mathbf{H}) \Rightarrow$$
$$\nabla\times(\nabla\times\mathbf{H}) + (k_- - k_+)\nabla\times\mathbf{H} - k_+k_-\mathbf{H} = \mathbf{0}$$

In brief,

$$\begin{cases} \nabla\times\left(\nabla\times\begin{Bmatrix}\mathbf{E}\\\mathbf{H}\end{Bmatrix}\right) + (k_- - k_+)\nabla\times\begin{Bmatrix}\mathbf{E}\\\mathbf{H}\end{Bmatrix} - k_+k_-\begin{Bmatrix}\mathbf{E}\\\mathbf{H}\end{Bmatrix} = \mathbf{0} \\ \nabla\times\left(\nabla\times\begin{Bmatrix}\mathbf{E}\\\mathbf{H}\end{Bmatrix}\right) - 2\omega\kappa\nabla\times\begin{Bmatrix}\mathbf{E}\\\mathbf{H}\end{Bmatrix} - \omega^2(\varepsilon\mu - \kappa^2)\begin{Bmatrix}\mathbf{E}\\\mathbf{H}\end{Bmatrix} = \mathbf{0} \end{cases}. \quad (S3.23)$$

Even in these forms, both $\{\mathbf{E},\mathbf{H}\}$ contain a pair, say, $\{\exp(ik_+x), \exp(ik_-x)\}$, of distinct propagation factors. It is hence incorrect as in [S2] to assume that $\{\mathbf{E},\mathbf{H}\} \propto \exp(i\mathbf{k}\cdot\mathbf{r})$ with a single generic wave vector $\mathbf{k}$, where $\mathbf{r}$ is a position vector in a three-dimensional space.

$$\begin{cases} \nabla\times(\nabla\times\mathbf{E}) + (k_- - k_+)\nabla\times\mathbf{E} - k_+k_-\mathbf{E} = \mathbf{0} \\ \nabla\times(\nabla\times\mathbf{H}) + (k_- - k_+)\nabla\times\mathbf{H} - k_+k_-\mathbf{H} = \mathbf{0} \end{cases}$$
$$\begin{cases} \nabla\times(\nabla\times\mathbf{E}) - 2\omega\kappa\nabla\times\mathbf{E} - \omega^2(\varepsilon\mu - \kappa^2)\mathbf{E} = \mathbf{0} \\ \nabla\times(\nabla\times\mathbf{H}) - 2\omega\kappa\nabla\times\mathbf{H} - \omega^2(\varepsilon\mu - \kappa^2)\mathbf{H} = \mathbf{0} \end{cases} \quad (S3.24)$$
$$\begin{Bmatrix}\mathbf{E}\\\mathbf{H}\end{Bmatrix} \propto \exp(i\mathbf{k}\cdot\mathbf{r}) \Rightarrow i\mathbf{k}\times\left(i\mathbf{k}\times\begin{Bmatrix}\mathbf{E}\\\mathbf{H}\end{Bmatrix}\right) - 2\omega\kappa i\mathbf{k}\times\begin{Bmatrix}\mathbf{E}\\\mathbf{H}\end{Bmatrix} - \omega^2(\varepsilon\mu - \kappa^2)\begin{Bmatrix}\mathbf{E}\\\mathbf{H}\end{Bmatrix} = \mathbf{0}$$

This equation is in a grave error!

Let us employ a generic vector identity $\mathbf{A} \in \mathbb{C}^3$ for $\nabla\times(\nabla\times\mathbf{A}) = \nabla(\nabla\cdot\mathbf{A}) - \nabla^2\mathbf{A}$. We then have the following.

$$\begin{cases} \nabla^2 \begin{Bmatrix} \mathbf{E} \\ \mathbf{H} \end{Bmatrix} + (k_+ - k_-)\nabla \times \begin{Bmatrix} \mathbf{E} \\ \mathbf{H} \end{Bmatrix} + k_+ k_- \begin{Bmatrix} \mathbf{E} \\ \mathbf{H} \end{Bmatrix} = \mathbf{0} \\ \nabla^2 \begin{Bmatrix} \mathbf{E} \\ \mathbf{H} \end{Bmatrix} + 2\omega\kappa\nabla \times \begin{Bmatrix} \mathbf{E} \\ \mathbf{H} \end{Bmatrix} + \omega^2(\varepsilon\mu - \kappa^2)\begin{Bmatrix} \mathbf{E} \\ \mathbf{H} \end{Bmatrix} = \mathbf{0} \end{cases} . \tag{S3.25}$$

Here, we implemented the divergence-free conditions $\nabla \cdot \mathbf{E} = \nabla \cdot \mathbf{H} = 0$. In brief, the Helmholtz equations are readily derivable from the set of four Maxwell equations. In comparison to the Maxwell equations, these Helmholtz equations are more convenient for seeking solutions to the Maxwell equations or the decay rates in case with evanescent waves.

**Section S4. Determination of the dispersion relation for eigenvalues**

The relative permittivity $\varepsilon_M$ of a loss-free metal is expressed below in terms of a plasma frequency $\omega_p$.

$$\varepsilon_M = 1 - \frac{\omega_p^2}{\omega^2} . \tag{S4.1}$$

In conformity to the solutions in the chiral medium, we assumed hybrid TM-TE waves in the half-space of metal with all three nonzero components. We found the solutions by following a direct (or inductive) way.

$$\Pi_M \equiv \exp(ik_x x + \gamma_M z - i\omega t) \implies \begin{cases} \mathbf{E} \equiv \frac{1}{\sqrt{2}}\frac{1}{k_x}\left(G_{Ex}\hat{\mathbf{x}} + G_{Ey}\hat{\mathbf{y}} + G_{Ez}\hat{\mathbf{z}}\right)\Pi_M \\ \mathbf{H} \equiv \frac{1}{\sqrt{2}}\frac{1}{k_x}\left(G_{Hx}\hat{\mathbf{x}} + G_{Hy}\hat{\mathbf{y}} + G_{Hz}\hat{\mathbf{z}}\right)\Pi_M \end{cases} . \tag{S4.2}$$

Following our approach, we just present below the so-found solutions.

$$\begin{cases} \sqrt{2}k_x \mathbf{E} = \left(G_{Ex}\hat{\mathbf{x}} + G_{Ey}\hat{\mathbf{y}} - i\frac{k_x}{\gamma_M}G_{Ex}\hat{\mathbf{z}}\right)\Pi_M \\ \sqrt{2}k_x \mathbf{H} = \left(i\frac{\gamma_M}{\omega\mu_M}G_{Ey}\hat{\mathbf{x}} + i\frac{\omega\varepsilon_M}{\gamma_M}G_{Ex}\hat{\mathbf{y}} + \frac{k_x}{\omega\mu_M}G_{Ey}\hat{\mathbf{z}}\right)\Pi_M \end{cases} . \tag{S4.3}$$

To obtain this pair from the preceding pair, we have employed both Faraday law $\nabla \times \mathbf{E} = i\omega\mu_M \mathbf{H}$ and Ampère law $\nabla \times \mathbf{H} = -i\omega\varepsilon_M \mathbf{E}$ in addition to $\nabla \cdot \mathbf{E} = \nabla \cdot \mathbf{H} = 0$.

Let us prove that the above solutions satisfy the divergence-free conditions.

$$\begin{cases} \sqrt{2}k_x \nabla \cdot \mathbf{E} = \left( ik_x G_{Ex} - \gamma_M i \dfrac{k_x}{\gamma_M} G_{Ex} \right) \Pi_M = 0 \\ \sqrt{2}k_x \nabla \cdot \mathbf{H} = \left( ik_x i \dfrac{\gamma_M}{\omega\mu_M} G_{Ey} + \gamma_M \dfrac{k_x}{\omega\mu_M} G_{Ey} \right) \Pi_M = 0 \end{cases} \qquad (S4.4)$$

Let us prove the Maxwell equations. Firstly, we prove the Faraday law $\nabla \times \mathbf{E} = i\omega\mu_M \mathbf{H}$.

$$\begin{aligned}
\sqrt{2}k_x \nabla \times \mathbf{E} &= \left( \dfrac{\partial G_{Ez}}{\partial y} - \dfrac{\partial G_{Ey}}{\partial z} \right)\hat{\mathbf{x}} + \left( \dfrac{\partial G_{Ex}}{\partial z} - \dfrac{\partial G_{Ez}}{\partial x} \right)\hat{\mathbf{y}} + \left( \dfrac{\partial G_{Ey}}{\partial x} - \dfrac{\partial G_{Ex}}{\partial y} \right)\hat{\mathbf{z}} \\
&= -\dfrac{\partial G_{Ey}}{\partial z}\hat{\mathbf{x}} + \left( \dfrac{\partial G_{Ex}}{\partial z} - \dfrac{\partial G_{Ez}}{\partial x} \right)\hat{\mathbf{y}} + \dfrac{\partial G_{Ey}}{\partial x}\hat{\mathbf{z}} \\
\dfrac{\sqrt{2}k_x \nabla \times \mathbf{E}}{\Pi_M} &= -\gamma_M G_{Ey}\hat{\mathbf{x}} + \left[ \gamma_M G_{Ex} - ik_x\left( -i\dfrac{k_x}{\gamma_M}G_{Ex} \right) \right]\hat{\mathbf{y}} + ik_x G_{Ey}\hat{\mathbf{z}} \\
&= -\gamma_M G_{Ey}\hat{\mathbf{x}} + \dfrac{G_{Ex}}{\gamma_M}\left( \gamma_M^2 - k_x^2 \right)\hat{\mathbf{y}} + ik_x G_{Ey}\hat{\mathbf{z}} = -\gamma_M G_{Ey}\hat{\mathbf{x}} - \dfrac{G_{Ex}}{\gamma_M}\omega^2\varepsilon_M\mu_M\hat{\mathbf{y}} + ik_x G_{Ey}\hat{\mathbf{z}} \\
&= i\omega\mu_M\left( i\dfrac{\gamma_M}{\omega\mu_M}G_{Ey}\hat{\mathbf{x}} + i\dfrac{\omega\varepsilon_M}{\gamma_M}G_{Ex}\hat{\mathbf{y}} + \dfrac{k_x}{\omega\mu_M}G_{Ey}\hat{\mathbf{z}} \right) = \sqrt{2}k_x\dfrac{i\omega\mu_M \mathbf{H}}{\Pi_M}
\end{aligned} \qquad (S4.5)$$

Here, we have utilized the relation for the decay rate $\gamma_M^2 \equiv k_x^2 - \omega^2\varepsilon_M\mu_M$ within a metal. Secondly, we prove the Ampère law $\nabla \times \mathbf{H} = -i\omega\varepsilon_M \mathbf{E}$ in an analogous fashion.

$$\begin{aligned}
\sqrt{2}k_x \nabla \times \mathbf{H} &= \left( \dfrac{\partial G_{Hz}}{\partial y} - \dfrac{\partial G_{Hy}}{\partial z} \right)\hat{\mathbf{x}} + \left( \dfrac{\partial G_{Hx}}{\partial z} - \dfrac{\partial G_{Hz}}{\partial x} \right)\hat{\mathbf{y}} + \left( \dfrac{\partial G_{Hy}}{\partial x} - \dfrac{\partial G_{Hx}}{\partial y} \right)\hat{\mathbf{z}} \\
&= -\dfrac{\partial G_{Hy}}{\partial z}\hat{\mathbf{x}} + \left( \dfrac{\partial G_{Hx}}{\partial z} - \dfrac{\partial G_{Hz}}{\partial x} \right)\hat{\mathbf{y}} + \dfrac{\partial G_{Hy}}{\partial x}\hat{\mathbf{z}} \\
\dfrac{\sqrt{2}k_x \nabla \times \mathbf{H}}{\Pi_M} &= -\gamma_M i\dfrac{\omega\varepsilon_M}{\gamma_M}G_{Ex}\hat{\mathbf{x}} + \left( \gamma_M i\dfrac{\gamma_M}{\omega\mu_M}G_{Ey} - ik_x\dfrac{k_x}{\omega\mu_M}G_{Ey} \right)\hat{\mathbf{y}} + ik_x i\dfrac{\omega\varepsilon_M}{\gamma_M}G_{Ex}\hat{\mathbf{z}} \\
&= -\gamma_M i\dfrac{\omega\varepsilon_M}{\gamma_M}G_{Ex}\hat{\mathbf{x}} + i\dfrac{1}{\omega\mu_M}\left( \gamma_M^2 - k_x^2 \right)G_{Ey}\hat{\mathbf{y}} - k_x\dfrac{\omega\varepsilon_M}{\gamma_M}G_{Ex}\hat{\mathbf{z}} \\
&= -i\omega\varepsilon_M G_{Ex}\hat{\mathbf{x}} - i\omega\varepsilon_M G_{Ey}\hat{\mathbf{y}} - k_x\dfrac{\omega\varepsilon_M}{\gamma_M}G_{Ex}\hat{\mathbf{z}} \\
&= -i\omega\varepsilon_M\left( G_{Ex}\hat{\mathbf{x}} + G_{Ey}\hat{\mathbf{y}} - i\dfrac{k_x}{\gamma_M}G_{Ex}\hat{\mathbf{z}} \right) = \sqrt{2}k_x\dfrac{-i\omega\varepsilon_M \mathbf{E}}{\Pi_M}
\end{aligned} \qquad (S4.6)$$

In making analysis of electromagnetic waves propagating through a chiral medium, another important branch is to handle spatially standing waves as occur in an optical resonator (cavity) [S3]. Such standing waves are related to the configurations often considered by quantum optics.

On the chiral-dielectric side, we need the field values at the planar interface at $z = 0$ as follows.

$$z=0 \;\Rightarrow\; \Pi_{\pm}(z=0)=\exp(ik_x x - i\omega t),\; \gamma_{\pm}^2 \equiv k_x^2 - \omega^2\left(\sqrt{\varepsilon_D \mu_D}\pm\kappa\right) \;\Rightarrow$$

$$\begin{cases} \sqrt{2}k_x \mathbf{E}(z=0) = \begin{bmatrix} Q_+\left(\gamma_+\hat{\mathbf{x}}+k_+\hat{\mathbf{y}}+ik_x\hat{\mathbf{z}}\right) \\ -P_-\left(\gamma_-\hat{\mathbf{x}}-k_-\hat{\mathbf{y}}+ik_x\hat{\mathbf{z}}\right) \end{bmatrix}\exp(ik_x x - i\omega t) \\ \sqrt{2}k_x \mathbf{H}(z=0) = \left(-iZ_D^{-1}\right)\begin{bmatrix} Q_+\left(\gamma_+\hat{\mathbf{x}}+k_+\hat{\mathbf{y}}+ik_x\hat{\mathbf{z}}\right) \\ +P_-\left(\gamma_-\hat{\mathbf{x}}-k_-\hat{\mathbf{y}}+ik_x\hat{\mathbf{z}}\right) \end{bmatrix}\exp(ik_x x - i\omega t) \end{cases} \quad . \quad \text{(S4.7)}$$

Likewise, on the metallic side, we need the field values at the planar interface at $z=0$ as follows.

$$z=0 \;\Rightarrow\; \Pi_M(z=0)=\exp(ik_x x - i\omega t),\; \gamma_M^2 \equiv k_x^2 - \omega^2 \varepsilon_M \mu_M \;\Rightarrow$$

$$\begin{cases} \sqrt{2}k_x \mathbf{E} = \left(G_{Ex}\hat{\mathbf{x}}+G_{Ey}\hat{\mathbf{y}}-i\dfrac{k_x}{\gamma_M}G_{Ex}\hat{\mathbf{z}}\right)\exp(ik_x x - i\omega t) \\ \sqrt{2}k_x \mathbf{H} = \left(i\dfrac{\gamma_M}{\omega\mu_M}G_{Ey}\hat{\mathbf{x}}+i\dfrac{\omega\varepsilon_M}{\gamma_M}G_{Ex}\hat{\mathbf{y}}+\dfrac{k_x}{\omega\mu_M}G_{Ey}\hat{\mathbf{z}}\right)\exp(ik_x x - i\omega t) \end{cases} \quad . \quad \text{(S4.8)}$$

By the continuity conditions of the tangential components in the $\{x,y\}$-directions, we arrive at four relations homogeneous in the four so-far undetermined coefficients $\{Q_+, P_-, G_{Ex}, G_{Ey}\}$. Here, $P_- \equiv iZ_D Q_-$

$$\begin{cases} \dfrac{\gamma_+}{k_x}Q_+ - \dfrac{\gamma_-}{k_x}P_- = \dfrac{1}{k_x}G_{Ex} \\ \dfrac{k_+}{k_x}Q_+ + \dfrac{k_-}{k_x}P_- = \dfrac{1}{k_x}G_{Ey} \end{cases}, \quad \begin{cases} -i\dfrac{1}{Z_D}\left(\dfrac{\gamma_+}{k_x}Q_+ + \dfrac{\gamma_-}{k_x}P_-\right) = i\dfrac{\gamma_M}{\omega\mu_M}\dfrac{1}{k_x}G_{Ey} \\ -i\dfrac{1}{Z_D}\left(\dfrac{k_+}{k_x}Q_+ - \dfrac{k_-}{k_x}P_-\right) = i\dfrac{\omega\varepsilon_M}{\gamma_M}\dfrac{1}{k_x}G_{Ex} \end{cases}$$

$$\begin{pmatrix} \gamma_+ & -\gamma_- & -1 & 0 \\ k_+ & k_- & 0 & -1 \\ \gamma_+ & \gamma_- & 0 & \dfrac{Z_D \gamma_M}{\omega\mu_M} \\ k_+ & -k_- & \dfrac{Z_D \omega\varepsilon_M}{\gamma_M} & 0 \end{pmatrix}\begin{Bmatrix} Q_+ \\ P_- \\ G_{Ex} \\ G_{Ey} \end{Bmatrix} = \begin{Bmatrix} 0 \\ 0 \\ 0 \\ 0 \end{Bmatrix} \;\Rightarrow\; \hat{\Upsilon}\boldsymbol{\Xi}=\mathbf{0}$$

$$\hat{\Upsilon}\equiv \begin{pmatrix} \gamma_+ & -\gamma_- & -1 & 0 \\ k_+ & k_- & 0 & -1 \\ \gamma_+ & \gamma_- & 0 & \dfrac{Z_D \gamma_M}{\omega\mu_M} \\ k_+ & -k_- & \dfrac{Z_D \omega\varepsilon_M}{\gamma_M} & 0 \end{pmatrix},\; \boldsymbol{\Xi}\equiv\begin{Bmatrix} Q_+ \\ P_- \\ G_{Ex} \\ G_{Ey} \end{Bmatrix} \quad . \quad \text{(S4.9)}$$

Let us take a determinant of the 4-by-4 matrix on the left-hand side, which we subsequently expand into a pair of 3-by-3 determinants.

$$\begin{vmatrix} \gamma_+ & -\gamma_- & -1 & 0 \\ k_+ & k_- & 0 & -1 \\ \gamma_+ & \gamma_- & 0 & \dfrac{Z_D \gamma_M}{\omega \mu_M} \\ k_+ & -k_- & \dfrac{Z_D \omega \varepsilon_M}{\gamma_M} & 0 \end{vmatrix} = 0 \quad \Rightarrow$$

$$\begin{vmatrix} \gamma_+ & -\gamma_- & -1 \\ \gamma_+ & \gamma_- & 0 \\ k_+ & -k_- & \dfrac{Z_D \omega \varepsilon_M}{\gamma_M} \end{vmatrix} + \dfrac{Z_D \gamma_M}{\omega \mu_M} \begin{vmatrix} \gamma_+ & -\gamma_- & -1 \\ k_+ & k_- & 0 \\ k_+ & -k_- & \dfrac{Z_D \omega \varepsilon_M}{\gamma_M} \end{vmatrix} = 0$$

(S4.10)

We expand this set of 3-by-3 determinants into four of 2-by-2 determinants to finally establish a dispersion relation.

$$-\begin{vmatrix} \gamma_+ & \gamma_- \\ k_+ & -k_- \end{vmatrix} + \dfrac{Z_D \omega \varepsilon_M}{\gamma_M}\begin{vmatrix} \gamma_+ & -\gamma_- \\ \gamma_+ & \gamma_- \end{vmatrix}$$
$$-\dfrac{Z_D \gamma_M}{\omega \mu_M}\begin{vmatrix} k_+ & k_- \\ k_+ & -k_- \end{vmatrix} + \dfrac{Z_D \gamma_M}{\omega \mu_M}\dfrac{Z_D \omega \varepsilon_M}{\gamma_M}\begin{vmatrix} \gamma_+ & -\gamma_- \\ k_+ & k_- \end{vmatrix} = 0$$
$$\gamma_+ k_- + \gamma_- k_+ + 2\dfrac{Z_D \omega \varepsilon_M}{\gamma_M}\gamma_+\gamma_- + 2\dfrac{Z_D \gamma_M}{\omega \mu_M}k_+ k_- + \dfrac{Z_D \gamma_M}{\omega \mu_M}\dfrac{Z_D \omega \varepsilon_M}{\gamma_M}(\gamma_+ k_- + \gamma_- k_+) = 0$$
$$\left(1 + Z_D^2 \dfrac{\varepsilon_M}{\mu_M}\right)(\gamma_+ k_- + \gamma_- k_+) + 2Z_D\left(\dfrac{\omega \varepsilon_M}{\gamma_M}\gamma_+\gamma_- + \dfrac{\gamma_M}{\omega \mu_M}k_+ k_-\right) = 0$$

(S4.11)

Via $Z_D \equiv \sqrt{\mu_D/\varepsilon_D}$, we end up with what we call the 'chiral dispersion relation'.

$$f(k_x, \omega, \kappa) \equiv \left(1 + \dfrac{\mu_D}{\varepsilon_D}\dfrac{\varepsilon_M}{\mu_M}\right)(k_+\gamma_- + k_-\gamma_+) + 2\sqrt{\dfrac{\mu_D}{\varepsilon_D}}\left(\dfrac{\omega \varepsilon_M}{\gamma_M}\gamma_+\gamma_- + \dfrac{\gamma_M}{\omega \mu_M}k_+ k_-\right) = 0$$

$$\left(1 + Z_D^2 \dfrac{\varepsilon_M}{\mu_M}\right)(k_+\gamma_- + k_-\gamma_+) + 2Z_D\left(\dfrac{\omega \varepsilon_M}{\gamma_M}\gamma_+\gamma_- + \dfrac{\gamma_M}{\omega \mu_M}k_+ k_-\right) = 0 \qquad \text{.(S4.12)}$$

$$\begin{cases} u \equiv \dfrac{\omega \varepsilon_M}{\gamma_M} \\ v \equiv \dfrac{\gamma_M}{\omega \mu_M} \end{cases} \Rightarrow \left(1 + Z_D^2 uv\right)(k_+\gamma_- + k_-\gamma_+) + 2Z_D(u\gamma_+\gamma_- + vk_+ k_-) = 0$$

We have other implicit dependence on the properties $\{\varepsilon_D, \mu_D, \varepsilon_M, \mu_M\}$.

Suppose that $\varepsilon_M$ is replaced with another dielectric property $\varepsilon_{D2} > 0$.

$$f(k_x, \omega, \kappa) \equiv \left(1 + \frac{\mu_D}{\varepsilon_D} \frac{\varepsilon_M}{\mu_M}\right)(k_+ \gamma_- + k_- \gamma_+)$$
$$+ 2\sqrt{\frac{\mu_D}{\varepsilon_D}} \left(\frac{\omega \varepsilon_M}{\gamma_M} \gamma_+ \gamma_- + \frac{\gamma_M}{\omega \mu_M} k_+ k_-\right) = 0 \quad \text{(S4.13)}$$

All terms on the left-hand side of this modified dispersion relation become positive, thus admitting no eigenvalues. Consequently, surface plasmon waves on resonance are not supported across an interface between a chiral medium and an achiral dielectric.

Meanwhile, recall the limiting forms for several relevant parameters.

$$\lim_{\kappa \to 0} c_\pm = c_D \equiv \frac{1}{\sqrt{\varepsilon_D \mu_D}} \Rightarrow \lim_{\kappa \to 0} k_\pm = k_D \equiv \frac{\omega}{c_D}, \quad \lim_{\kappa \to 0} \gamma_\pm = \gamma_D \equiv \sqrt{k_x^2 - k_D^2}$$
$$\varepsilon_M \equiv -|\varepsilon_M| < 0 \quad \text{(S4.14)}$$

Therefore, we can take a limit of the above chiral dispersion relation as $\kappa \to 0$.

$$\lim_{\kappa \to 0} f(k_x, \omega, \kappa) = 0 \Rightarrow$$
$$\left(1 + \frac{\mu_D}{\varepsilon_D} \frac{\varepsilon_M}{\mu_M}\right) k_D \gamma_D + \sqrt{\frac{\mu_D}{\varepsilon_D}} \left(\frac{\omega \varepsilon_M}{\gamma_M} \gamma_D^2 + \frac{\gamma_M}{\omega \mu_M} k_D^2\right) = 0 \Rightarrow$$
$$\left[\left(1 - \frac{\mu_D}{\varepsilon_D} \frac{|\varepsilon_M|}{\mu_M}\right) k_D \gamma_D + \sqrt{\frac{\mu_D}{\varepsilon_D}} \left(\frac{\gamma_M}{\omega \mu_M} k_D^2 - \frac{\omega |\varepsilon_M|}{\gamma_M} \gamma_D^2\right) = 0\right](\omega \varepsilon_D \mu_M \gamma_M) \quad \text{(S4.15)}$$
$$\omega(\varepsilon_D \mu_M - \mu_D |\varepsilon_M|) k_D \gamma_M \gamma_D + \sqrt{\varepsilon_D \mu_D} \left(\gamma_M^2 k_D^2 - \omega^2 |\varepsilon_M| \mu_M \gamma_D^2\right) = 0$$

Th decay rates can be further cast into much simpler forms, when if introduce the pair $\{k_D^2, k_M^2\}$.

$$\begin{cases} k_D^2 \equiv \omega^2 \varepsilon_D \mu_{DM} \\ k_M^2 \equiv \omega^2 |\varepsilon_M| \mu_{DM} \end{cases} \Rightarrow \begin{cases} \gamma_D \equiv \sqrt{k_x^2 - \omega^2 \varepsilon_D \mu_D} \equiv \sqrt{k_x^2 - k_D^2} \\ \gamma_M \equiv \sqrt{k_x^2 + \omega^2 |\varepsilon_M| \mu_M} \equiv \sqrt{k_x^2 + k_M^2} \end{cases} \Rightarrow$$
$$\gamma_D \gamma_M \omega^2 (\varepsilon_D \mu_M - \mu_D |\varepsilon_M|) + \gamma_M^2 k_D^2 - \omega^2 |\varepsilon_M| \mu_M \gamma_D^2 = 0 \quad \text{(S4.16)}$$

For convenience, we set both relative magnetic permeabilities to be equal, namely, $\mu_{DM} \equiv \mu_D = \mu_M$, as we did for the achiral case. We then take squares of both sides.

$$\mu_{DM} \equiv \mu_D = \mu_M \Rightarrow \begin{cases} k_D^2 \equiv \omega^2 \varepsilon_D \mu_{DM} \\ k_M^2 \equiv \omega^2 |\varepsilon_M| \mu_{DM} \end{cases} \Rightarrow$$
$$\sqrt{k_x^2 - k_D^2} \sqrt{k_x^2 + k_M^2} (k_D^2 - k_M^2) + (k_x^2 + k_M^2) k_D^2 - k_M^2 (k_x^2 - k_D^2) = 0 \Rightarrow$$
$$(k_x^2 - k_D^2)(k_x^2 + k_M^2)(k_D^2 - k_M^2)^2 \quad \text{(S4.17)}$$
$$= k_M^4 (k_x^2 - k_D^2)^2 + k_D^4 (k_x^2 + k_M^2)^2 - 2 k_M^2 k_D^2 (k_x^2 - k_D^2)(k_x^2 + k_M^2) \Rightarrow$$
$$(k_x^2 - k_D^2)(k_x^2 + k_M^2)\left[(k_D^2 - k_M^2)^2 + 2 k_M^2 k_D^2\right] = k_M^4 (k_x^2 - k_D^2)^2 + k_D^4 (k_x^2 + k_M^2)^2$$

Therefore, we get a quartic equation in $k_x$ as follows.

$$\left[k_x^4+\left(k_M^2-k_D^2\right)k_x^2-k_M^2k_D^2\right]\left(k_D^4+k_M^4-2k_M^2k_D^2+2k_x^2k_D^2\right)$$
$$=k_M^4\left(k_x^4+k_D^4-2k_x^2k_D^2\right)+k_D^4\left(k_x^4+k_M^4+2k_x^2k_M^2\right)\;\Rightarrow$$
$$\left(k_D^4+k_M^4\right)\left[k_x^4+\left(k_M^2-k_D^2\right)k_x^2-k_M^2k_D^2\right]$$
$$=\left(k_D^4+k_M^4\right)k_x^4+2k_x^2\left(k_D^4k_M^2-k_M^4k_D^2\right)+2k_D^4k_M^4\;\Rightarrow$$
$$\left(k_D^4+k_M^4\right)\left(k_M^2-k_D^2\right)k_x^2-k_M^2k_D^2\left(k_D^4+k_M^4\right)=2k_x^2\left(k_D^4k_M^2-k_M^4k_D^2\right)+2k_D^4k_M^4\;\Rightarrow\quad\text{(S4.18)}$$
$$\left[\left(k_D^4+k_M^4\right)\left(k_M^2-k_D^2\right)+2\left(k_M^4k_D^2-k_D^4k_M^2\right)\right]k_x^2=k_D^2k_M^2\left(k_D^4+k_M^4\right)+2k_D^4k_M^4\;\Rightarrow$$
$$\left(k_D^4k_M^2+k_M^6-k_D^6-k_D^2k_M^4+2k_D^2k_M^4-2k_D^4k_M^2\right)k_x^2=k_D^2k_M^2\left(k_D^4+k_M^4+2k_D^2k_M^2\right)$$
$$\left(k_M^6-k_D^6+k_D^2k_M^4-k_D^4k_M^2\right)k_x^2=k_D^2k_M^2\left(k_D^2+k_M^2\right)^2\;\Rightarrow$$
$$\left(k_M^2+k_D^2\right)\left(k_M^4-k_D^4\right)k_x^2=k_D^2k_M^2\left(k_D^2+k_M^2\right)^2$$

If further simplified, we recover the achiral dispersion relation under the same assumption $\mu_{DM}\equiv\mu_D=\mu_M$ as follows.

$$\left(k_M^2+k_D^2\right)^2\left(k_M^2-k_D^2\right)k_x^2=k_D^2k_M^2\left(k_D^2+k_M^2\right)^2\;\Rightarrow\;k_x^2=\frac{k_D^2k_M^2}{k_M^2-k_D^2}$$
$$k_x>0\;\Rightarrow\;k_x=\frac{k_Dk_M}{\sqrt{k_M^2-k_D^2}}=\omega\frac{\sqrt{\varepsilon_D\mu_{DM}}\sqrt{|\varepsilon_M|\mu_{DM}}}{\sqrt{|\varepsilon_M|\mu_{DM}-\varepsilon_D\mu_{DM}}}=\omega\frac{\sqrt{\varepsilon_D|\varepsilon_M|}}{\sqrt{|\varepsilon_M|-\varepsilon_D}}\quad\text{(S4.19)}$$

From a physical viewpoint, this reduction of the chiral case to an achiral case in the achiral limit is nothing curious.

**Section S5. Properties of the dispersion relation and eigenvectors**

We exploit the series of symmetry properties: $k_\pm(-\kappa)\equiv k_\mp(\kappa)$, $\gamma_\pm(-\kappa)\equiv\gamma_\mp(\kappa)$, and $\gamma_\pm(-k_x)\equiv\gamma_\pm(k_x)$ in showing the symmetry property of the chiral dispersion relation with respect either $k_x$ and to $\kappa$ in the following manner.

$$k_{\pm} \equiv \omega\left(\sqrt{\varepsilon_D \mu_D} \pm \kappa\right) \Rightarrow k_{\pm}(-\kappa) \equiv k_{\mp}(\kappa)$$

$$\begin{cases} \gamma_{\pm} \equiv \sqrt{k_x^2 - \omega^2\left(\sqrt{\varepsilon\mu} \pm \kappa\right)^2} \equiv \sqrt{k_x^2 - k_{\pm}^2} \\ k_x^2 \equiv \gamma_{\pm}^2 + k_{\pm}^2 \end{cases} \Rightarrow \begin{cases} \gamma_{\pm}(-\kappa) \equiv \gamma_{\mp}(\kappa) \\ \gamma_{\pm}(-k_x) \equiv \gamma_{\pm}(k_x) \end{cases} \Rightarrow$$

$$f(k_x, \omega, \kappa) \equiv \left(1 + \frac{\mu_D}{\varepsilon_D}\frac{\varepsilon_M}{\mu_M}\right)(k_+\gamma_- + k_-\gamma_+) + 2\sqrt{\frac{\mu_D}{\varepsilon_D}}\left(\frac{\omega\varepsilon_M}{\gamma_M}\gamma_+\gamma_- + \frac{\gamma_M}{\omega\mu_M}k_+k_-\right) \quad . \quad \text{(S5.1)}$$

$$\begin{cases} f(-k_x, \omega, \kappa) = f(k_x, \omega, \kappa) \\ f(k_x, \omega, -\kappa) = \left(1 + \frac{\mu_D}{\varepsilon_D}\frac{\varepsilon_M}{\mu_M}\right)(k_-\gamma_+ + k_+\gamma_-) + 2\sqrt{\frac{\mu_D}{\varepsilon_D}}\left(\frac{\omega\varepsilon_M}{\gamma_M}\gamma_-\gamma_+ + \frac{\gamma_M}{\omega\mu_M}k_-k_+\right) \\ \equiv f(k_x, \omega, \kappa) \end{cases}$$

We find that the solutions to $k_x$ are symmetric with respect to $\kappa$, namely, $\omega(+\kappa) = \omega(-\kappa)$ and $k_x(+\kappa) = k_x(-\kappa)$. Therefore, we normally present our dispersion curves only in the ranges $k_x, \omega > 0$.

Let us explain how we have obtained numerical solutions to the chiral dispersion relation. To this goal, we resorted to a commercial software Mathematica®. We called upon the 'FindRoot' function, which is a root-finder of a Newton-Raphson type. Since $0 < k_x < \infty$ for our choice of $\varepsilon_D = \mu_D = 1$, we take a certain value of $k_x$ as a parameter in $f(k_x, \omega, \kappa)$. We then choose an initial guess for $\omega$ to solve $f(k_x, \omega, \kappa) = 0$ with $\omega$ as the unknown. Normally after a couple of second, a converged root is produced on a computer. We have checked the residual $|f(k_x, \omega, \kappa)|$, which lies below a certain small value, say, $|f(k_x, \omega, \kappa)| < 10^{-10}$. The whole process is repeated for varying values of $k_x$ over $0 < k_x < \infty$.

Both black, blue, and red curves on Fig. 2 in Primary Document have been obtained in this way by solving the achiral and chiral dispersion relations. The disallowed (or cut-off or forbidden) portion for each curve is simply determined by comparing $\omega$ as a solution to $f(k_x, \omega, \kappa) = 0$ with the tilted straight line $c_+(\kappa)k_x$ such that $\omega > c_+(\kappa)k_x$. Notice over $\omega > c_+(\kappa)k_x$ that the decay rate is undefined or the argument of the square root is imaginary, namely, $k_x^2 < k_+^2$ for $\gamma_+ \equiv \sqrt{k_x^2 - k_+^2}$, where $k_+(\omega, \kappa) \equiv \omega\left(\sqrt{\varepsilon_D \mu_D} + \kappa\right)$ with $\kappa > 0$. Straight lines are given by $\omega = c_\pm k_x$ and $\omega = c_D k_x$. Below given are a couple of figures that help better understand Fig. 2 of Primary Document.

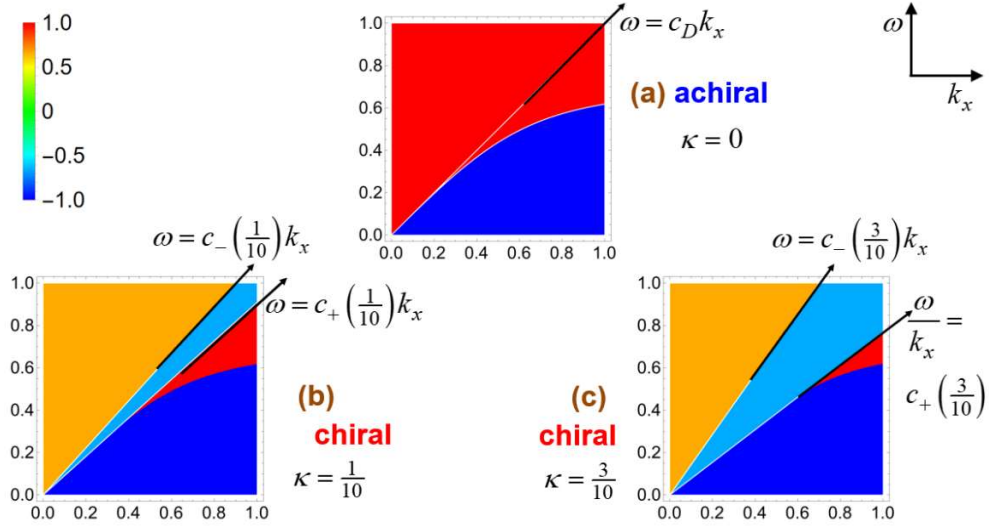

Fig. S1. Domains and reference lines on the parameter plane of $\{k_x, \omega\}$. The number within ( ) means the value of the assigned $\kappa$, for which two of $\kappa = \frac{1}{10}, \frac{3}{10}$ are chosen for the chirality parameter. The pairs $\{c_+\left(\frac{1}{10}\right), c_-\left(\frac{1}{10}\right)\}$ and $\{c_+\left(\frac{3}{10}\right), c_-\left(\frac{3}{10}\right)\}$ are constants for a given set of optical properties $\{\varepsilon, \mu, \kappa\}$ of media. Given data includes $\varepsilon_D = \mu_D = \mu_M = 1$ and $\omega_p = 1$.

Figure S1(a) shows various regions delineated for the achiral case so that the well-known dispersion relation $\varepsilon_D \gamma_M = |\varepsilon_M| \gamma_D$ for the achiral case is examined. Its solution $\omega(k_x)$ corresponds to the boundary between the upper red zone and the lower blue zone. The light line $\omega = c_D k_x$ with $c_D = 1$ is a straight line with a tilt angle of $45^o$. The zone above $\omega = c_D k_x$ corresponds to phase speeds larger than the light speed (superluminal), whereas the zone below $\omega = c_D k_x$ corresponds to phase speeds smaller than the light speed (subluminal). Consequently, the subluminal solution curve $\omega(k_x)$ is acceptable.

Figure S1(b) and S1(c) are obtained for the chiral case. Of course, the two lines $\omega = c_\pm\left(\frac{1}{10}\right)k_x$ are closer to each other than $\omega = c_\pm\left(\frac{3}{10}\right)k_x$ are. These two lines will collapse to the single light line $\omega = c_D k_x$ as $\kappa \to 0$. The solution curve $\omega(k_x)$ is given again by the boundary between the upper red zone and the blue lower zone. These two solution curves on Figs. S1(a) and S1(b) are almost indistinguishable from each other, but they are distinct.

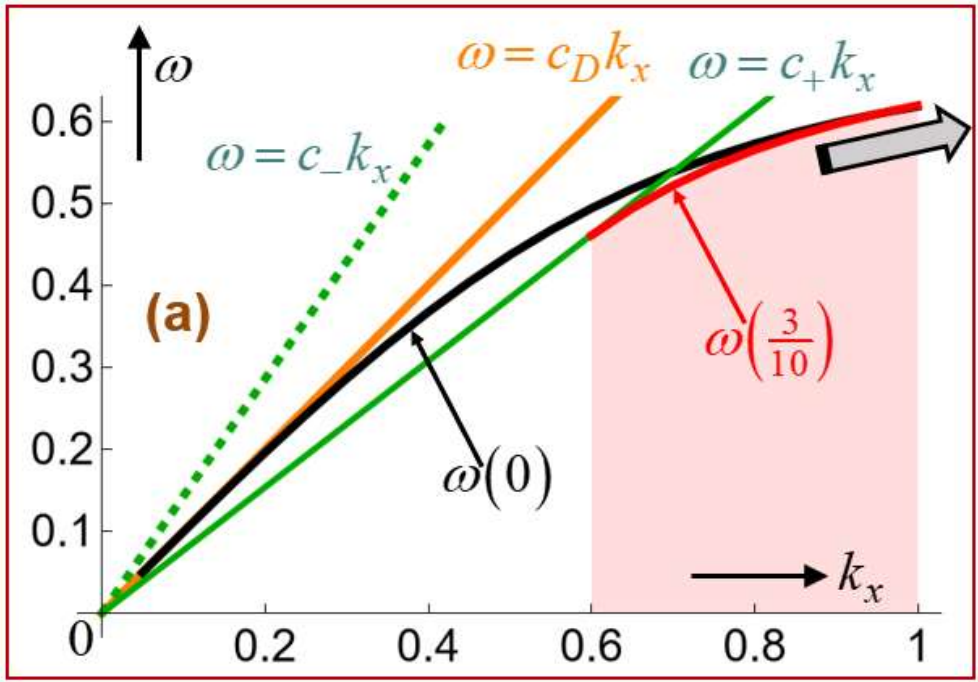

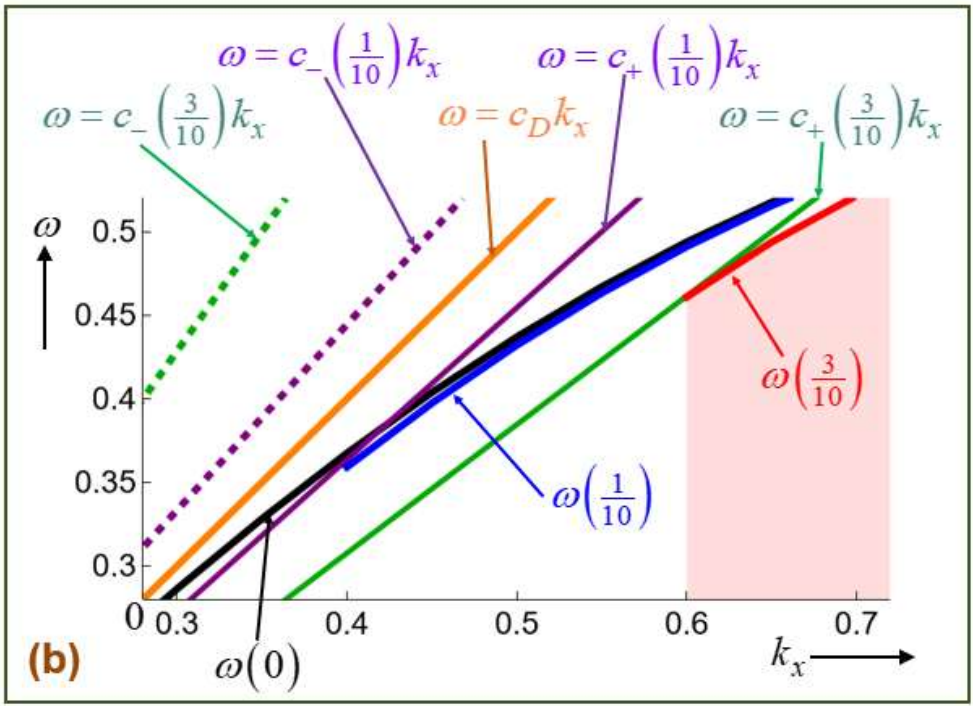

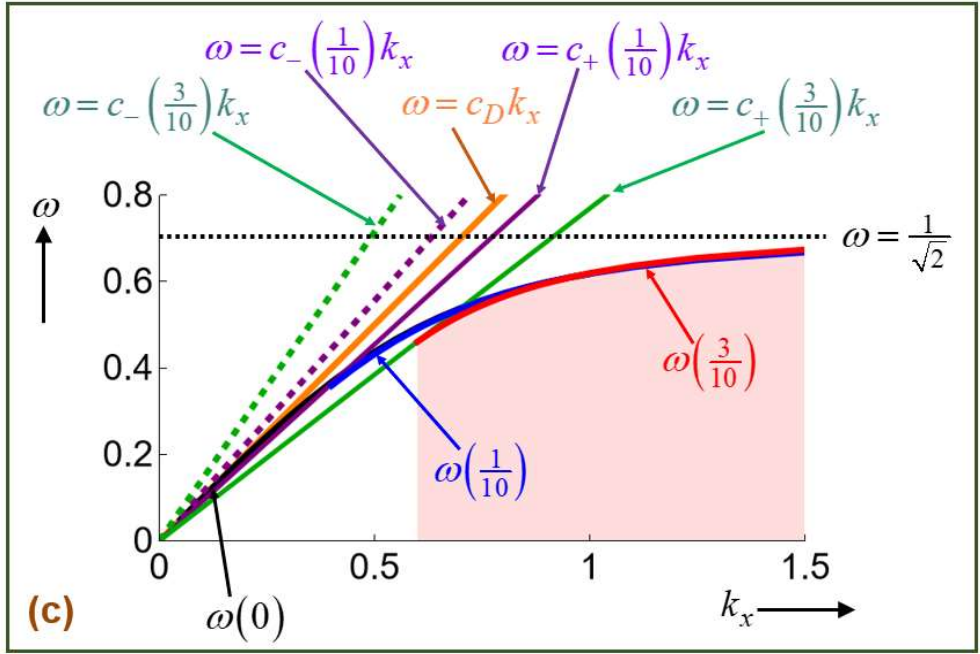

Fig. S2. (a) Fig. 2 in Primary Document reproduced. (b) A zoomed-in version of Fig. 2 in Primary Document over a reduced interval $k_x$. (c) Fig. 2 in Primary Document reproduced on a wider range of $0 \leq k_x \leq 1.5$, where the asymptotic value $\omega = \frac{1}{\sqrt{2}}$ is given by the dotted horizontal line. All panles are plotted on the parameter plane of $\{k_x, \omega\}$. The curves $\omega(k_x)$ are solutions to the dispersion relation over the interval $0 < k_x < \infty$. Given data includes $\varepsilon_D = \mu_D = \mu_M = 1$ and $\omega_p = 1$. Hence, $c_D = 1$. As a reference, straight lines are given respectively by $\omega = c_\pm k_x$ and $\omega = c_D k_x$.

From the zoomed-in Fig. S2(b), the cut-offs on $\omega(k_x)$ are more discernible. They are differentiated by the number within ( ) means the value of the assigned $\kappa$ such that $\omega(0), \omega\left(\frac{1}{10}\right), \omega\left(\frac{3}{10}\right)$. Hence, two of $\kappa = \frac{1}{10}, \frac{3}{10}$ are chosen for the chirality parameter.

As a reference, we list below several key parameters.

$$\begin{cases} c_D^{-1} \equiv \sqrt{\varepsilon_D \mu_D} \\ c_\pm^{-1} \equiv \sqrt{\varepsilon_D \mu_D} \pm \kappa \end{cases}, \quad k_\pm \equiv \frac{\omega}{c_\pm}, \quad \gamma_\pm \equiv \sqrt{k_x^2 - k_\pm^2}. \quad (S5.2)$$

Like $\{\Pi_D, \Pi_M\}$ for an achiral case, we define $\Pi_\pm \equiv \exp(ik_x x - \gamma_\pm z)$ with $\gamma_\pm > 0$ for a chiral medium. Here, $c_\pm$ are the respective light speeds. The dimensionless light speed $c_D$ for an achiral case is recovered by setting $\kappa = 0$ in $c_\pm$. Furthermore, $\{k_\pm, \gamma_\pm\}$ are wave numbers and decay rates.

Figure S2(a) is reproduced from Fig. 2 in Primary Document. On the other hand, Fig. S2(b) is a zoomed-in version Fig. 2 of Primary Document over a reduced interval $k_x$. Here, the black solid curve pointed out by $\omega(0)$ displays the dispersion relation $\varepsilon_D \gamma_M = |\varepsilon_M| \gamma_D$. Certainly, $\omega(0)$ on resonance increases monotonically with $k_x$. Besides, the straight line along $\omega = c_D k_D$ indicates the light line.

All panels of Fig. S2 are plotted on the $\{k_x, \omega\}$-parameter plane. Figure S2(b) shows an additional set of curves for $\kappa = \frac{1}{10}$. Together, the solution curves are drawn as solid curves denoted by $\omega\left(\frac{1}{10}\right)$ in blue and $\omega\left(\frac{3}{10}\right)$ in red. The frequencies on both curves exhibit monotonically increasing behaviors with $k_x$. Both solid curves for the chiral case are lying below the black solid curve obtained for the achiral case. In addition, $\omega\left(\frac{3}{10}\right)$ lies uniformly below $\omega\left(\frac{1}{10}\right)$. The orange solid line is the light line in vacuum indicated by $\omega = c_D k_x$. In addition, there are four straight lines $\omega = c_\pm \left(\frac{1}{10}\right) k_x$ and $\omega = c_\pm \left(\frac{3}{10}\right) k_x$. The superluminal curves $\omega = c_- \left(\frac{1}{10}\right) k_x$ are steeper than the subluminal curves $\omega = c_+ \left(\frac{1}{10}\right) k_x$.

It is seen here that the blue solution curve $\omega\left(\frac{1}{10}\right)$ for the chiral case lies barely below the solution curve $\omega(\kappa = 0)$ obtained for the achiral case. The portion of the blue curve $\omega\left(\frac{1}{10}\right)$ lying below the purple light line $\omega = c_+ \left(\frac{1}{10}\right) k_x$ is discarded. Likewise, the portion of the red curve $\omega\left(\frac{3}{10}\right)$ lying below the green light line $\omega = c_+ \left(\frac{3}{10}\right) k_x$ is discarded. The pink shaded area under the blue solid curve refers to the allowed frequency range for the chiral case with $\kappa = \frac{3}{10}$.

The bigger arrow in grey color lying on the top right corner on Fig. S2(a) points to the usual asymptotic frequency $\omega = \frac{1}{\sqrt{2}} = 0.707$ in case of a vacuum, which is easily computed from the resonance condition for the achiral case as follows.

$$\begin{cases} k_x > 0 \\ \mu_D = \mu_M > 0 \end{cases}, \quad \varepsilon_M = 1 - \frac{1}{\omega^2} < 0 \quad \Rightarrow \quad \frac{k_x}{k_D} \equiv \frac{k_x}{\omega}\sqrt{\varepsilon_D \mu_D} = \sqrt{\frac{1-\omega^2}{1-(1+\varepsilon_D)\omega^2}}$$

$$\lim_{(1+\varepsilon_D)\omega^2 \to 1} \frac{k_x}{k_D} \equiv \lim_{(1+\varepsilon_D)\omega^2 \to 1} \frac{k_x}{\omega}\sqrt{\varepsilon_D \mu_D} = \lim_{(1+\varepsilon_D)\omega^2 \to 1} \sqrt{\frac{1-\omega^2}{1-(1+\varepsilon_D)\omega^2}} \to \infty$$

(S5.3)

Therefore, taking a limit as $(1+\varepsilon_D)\omega^2 \to 1$ gives us $k_x \to \infty$. This limit corresponds to the famous high-frequency cut-off $\omega = \frac{1}{\sqrt{2}} = 0.707$ in case with a vacuum with $\varepsilon_D = \mu_D = 1$. Let us revert to the dimensional form of the metallic property $\varepsilon_M = 1 - \omega^{-2} = 1 - \left(\bar{\omega}_p/\bar{\omega}\right)^{-2} < 0$ with $\bar{\omega}_p$ being a prescribed plasma frequency. The high-frequency gap $\frac{1}{\sqrt{2}} = 0.707 < \omega < 1$ or $\frac{1}{\sqrt{2}}\bar{\omega}_p = 0.707\bar{\omega}_p < \bar{\omega} < \bar{\omega}_p$ signifies an insufficient metallic support for the surface plasmon waves. In other words, the additional high-frequency range $\frac{1}{\sqrt{2}} = 0.707 < \omega < 1$ is disallowed.

Consequently, we can easily derive the allowed range of $|\kappa|$ for surface plasmon resonances in the following fashion.

$$\gamma_\pm \equiv \sqrt{k_x^2 - k_\pm^2} \equiv \sqrt{k_x^2 - \omega^2\left(\sqrt{\varepsilon_D \mu_D} \pm \kappa\right)^2} \to 0 \Rightarrow k_x^2 = \omega^2\left(\sqrt{\varepsilon_D \mu_D} + |\kappa|\right)^2$$
$$\varepsilon_D = \mu_D = 1 \Rightarrow |\kappa| = \frac{k_x}{\omega} - \sqrt{\varepsilon_D \mu_D} = \sqrt{2}k_x - \sqrt{\varepsilon_D \mu_D} \quad \text{(S5.4)}$$

Therefore, $|\kappa|$ is constrained by $|\kappa| = \sqrt{2}k_x - \sqrt{\varepsilon_D \mu_D}$ in the limit as $k_x \to \infty$. Since $|\kappa| \approx 0.01$ or smaller in practice, the allowed range $|\kappa|$ is practically infinite.

Figure S2 shows that only the solution curves below $\omega(0) \equiv \omega(\kappa = 0)$ are allowed as indicated by the red and blue curves. This low-frequency cut-off or the low-wave-number cut-off phenomenon is one of key findings of this study. It is numerically found that $0.4 < k_x < \infty$ with $0.359 < \omega < \frac{1}{\sqrt{2}} = 0.707$ and $0.6 < k_x < \infty$ with $0.461 < \omega < \frac{1}{\sqrt{2}} = 0.707$ are the allowed ranges respectively for $\kappa = \frac{1}{10}, \frac{3}{10}$. In terms of the relative electric permittivity $\varepsilon_M = 1 - \left(\omega_p^2/\omega^2\right)$ with $\omega_p = 1$ for a metal, these allowed ranges translate respectively into $-6.76 < \varepsilon_M < -1$ and $-3.71 < \varepsilon_M < -1$, thereby disallowing weaker metals from supporting surface plasmon resonances.

In brief,

$$\begin{cases} \kappa = \frac{1}{10}: & 0.4 < k_x < \infty \\ \kappa = \frac{3}{10}: & 0.6 < k_x < \infty \end{cases} \Rightarrow \begin{cases} 0.359 < \omega < \frac{1}{\sqrt{2}} = 0.707, & -6.76 < \varepsilon_M < -1 \\ 0.461 < \omega < \frac{1}{\sqrt{2}} = 0.707, & -3.71 < \varepsilon_M < -1 \end{cases}. \quad \text{(S5.5)}$$

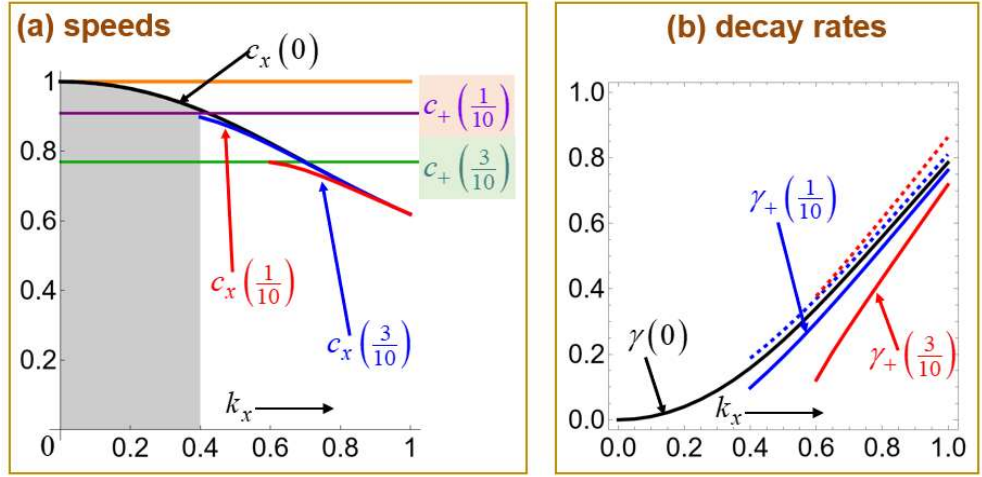

Fig. S3. (a) Deceleration shown by the phase speeds $c_x$, and (b) confinements indicated by the decay rates $\gamma_\pm$. Both are plotted against the longitudinal wave number $k_x$ on the horizontal axis.

Based on the dispersion curves $\{\omega, k_x\}$ displayed on Fig. S2, Fig. S3(a) displays various phase speeds, whereas Fig. S3(b) displays the decay rates $\gamma_\pm$ for the chiral case. All results displayed on Figs. S2 and S3 have been obtained $\kappa > 0$ for fixedness. Notice that the dispersion curves on $\{\omega, k_x\}$ are symmetric with respect to $k_x$, namely, $\omega(k_x) = \omega(-k_x)$.

On Fig. S3(a), $c_x(\kappa) \equiv \omega/k_x(\kappa)$ obtained from the pair $\{\omega, k_x(\kappa)\}$, which are in turn the solution to the chiral dispersion relation. The achiral case gives rise to $c_x(0)$, whereas the chiral case provides two curves $\{c_x(\tfrac{1}{10}), c_x(\tfrac{3}{10})\}$ with a pair of medium chirality $\kappa = \tfrac{1}{10}, \tfrac{3}{10}$. There are additional straight lines as guides for viewing. Here, only two subluminal curves $c_+(\kappa) < 1$ are displayed as horizontal lines. Notice that $c_+(\tfrac{3}{10}) < c_+(\tfrac{1}{10}) < c(0) < 1 \equiv c_D$, where $c_D = (\varepsilon_D \mu_D)^{-1/2} = 1$ for our choice $\varepsilon_D = \mu_D = 1$. In comparison, the superluminal curves are not shown here since $c_-(\kappa) > 1$. Hence, a larger medium chirality leads to a smaller phase speed. Note that $c_x(0)$ on Fig. S3(a) and $\gamma(0)$ on Fig. S3(b) are obtained for the achiral case on surface plasmon resonance along a metal-dielectric (M-D) interface.

Figure S3(b) presents the decay rates $\gamma_\pm \equiv \sqrt{k_x^2 - k_\pm^2}$ in the chiral case. Only the subluminal solid curves $\gamma_+$ are of interests to us. Both curves lie below the curve for the achiral case, viz., $\gamma_+\left(\frac{3}{10}\right) < \gamma_+\left(\frac{1}{10}\right) < \gamma(0)$. In other words, a larger medium chirality leads to a smaller decay rate (a looser confinement) in the chiral medium. Notice that a larger medium chirality leads to a smaller phase speed. The disallowed range, for $\kappa = \frac{1}{10}$, is marked by a gray filling on Fig. S3(a). Such a trend is characterized by the steeper gradients on Fig. S2 from curve $\omega(0)$ through line $\omega = c_+ k_x$ with $\kappa = \frac{1}{10}$ to line $\omega = c_+ k_x$ with $\kappa = \frac{3}{10}$. The frequency cut-offs discussed for Figs. S2 and S3 stem from confinement or evanescence since $\gamma_\pm > 0$. This frequency cut-offs can be certainly implemented for high-frequency pass filters.

From measurement perspectives, we can make use of the two disparate decay rates $\exp(-\gamma_\pm z)$ for the chiral case. Such double decay rates lead in turn to three distinct decay rates in the field intensities, namely, $\exp(-2\gamma_\pm z)$ and $\exp[-(\gamma_+ + \gamma_-)z]$ arising from interferences. This triple decay rates will have great implication in sensing applications. Symbolically, we could express such dependencies as follows.

$$\begin{cases} A, B \in \mathbb{C} \\ \left|A\exp(-\gamma_+ z) + B\exp(-\gamma_- z)\right|^2 \end{cases} \Rightarrow \begin{cases} \exp(-2\gamma_\pm z) \\ \exp[-(\gamma_+ + \gamma_-)z] \end{cases}. \quad (S5.6)$$

From the matrix equation determining the eigenvalues and eigenvectors, we obtain the relationship among the components $\{P_+ \equiv Q_+, P_-, G_{Ex}, G_{Ey}\}$ of the eigenvector. What we need is only the ratio between $\{P_+ \equiv Q_+, P_-\}$ in the following manner since we are currently concerned with the fields established in a chiral medium.

$$\begin{pmatrix} \gamma_+ & -\gamma_- & -1 & 0 \\ k_+ & k_- & 0 & -1 \\ \gamma_+ & \gamma_- & 0 & \frac{Z_D \gamma_M}{\omega \mu_M} \\ k_+ & -k_- & \frac{Z_D \omega \varepsilon_M}{\gamma_M} & 0 \end{pmatrix} \begin{Bmatrix} Q_+ \\ P_- \\ G_{Ex} \\ G_{Ey} \end{Bmatrix} = \begin{Bmatrix} 0 \\ 0 \\ 0 \\ 0 \end{Bmatrix} \Rightarrow$$

$$\begin{cases} \gamma_+ Q_+ - \gamma_- P_- = G_{Ex} \\ k_+ Q_+ - k_- P_- = -\frac{Z_D \omega \varepsilon_M}{\gamma_M} G_{Ex} \end{cases}, \begin{cases} \gamma_+ \frac{Z_D \omega \varepsilon_M}{\gamma_M} Q_+ - \gamma_- \frac{Z_D \omega \varepsilon_M}{\gamma_M} P_- = \frac{Z_D \omega \varepsilon_M}{\gamma_M} G_{Ex} \\ k_+ Q_+ - k_- P_- = -\frac{Z_D \omega \varepsilon_M}{\gamma_M} G_{Ex} \end{cases}. \quad (S5.7)$$

$$\Rightarrow \left(k_+ + \gamma_+ \frac{Z_D \omega \varepsilon_M}{\gamma_M}\right) Q_+ = \left(k_- + \gamma_- \frac{Z_D \omega \varepsilon_M}{\gamma_M}\right) P_-$$

We find convenient to introduce the intermediaries $\{\beta_\pm, \Gamma\}$, where $\beta_\pm \in \mathbb{R}$ is a ratio and $\Gamma \in \mathbb{C}$ is what we call a 'complex magnitude'.

$$\beta_{\pm} \equiv k_{\mp} + \gamma_{\mp} \frac{Z_D \omega \varepsilon_M}{\gamma_M}, \quad Y \equiv Z_D \Gamma \Rightarrow$$

$$iZ_D \beta_+ \beta_- \Gamma \equiv \beta_- Q_+ = \beta_+ P_- = iZ_D \beta_+ Q_- \Rightarrow \begin{cases} Q_+ \equiv iZ_D \beta_+ \Gamma \\ Q_- \equiv \beta_- \Gamma \end{cases}. \quad (S5.8)$$

$$\begin{cases} P_+ \equiv Q_+ \\ P_- \equiv iZ_D Q_- \end{cases} \Rightarrow \quad \color{red}{P_\pm \equiv iZ_D \beta_\pm \Gamma}$$

Hence, we arrived at a neat pair of the eigenvector components $\{P_+, P_-\}$ as follows.

$$\frac{P_+}{P_-} \equiv \frac{Q_+}{iZ_D Q_-} = \frac{iZ_D \left( k_- + \gamma_- \frac{Z_D \omega \varepsilon_M}{\gamma_M} \right) \Gamma}{iZ_D \left( k_+ + \gamma_+ \frac{Z_D \omega \varepsilon_M}{\gamma_M} \right) \Gamma}$$

$$= \frac{k_- + \gamma_- \frac{Z_D \omega \varepsilon_M}{\gamma_M}}{k_+ + \gamma_+ \frac{Z_D \omega \varepsilon_M}{\gamma_M}} \equiv \frac{\beta_+}{\beta_-} = \frac{\gamma_M k_- - \gamma_- Z_D \omega |\varepsilon_M|}{\gamma_M k_+ - \gamma_+ Z_D \omega |\varepsilon_M|} \equiv \beta_-^+ \quad (S5.9)$$

In particular, the real value $P_+/P_- \equiv Q_+/(iZ_D Q_-)$ is the left-right mixture ratio, which is essentially the TM-TE mixture ratio. The fact that $P_+, P_- \in \mathbb{R}$ corresponds to both pairs $\{E_z, E_x\}$ and $\{E_y, E_z\}$ standing respectively out of phase by $\pm 90^o$.

Because $\varepsilon_M < 0$, this ratio $\beta_-^+ \equiv \beta_+/\beta_- \equiv P_+/P_- \equiv Q_+/iZ_D Q_-$ plays a crucial role in determining the depth-wise behaviors of various bilinear parameters. This ratio depending on the solution to the chiral dispersion relation could lead to sign flips. Only at two frequencies for the last couple of relations, a TE wave is separable from a TM wave.

$$k_x(\omega): \begin{cases} \dfrac{P_+}{P_-} = 0 \Rightarrow k_- = \gamma_- \dfrac{Z_D \omega |\varepsilon_M|}{\gamma_M} \\ \dfrac{P_+}{P_-} = \infty \Rightarrow k_+ = \gamma_+ \dfrac{Z_D \omega |\varepsilon_M|}{\gamma_M} \end{cases}. \quad (S5.10)$$

Each of these conditions is hardly satisfied since $k_x(\omega)$ should be satisfied on resonance as well.

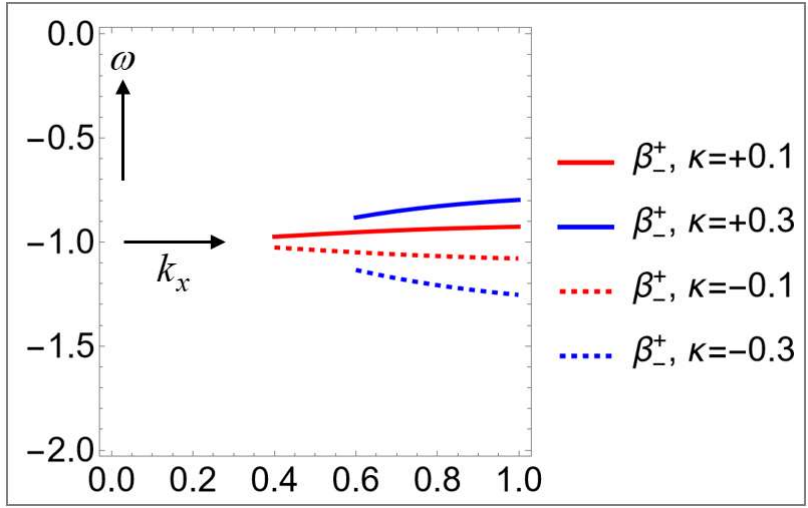

Fig. S4. The dispersion ratio $\beta_-^+ \equiv \beta_+/\beta_- \equiv P_+/P_- \equiv Q_+/(iZ_D Q_-)$ plotted against $k_x$ over the interval $0 < k_x < \infty$. Two of $\kappa = \frac{1}{10}, \frac{3}{10}$ are chosen for the chirality parameter. Given data includes $\varepsilon_D = \mu_D = \mu_M = 1$ and $\omega_p = 1$.

Figure S4 shows the dispersion ratio $\beta_-^+ \equiv \beta_+/\beta_- \equiv P_+/P_- \equiv Q_+/(iZ_D Q_-)$ plotted against $k_x$ over the interval $0 < k_x < \infty$. It turns out that $-1 < \beta_-^+ < -\frac{1}{2}$ for $\kappa > 0$, whereas $-\frac{3}{2} < \beta_-^+ < -1$ for $\kappa < 0$. Besides, this ratio is symmetric across the horizontal line at $\beta_-^+ = -1$, namely, $\beta_-^+(\kappa) + \beta_-^+(-\kappa) = -2$. Besides, $\beta_-^+ < 0$ corresponds to $\beta_+ \beta_- < 0$. In brief, $\beta_-^+ < 0$ as seen from numerical results.

For upcoming tests of symmetry or anti-symmetry, it is useful to examine the following.

$$\begin{cases} k_\pm \equiv \omega\left(\sqrt{\varepsilon\mu} \pm \kappa\right) & \Rightarrow \quad k_\pm(-\kappa) = k_\mp(\kappa) \\ \gamma_\pm \equiv \sqrt{k_x^2 - k_\pm^2} & \Rightarrow \quad \gamma_\pm(-\kappa) = \gamma_\mp(\kappa) \\ \beta_\pm \equiv k_\mp + \gamma_\mp \dfrac{Z_D \omega \varepsilon_M}{\gamma_M} & \Rightarrow \quad \beta_\pm(-\kappa) = \beta_\mp(\kappa) \end{cases} \Rightarrow \qquad (S5.11)$$

Here, we included the sign flips and reversals in $k_\pm(-\kappa) = k_\mp(\kappa)$ and $\gamma_\pm(-\kappa) = \gamma_\mp(\kappa)$ as seen from construction.

There is one more concern of the sign in $\gamma_+ - \gamma_-$ in the following manner.

$$\begin{cases} k_+ \equiv \omega\left(\sqrt{\varepsilon\mu}+\kappa\right) \equiv k_x \sin\theta_+ \\ k_- \equiv \omega\left(\sqrt{\varepsilon\mu}-\kappa\right) \equiv k_x \sin\theta_- \end{cases} \Rightarrow \begin{cases} \gamma_+ \equiv \sqrt{k_x^2 - k_+^2} = k_x \cos\theta_+ \\ \gamma_- \equiv \sqrt{k_x^2 - k_-^2} = k_x \cos\theta_- \end{cases} \Rightarrow$$

$$\gamma_+ - \gamma_- \equiv \sqrt{k_x^2 - k_+^2} - \sqrt{k_x^2 - k_-^2} \equiv \sqrt{k_x^2 - \omega^2\left(\sqrt{\varepsilon\mu}+\kappa\right)^2} - \sqrt{k_x^2 - \omega^2\left(\sqrt{\varepsilon\mu}-\kappa\right)^2} \quad \text{. (S5.12)}$$

$$\begin{cases} \kappa > 0: \quad k_+^2 > k_-^2 \Rightarrow \gamma_+ < \gamma_- \\ \kappa < 0: \quad k_+^2 < k_-^2 \Rightarrow \gamma_+ > \gamma_- \end{cases}$$

**Section S6. Electric-field intensity**

There are two ways of writing down the circular vectors $\{\mathbf{Q}_+, \mathbf{Q}_-\}$.

$$Z_D \equiv \sqrt{\frac{\mu}{\varepsilon}} \Rightarrow \begin{cases} \mathbf{E} = \mathbf{Q}_+ - iZ_D\mathbf{Q}_- \\ \mathbf{H} = -iZ_D^{-1}\mathbf{Q}_+ + \mathbf{Q}_- \end{cases} \Rightarrow \begin{cases} \mathbf{E} = \mathbf{Q}_+ - iZ_D\mathbf{Q}_- \\ iZ_D\mathbf{H} = \mathbf{Q}_+ + iZ_D\mathbf{Q}_- \end{cases}$$

$$\begin{cases} \mathbf{P}_+ \equiv \mathbf{Q}_+ \\ \mathbf{P}_- \equiv iZ_D\mathbf{Q}_- \end{cases} \Rightarrow \begin{cases} \mathbf{E} = \mathbf{P}_+ - \mathbf{P}_- \\ iZ_D\mathbf{H} = \mathbf{P}_+ + \mathbf{P}_- \end{cases} \quad \text{(S6.1)}$$

Let us evaluate the electric-field intensity $|\mathbf{E}|^2 \equiv \mathbf{E}^* \cdot \mathbf{E}$. We are mainly concerned only with the chiral medium (not a metal) that fills the semi-infinite space with $z > 0$. Hence, any notation such as the subscripted $\mathbf{E}_D$ will not be employed so that we just employ the simpler $\mathbf{E}$. Let us express the electric-field intensity in terms of the circular vectors $\{\mathbf{Q}_+, \mathbf{Q}_-\}$.

$$\begin{aligned} |\mathbf{E}|^2 &\equiv \mathbf{E}^* \cdot \mathbf{E} \equiv \mathbf{E} \cdot \mathbf{E}^* \\ &= (\mathbf{Q}_+ - iZ_D\mathbf{Q}_-) \cdot (\mathbf{Q}_+ - iZ_D\mathbf{Q}_-)^* = (\mathbf{Q}_+ - iZ_D\mathbf{Q}_-) \times (\mathbf{Q}_+^* + iZ_D\mathbf{Q}_-^*) \\ &= \mathbf{Q}_+ \cdot \mathbf{Q}_+^* + Z_D^2\mathbf{Q}_- \cdot \mathbf{Q}_-^* - iZ_D\mathbf{Q}_- \cdot \mathbf{Q}_+^* + iZ_D\mathbf{Q}_+ \cdot \mathbf{Q}_-^* \\ &= \mathbf{Q}_+ \cdot \mathbf{Q}_+^* + Z_D^2\mathbf{Q}_- \cdot \mathbf{Q}_-^* - iZ_D\mathbf{Q}_- \cdot \mathbf{Q}_+^* + iZ_D\mathbf{Q}_-^* \cdot \mathbf{Q}_+ \\ &= |\mathbf{Q}_+|^2 + Z_D^2|\mathbf{Q}_-|^2 + 2Z_D \operatorname{Im}(\mathbf{Q}_- \cdot \mathbf{Q}_+^*) = |\mathbf{P}_+|^2 + |\mathbf{P}_-|^2 + 2\operatorname{Re}(\mathbf{P}_- \cdot \mathbf{P}_+^*) \end{aligned} \quad \text{. (S6.2)}$$

$$Z_D^2|\mathbf{H}|^2 = |\mathbf{Q}_+|^2 + Z_D^2|\mathbf{Q}_-|^2 - 2Z_D \operatorname{Im}(\mathbf{Q}_- \cdot \mathbf{Q}_+^*) = |\mathbf{P}_+|^2 + |\mathbf{P}_-|^2 - 2\operatorname{Re}(\mathbf{P}_- \cdot \mathbf{P}_+^*)$$

Here, the self-dot-products $|\mathbf{Q}_\pm|^2 \equiv \mathbf{Q}_\pm \cdot \mathbf{Q}_\pm^*$ are employed. The complex magnitudes $\{Q_+, Q_-\}$ are not independent of each other. Rather, they are related to each other by the solution of the chiral dispersion relation obtained on surface plasmon resonance, as we have done in the preceding section. The magnetic-field intensity is analogously evaluated, while the sole difference is the negative sign in the interference term and the pre-multiplier $Z_D^2 \equiv \mu/\varepsilon$.

To evaluate each of $\{|\mathbf{Q}_+|^2, |\mathbf{Q}_-|^2, \text{Im}(\mathbf{Q}_-\cdot\mathbf{Q}_+^*)\}$, we need to find the following set of dot-products and vector products.

$$\begin{cases} k_\pm \equiv \omega\left(\sqrt{\varepsilon_D \mu_D} \pm \kappa\right) \\ \gamma_\pm \equiv \sqrt{k_x^2 - k_\pm^2} \end{cases} \Rightarrow \begin{cases} k_\pm(-\kappa) = k_\mp(\kappa) \\ \gamma_\pm(-\kappa) = \gamma_\mp(\kappa) \end{cases}$$

$$\Pi_\pm \equiv \exp(ik_x x - \gamma_\pm z), \quad \mathbf{Q}_\pm = Q_\pm \frac{1}{\sqrt{2}} \frac{1}{k_x}(\gamma_\pm \hat{\mathbf{x}} \pm k_\pm \hat{\mathbf{y}} + ik_x \hat{\mathbf{z}}) \Pi_\pm \Rightarrow$$

$$\frac{2k_x^2 \mathbf{Q}_\pm \cdot \mathbf{Q}_\pm^* \equiv 2k_x^2 |\mathbf{Q}_\pm|^2}{|Q_\pm|^2 \exp(-2\gamma_\pm z)} = (\gamma_\pm \hat{\mathbf{x}} \pm k_\pm \hat{\mathbf{y}} + ik_x \hat{\mathbf{z}}) \cdot (\gamma_\pm \hat{\mathbf{x}} \pm k_\pm \hat{\mathbf{y}} - ik_x \hat{\mathbf{z}})$$

$$= \gamma_\pm^2 + k_\pm^2 + k_x^2 = k_x^2 - k_\pm^2 + k_\pm^2 + k_x^2 = 2k_x^2 \Rightarrow |\mathbf{Q}_\pm|^2 = |Q_\pm|^2 \exp(-2\gamma_\pm z). \quad (S6.3)$$

$$\frac{2k_x^2 \text{Im}(\mathbf{Q}_-\cdot\mathbf{Q}_+^*)}{Q_-Q_+^* \exp[-(\gamma_+ + \gamma_-)z]} = (\gamma_- \hat{\mathbf{x}} - k_- \hat{\mathbf{y}} + ik_x \hat{\mathbf{z}}) \cdot (\gamma_+ \hat{\mathbf{x}} + k_+ \hat{\mathbf{y}} + ik_x \hat{\mathbf{z}})^*$$

$$= (\gamma_- \hat{\mathbf{x}} - k_- \hat{\mathbf{y}} + ik_x \hat{\mathbf{z}}) \cdot (\gamma_+ \hat{\mathbf{x}} + k_+ \hat{\mathbf{y}} - ik_x \hat{\mathbf{z}}) = \gamma_+\gamma_- - k_+k_- + k_x^2 \Rightarrow$$

$$2k_x^2 \text{Im}(\mathbf{Q}_-\cdot\mathbf{Q}_+^*) = \text{Im}(Q_-Q_+^*)(\gamma_+\gamma_- - k_+k_- + k_x^2)\exp[-(\gamma_+ + \gamma_-)z]$$

It is worth noting that $\{\mathbf{Q}_+, \mathbf{Q}_-\}$ are $z$-dependent evanescent vectors, whereas $\{Q_+, Q_-\}$ are just spatially homogeneous complex scalars. This cross-coupling factor $\text{Im}(Q_-Q_+^*)$ will play a great role in the ensuing development. Besides, we have utilized $\gamma_\pm^2 \equiv k_x^2 - k_\pm^2$ during the process of proof.

Collecting the three terms for the electric-field intensity, we get the following.

$$k_x^2 |\mathbf{E}|^2 = k_x^2 |\mathbf{Q}_+|^2 + k_x^2 Z_D^2 |\mathbf{Q}_-|^2 + 2k_x^2 Z_D \text{Im}(\mathbf{Q}_-\cdot\mathbf{Q}_+^*)$$
$$= |Q_+|^2 k_x^2 \exp(-2\gamma_+ z) + Z_D^2 |Q_-|^2 k_x^2 \exp(-2\gamma_- z) \quad (S6.4)$$
$$\pm Z_D \text{Im}(Q_-Q_+^*)(\gamma_+\gamma_- - k_+k_- + k_x^2)\exp[-(\gamma_+ + \gamma_-)z]$$

Meanwhile, let us pause to examine the following sum-difference parameter, which admits the following geometric interpretation.

$$\begin{cases} k_+ \equiv k_x \sin\theta_+ \\ k_- \equiv k_x \sin\theta_- \end{cases} \Rightarrow \begin{cases} \gamma_+ \equiv \sqrt{k_x^2 - k_+^2} = k_x \cos\theta_+ \\ \gamma_- \equiv \sqrt{k_x^2 - k_-^2} = k_x \cos\theta_- \end{cases} \Rightarrow$$

$$\gamma_+\gamma_- - k_+k_- + k_x^2 \equiv \sqrt{k_x^2 - k_+^2}\sqrt{k_x^2 - k_-^2} - k_+k_- + k_x^2$$
$$= k_x^2 \cos\theta_+ \cos\theta_- - k_x^2 \sin\theta_+ \sin\theta_- + k_x^2 = k_x^2[1 - \cos(\theta_+ + \theta_-)] \geq 0. \quad (S6.5)$$

$$\Rightarrow |\mathbf{E}|^2 = |Q_+|^2 \exp(-2\gamma_+ z) + Z_D^2 |Q_-|^2 \exp(-2\gamma_- z)$$
$$+ Z_D \text{Im}(Q_-Q_+^*)[1 - \cos(\theta_+ + \theta_-)]\exp[-(\gamma_+ + \gamma_-)z]$$

This form $1 - \cos(\theta_+ + \theta_-)$ lends itself to a certain geometric interpretation for the cross (interaction) term.

To evaluate the field intensities, it is better to examine $\{\mathrm{Re}(Q_-^* Q_+), \mathrm{Im}(Q_-^* Q_+)\}$ separately based on the circular vectors.

$$\begin{cases} P_+ \equiv Q_+ \\ P_- \equiv iZ_D Q_- \end{cases}, \quad \beta_\pm \equiv k_\mp + \gamma_\mp \frac{Z_D \omega \varepsilon_M}{\gamma_M} \in \mathbb{R}, \quad \begin{cases} Y \equiv Z_D \Gamma \\ P_\pm \equiv iZ_D \beta_\pm \Gamma \equiv i\beta_\pm Y \end{cases} \Rightarrow$$

$$Q_-^* Q_+ = (\beta_-\Gamma)^* iZ_D \beta_+ \Gamma = \beta_-\Gamma^* iZ_D \beta_+ \Gamma = iZ_D \beta_+ \beta_- |\Gamma|^2 \Rightarrow \qquad (S6.6)$$

$$\begin{cases} \mathrm{Re}(Q_-^* Q_+) = 0 = \mathrm{Re}(Q_- Q_+^*) \\ \mathrm{Im}(Q_-^* Q_+) = Z_D |\Gamma|^2 \beta_+ \beta_- = -\mathrm{Im}(Q_- Q_+^*) \end{cases}$$

Let us finally express both electric- and magnetic field intensities in terms of the single complex magnitude $\Gamma$ in the following fashion.

$$\begin{aligned}
Z_D^{-2} |\Gamma|^{-2} |\mathbf{E}|^2 &\equiv |Y|^{-2} |\mathbf{E}|^2 = (\beta_+)^2 \exp(-2\gamma_+ z) + (\beta_-)^2 \exp(-2\gamma_- z) \\
&\quad - \beta_+ \beta_- [1 - \cos(\theta_+ + \theta_-)] \exp[-(\gamma_+ + \gamma_-)z] \\
&= (\beta_+)^2 \exp(-2\gamma_+ z) + (\beta_-)^2 \exp(-2\gamma_- z) \\
&\quad - \beta_+ \beta_- \frac{\gamma_+ \gamma_- - k_+ k_- + k_x^2}{k_x^2} \exp[-(\gamma_+ + \gamma_-)z] \\
Z_D^{-2} |\Gamma|^{-2} |\mathbf{H}|^2 &\equiv |Y|^{-2} |\mathbf{H}|^2 = (\beta_+)^2 \exp(-2\gamma_+ z) + (\beta_-)^2 \exp(-2\gamma_- z) \\
&\quad + \beta_+ \beta_- \frac{\gamma_+ \gamma_- - k_+ k_- + k_x^2}{k_x^2} \exp[-(\gamma_+ + \gamma_-)z]
\end{aligned} \qquad (S6.7)$$

Therefore, we obtained quite concise forms for the real-valued field energy densities (both electric and magnetic). We will find later that the sign reversal $\mp \beta_+ \beta_- (\circ)$ in the coupling terms of $\{|\mathbf{E}|^2, |\mathbf{H}|^2\}$ plays a crucial role in determining the even and/or odd properties of the spin density.

It is instructive to examine this formula for the reduction to the achiral case.

$$\kappa = 0 \Rightarrow \begin{cases} k_\pm = k_D \Rightarrow \gamma_\pm = \gamma_D = \sqrt{k_x^2 - k_D^2} \\ \gamma_+ \gamma_- - k_+ k_- + k_x^2 = (k_x^2 - k_D^2) - k_D^2 + k_x^2 = 2k_x^2 - 2k_D^2 \end{cases} \Rightarrow$$

$$\begin{aligned}
k_x^2 |\mathbf{E}|^2 &= |Q_+|^2 k_x^2 \exp(-2\gamma_+ z) + Z_D^2 |Q_-|^2 k_x^2 \exp(-2\gamma_- z) \\
&\quad + 2Z_D \mathrm{Im}(Q_- Q_+^*)(k_x^2 - k_D^2) \exp[-(\gamma_+ + \gamma_-)z]
\end{aligned} \qquad (S6.8)$$

Resultantly, we can make use of the following boundedness properties of the field intensities for the achiral case.

$$\kappa = 0 \Rightarrow \begin{cases} k_+ \equiv k_x \sin\theta_+ \\ k_- \equiv k_x \sin\theta_- \end{cases}, \quad \begin{cases} |\mathrm{Im}(Q_- Q_+^*)| \leq |Q_+||Q_-| \\ \gamma_+ \gamma_- - k_+ k_- + k_x^2 \equiv k_x^2 [1 - \cos(\theta_+ + \theta_-)] \geq 0 \end{cases} \Rightarrow$$

$$\begin{aligned}
B_\pm &\equiv |Q_+|^2 \exp(-2\gamma_+ z) + Z_D^2 |Q_-|^2 \exp(-2\gamma_- z) \pm 2Z_D |Q_+||Q_-| \exp[-(\gamma_+ + \gamma_-)z] \\
&= [|Q_+| \exp(-\gamma_+ z) \pm Z_D |Q_-| \exp(-\gamma_- z)]^2 \Rightarrow B_- \leq I_{EH}^\pm \leq B_+
\end{aligned} \qquad (S6.9)$$

Consequently, we have proved the maximum bound of the field intensities.

We are now ready to take a spatial integration of this intensity over the upper half-space.

$$Z_D^{-2}|\Gamma|^{-2}|\mathbf{E}|^2 \equiv |Y|^{-2}|\mathbf{E}|^2 = (\beta_+)^2 \exp(-2\gamma_+ z) + (\beta_-)^2 \exp(-2\gamma_- z)$$

$$-\beta_+\beta_- \frac{\gamma_+\gamma_- - k_+k_- + k_x^2}{k_x^2} \exp\left[-(\gamma_+ + \gamma_-)z\right] \Rightarrow \int|\mathbf{E}|^2 \equiv \int_0^\infty |\mathbf{E}|^2\, dz \Rightarrow \quad . \quad \text{(S6.10)}$$

$$\begin{cases} Z_D^{-2}|\Gamma|^{-2}\int|\mathbf{E}|^2 \equiv |Y|^{-2}\int|\mathbf{E}|^2 = \dfrac{(\beta_+)^2}{2\gamma_+} + \dfrac{(\beta_-)^2}{2\gamma_-} - \beta_+\beta_-\dfrac{\gamma_+\gamma_- - k_+k_- + k_x^2}{k_x^2(\gamma_+ + \gamma_-)} \\ Z_D^{-2}|\Gamma|^{-2}\int|\mathbf{H}|^2 \equiv |Y|^{-2}\int|\mathbf{H}|^2 = \dfrac{(\beta_+)^2}{2\gamma_+} + \dfrac{(\beta_-)^2}{2\gamma_-} + \beta_+\beta_-\dfrac{\gamma_+\gamma_- - k_+k_- + k_x^2}{k_x^2(\gamma_+ + \gamma_-)} \end{cases}$$

This depth-wise integrated electric-field intensity is not separable into two factors. Notice from numerical tests, we learned that $\beta_+\beta_- < 0$. Hence, we keep a possibility only of the magnetic-field intensity going down to zero, which turns out not to be the case.

By following the practice in determining the evenness and/or the oddness of the quantum wave functions, we exploit the series of symmetry properties: $k_\pm(-\kappa) \equiv k_\mp(\kappa)$, $\gamma_\pm(-\kappa) \equiv \gamma_\mp(\kappa)$, and $\beta_\pm(-\kappa) \equiv \beta_\mp(\kappa)$; $\gamma_\pm(-k_x) \equiv \gamma_\pm(k_x)$ and $\beta_\pm(-k_x) \equiv \beta_\pm(k_x)$. We can thus prove in a formal way that electric-field intensity is symmetric with respect to the medium chirality as follows.

$$\begin{cases} k_\pm(-\kappa) \equiv k_\mp(\kappa) \\ \gamma_\pm(-\kappa) \equiv \gamma_\mp(\kappa) \end{cases} \Rightarrow \beta_\pm(-\kappa) \equiv \beta_\mp(\kappa) \Rightarrow$$

$$Z_D^{-2}|\Gamma|^{-2}|\mathbf{E}|^2 \equiv |Y|^{-2}|\mathbf{E}|^2 \equiv u(\kappa, z) \Rightarrow$$

$$u(\kappa, z) \equiv (\beta_+)^2 \exp\left[-2\gamma_+(\kappa)z\right] + (\beta_-)^2 \exp\left[-2\gamma_-(\kappa)z\right]$$

$$-\beta_+\beta_- \frac{\gamma_+\gamma_- - k_+k_- + k_x^2}{k_x^2} \exp\left\{-\left[\gamma_+(\kappa) + \gamma_-(\kappa)\right]z\right\} \Rightarrow \qquad \text{(S6.11)}$$

$$u(-\kappa, z) \equiv (\beta_-)^2 \exp\left[-2\gamma_-(\kappa)z\right] + (\beta_+)^2 \exp\left[-2\gamma_+(\kappa)z\right]$$

$$-\beta_-\beta_+ \frac{\gamma_-\gamma_+ - k_-k_+ + k_x^2}{k_x^2} \exp\left\{-\left[\gamma_-(\kappa) + \gamma_+(\kappa)\right]z\right\} = u(\kappa, z)$$

Likewise, the magnetic-field intensity is symmetric as well. Physically speaking, such symmetry properties are expected since the interchange in the left and right waves leads to an energetically identical state. Proofs of symmetry and anti-symmetry can be performed by inspection for experienced readers.

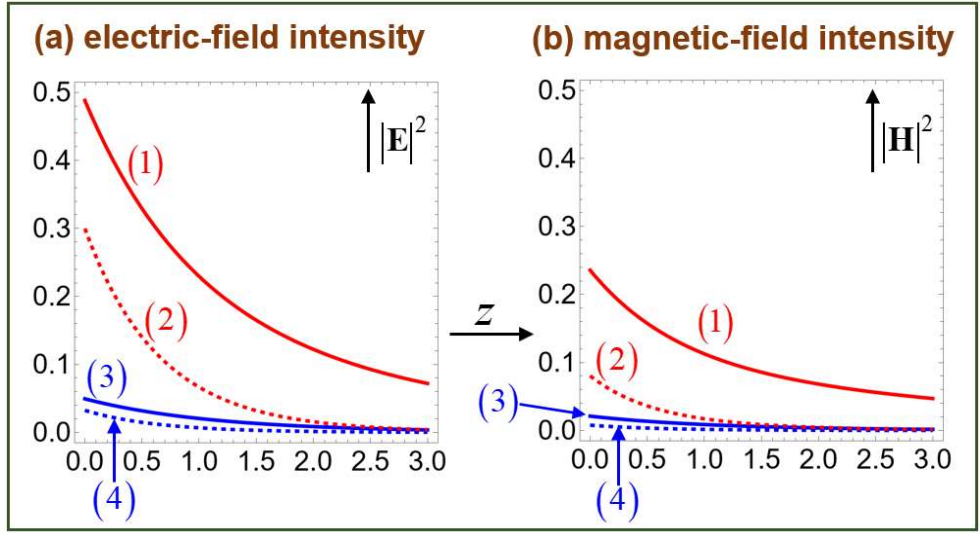

Fig. S5. (a) The electric-field intensity $Z_D^{-2} |\Gamma|^{-2} |\mathbf{E}|^2$ and (b) the magnetic-field intensity $Z_D^{-2} |\Gamma|^{-2} |\mathbf{H}|^2$ plotted against the depth-wise $z$-coordinate. It is specified that $Z_D |\Gamma| = 1$. Four curves on each panel are arranged according to the orders: two upper red curves (1,2) for $\kappa = \frac{3}{10}$ and the two lower blue curves (3,4) with $\kappa = \frac{1}{10}$. For each medium chirality, the solid curves refer to lower frequencies, whereas the dotted curves refer to higher frequencies.

In Fig. S5, the trio of $\{\kappa, k_x, \omega\}$ is hence specified from the top to the bottom: (1) $\{\frac{3}{10}, 0.7, 0.522\}$, (2) $\{\frac{3}{10}, 0.999, 0.619\}$, (3) $\{\frac{1}{10}, 0.7, 0.537\}$, and (4) $\{\frac{1}{10}, 0.999, 0.618\}$. Given data includes $\varepsilon_D = \mu_D = \mu_M = 1$ and $\omega_p = 1$. We learn from Fig. S5 that the electric-field intensity is roughly greater than the magnetic-field intensity over all depth-wise locations.

**Section S7. Spin density and its (anti-)symmetry**

We have seen in Eq. (S2.14) that the spin density $\text{Im}(\mathbf{E}^* \times \mathbf{E})$ represents the state of polarization. In that achiral case, there is a sole nonzero component $\left[\text{Im}(\mathbf{E}^* \times \mathbf{E})\right]_y$. In this chiral case, will find a two-component $\text{Im}(\mathbf{E}^* \times \mathbf{E})$. Let us prepare further dot- and cross-products between the circular vectors $\{\mathbf{Q}_+, \mathbf{Q}_-\}$.

Firstly, consider the cross-products.

$$Z_D \equiv \sqrt{\frac{\mu}{\varepsilon}} \Rightarrow \begin{cases} \mathbf{E} = \mathbf{Q}_+ - iZ_D\mathbf{Q}_- \\ \mathbf{H} = -iZ_D^{-1}\mathbf{Q}_+ + \mathbf{Q}_- \end{cases} \Rightarrow \begin{cases} \mathbf{E} = \mathbf{Q}_+ - iZ_D\mathbf{Q}_- \\ iZ_D\mathbf{H} = \mathbf{Q}_+ + iZ_D\mathbf{Q}_- \end{cases}$$

$$\begin{cases} \mathbf{P}_+ \equiv \mathbf{Q}_+ \\ \mathbf{P}_- \equiv iZ_D\mathbf{Q}_- \end{cases} \Rightarrow \begin{cases} \mathbf{E} = \mathbf{P}_+ - \mathbf{P}_- \\ iZ_D\mathbf{H} = \mathbf{P}_+ + \mathbf{P}_- \end{cases}$$

$$\mathbf{E}^* \times \mathbf{E} = (\mathbf{Q}_+ - iZ_D\mathbf{Q}_-)^* \times (\mathbf{Q}_+ - iZ_D\mathbf{Q}_-) = (\mathbf{Q}_+^* + iZ_D\mathbf{Q}_-^*) \times (\mathbf{Q}_+ - iZ_D\mathbf{Q}_-)$$
$$= \mathbf{Q}_+^* \times \mathbf{Q}_+ + Z_D^2 \mathbf{Q}_-^* \times \mathbf{Q}_- + iZ_D(\mathbf{Q}_-^* \times \mathbf{Q}_+ - \mathbf{Q}_+^* \times \mathbf{Q}_-) \qquad (S7.1)$$
$$= \mathbf{Q}_+^* \times \mathbf{Q}_+ + Z_D^2 \mathbf{Q}_-^* \times \mathbf{Q}_- + iZ_D(\mathbf{Q}_-^* \times \mathbf{Q}_+ + \mathbf{Q}_- \times \mathbf{Q}_+^*)$$
$$= \mathbf{Q}_+^* \times \mathbf{Q}_+ + Z_D^2 \mathbf{Q}_-^* \times \mathbf{Q}_- + 2iZ_D \operatorname{Re}(\mathbf{Q}_-^* \times \mathbf{Q}_+)$$
$$= \mathbf{P}_+^* \times \mathbf{P}_+ + \mathbf{P}_-^* \times \mathbf{P}_- - 2i\operatorname{Im}(\mathbf{P}_-^* \times \mathbf{P}_+)$$

By taking real and imaginary parts on both sides, we form physically crucial bilinear parameters as follows.

$$\omega\mathbf{M} \equiv \operatorname{Im}(\mathbf{E}^* \times \mathbf{E}) = \operatorname{Im}(\mathbf{Q}_+^* \times \mathbf{Q}_+) + Z_D^2 \operatorname{Im}(\mathbf{Q}_-^* \times \mathbf{Q}_-) + 2Z_D \operatorname{Re}(\mathbf{Q}_-^* \times \mathbf{Q}_+). \qquad (S7.2)$$

Going over to the circular vectors, the self-vector-products $\mathbf{Q}_\pm^* \times \mathbf{Q}_\pm$ are evaluated as follows.

$$\begin{cases} k_\pm \equiv \omega\left(\sqrt{\varepsilon_D\mu_D} \pm \kappa\right), \ \gamma_\pm \equiv \sqrt{k_x^2 - k_\pm^2} \\ \Pi_\pm \equiv \exp(ik_x x - \gamma_\pm z), \ \mathbf{Q}_\pm = Q_\pm \frac{1}{\sqrt{2}} \frac{1}{k_x}(\gamma_\pm \hat{\mathbf{x}} \pm k_\pm \hat{\mathbf{y}} + ik_x \hat{\mathbf{z}})\Pi_\pm \end{cases} \Rightarrow$$

$$\frac{2k_x^2 \mathbf{Q}_\pm^* \times \mathbf{Q}_\pm}{|Q_\pm|^2 \exp(-2\gamma_\pm z)} = (\gamma_\pm \hat{\mathbf{x}} \pm k_\pm \hat{\mathbf{y}} - ik_x \hat{\mathbf{z}}) \times (\gamma_\pm \hat{\mathbf{x}} \pm k_\pm \hat{\mathbf{y}} + ik_x \hat{\mathbf{z}})$$
$$= (\gamma_\pm \hat{\mathbf{x}} \pm k_\pm \hat{\mathbf{y}}) \times (ik_x \hat{\mathbf{z}}) + (-ik_x \hat{\mathbf{z}}) \times (\gamma_\pm \hat{\mathbf{x}} \pm k_\pm \hat{\mathbf{y}}) \qquad (S7.3)$$
$$= 2(\gamma_\pm \hat{\mathbf{x}} \pm k_\pm \hat{\mathbf{y}}) \times (ik_x \hat{\mathbf{z}}) = 2ik_x(\pm k_\pm \hat{\mathbf{x}} - \gamma_\pm \hat{\mathbf{y}}) \Rightarrow$$

$$\frac{2k_x^2 \operatorname{Im}(\mathbf{Q}_\pm^* \times \mathbf{Q}_\pm)}{|Q_\pm|^2 \exp(-2\gamma_\pm z)} = 2k_x(\pm k_\pm \hat{\mathbf{x}} - \gamma_\pm \hat{\mathbf{y}}) \Rightarrow$$

$$\operatorname{Im}(\mathbf{Q}_\pm^* \times \mathbf{Q}_\pm) = |Q_\pm|^2 \exp(-2\gamma_\pm z) \frac{\pm k_\pm \hat{\mathbf{x}} - \gamma_\pm \hat{\mathbf{y}}}{k_x} = -\operatorname{Im}(\mathbf{Q}_\pm \times \mathbf{Q}_\pm^*)$$

Notice that $\{\hat{\mathbf{x}}, \hat{\mathbf{y}}\}$ point respectively into the longitudinal and transverse (not the depth-wise) directions. We made use of the fact that $(\gamma_\pm \hat{\mathbf{x}} \pm k_\pm \hat{\mathbf{y}}) \times (\gamma_\pm \hat{\mathbf{x}} \pm k_\pm \hat{\mathbf{y}}) = \mathbf{0}$ and $\hat{\mathbf{z}} \times \hat{\mathbf{z}} = \mathbf{0}$ during the above derivations. The sign reversal $\operatorname{Im}(\mathbf{Q}_\pm^* \times \mathbf{Q}_\pm) = -\operatorname{Im}(\mathbf{Q}_\pm \times \mathbf{Q}_\pm^*)$ should be cared for.

Let us then evaluate the cross-vector-product $\operatorname{Re}(\mathbf{Q}_-^* \times \mathbf{Q}_+)$ that is necessary for the spin density.

$$\frac{2k_x^2 \mathbf{Q}_-^* \times \mathbf{Q}_+}{Q_-^* Q_+ \exp\left[-(\gamma_+ + \gamma_-)z\right]} = (\gamma_- \hat{\mathbf{x}} - k_- \hat{\mathbf{y}} - ik_x \hat{\mathbf{z}}) \times (\gamma_+ \hat{\mathbf{x}} + k_+ \hat{\mathbf{y}} + ik_x \hat{\mathbf{z}})$$
$$= (\gamma_- \hat{\mathbf{x}} - k_- \hat{\mathbf{y}}) \times (\gamma_+ \hat{\mathbf{x}} + k_+ \hat{\mathbf{y}}) + (\gamma_- \hat{\mathbf{x}} - k_- \hat{\mathbf{y}}) \times (ik_x \hat{\mathbf{z}}) + (-ik_x \hat{\mathbf{z}}) \times (\gamma_+ \hat{\mathbf{x}} + k_+ \hat{\mathbf{y}})$$
$$= (\gamma_- k_+ + \gamma_+ k_-) \hat{\mathbf{z}} + ik_x \left[(\gamma_- \hat{\mathbf{x}} - k_- \hat{\mathbf{y}}) \times \hat{\mathbf{z}} + (\gamma_+ \hat{\mathbf{x}} + k_+ \hat{\mathbf{y}}) \times \hat{\mathbf{z}}\right] \quad \quad (S7.4)$$
$$= (\gamma_- k_+ + \gamma_+ k_-) \hat{\mathbf{z}} + ik_x \left[(\gamma_+ + \gamma_-) \hat{\mathbf{x}} + (k_+ - k_-) \hat{\mathbf{y}}\right] \times \hat{\mathbf{z}}$$
$$= (\gamma_- k_+ + \gamma_+ k_-) \hat{\mathbf{z}} + ik_x \left[(k_+ - k_-) \hat{\mathbf{x}} - (\gamma_+ + \gamma_-) \hat{\mathbf{y}}\right]$$

Taking the real and imaginary parts, we obtain three nonzero components as follows from naïve standpoints.

$$\zeta \equiv \frac{2k_x^2 \mathbf{Q}_-^* \times \mathbf{Q}_+}{Q_-^* Q_+ \exp\left[-(\gamma_+ + \gamma_-)z\right]} = (\gamma_- k_+ + \gamma_+ k_-) \hat{\mathbf{z}} + ik_x \left[(k_+ - k_-) \hat{\mathbf{x}} - (\gamma_+ + \gamma_-) \hat{\mathbf{y}}\right]$$
$$\begin{cases} \mathrm{Re}(Q_-^* Q_+) = 0 = \mathrm{Re}(Q_- Q_+^*) \\ \mathrm{Im}(Q_-^* Q_+) = Z_D \beta_+ \beta_- |\Gamma|^2 = -\mathrm{Im}(Q_- Q_+^*) \end{cases} \Rightarrow$$
$$\mathrm{Re}(\zeta) = \mathrm{Re}(Q_-^* Q_+)(\gamma_- k_+ + \gamma_+ k_-) \hat{\mathbf{z}} - \mathrm{Im}(Q_-^* Q_+) k_x \left[(k_+ - k_-) \hat{\mathbf{x}} - (\gamma_+ + \gamma_-) \hat{\mathbf{y}}\right]$$
$$\quad - \mathrm{Im}(Q_-^* Q_+) k_x \left[(k_+ - k_-) \hat{\mathbf{x}} - (\gamma_+ + \gamma_-) \hat{\mathbf{y}}\right] \quad \quad (S7.5)$$
$$= -Z_D \beta_+ \beta_- |\Gamma|^2 k_x \left[(k_+ - k_-) \hat{\mathbf{x}} - (\gamma_+ + \gamma_-) \hat{\mathbf{y}}\right]$$
$$\mathrm{Im}(\zeta) = \mathrm{Im}(Q_-^* Q_+)(\gamma_- k_+ + \gamma_+ k_-) \hat{\mathbf{z}} + \mathrm{Re}(Q_-^* Q_+) k_x \left[(k_+ - k_-) \hat{\mathbf{x}} - (\gamma_+ + \gamma_-) \hat{\mathbf{y}}\right]$$
$$= \mathrm{Im}(Q_-^* Q_+)(\gamma_- k_+ + \gamma_+ k_-) \hat{\mathbf{z}} = Z_D \beta_+ \beta_- |\Gamma|^2 (\gamma_- k_+ + \gamma_+ k_-) \hat{\mathbf{z}}$$

It is stressed in general that the depth-wise components appear to survive. Recall in Eq. (S6.6) that we have already evaluated $\mathrm{Re}(Q_-^* Q_+) = 0$ and $\mathrm{Im}(Q_-^* Q_+) = Z_D \beta_+ \beta_- |\Gamma|^2$ separately based on the circular vectors. Notice that $\{Q_+, Q_-\}$ are out of phase by $90^o$. It surprises us that the four scalars $\{|\mathbf{E}|^2, |\mathbf{H}|^2, M_x, M_y\}$ contain an interference factor $Q_- Q_+^*$ during the derivations.

Consequently, portions of the spin density are evaluated below based on the circular vectors.

$$\begin{cases} P_+ \equiv Q_+ \\ P_- \equiv iZ_D Q_- \end{cases}, \quad \beta_\pm \equiv k_\mp + \gamma_\mp \frac{Z_D \omega \varepsilon_M}{\gamma_M} \in \mathbb{R}, \quad \begin{cases} Y \equiv Z_D \Gamma \\ P_\pm \equiv iZ_D \beta_\pm \Gamma \equiv i\beta_\pm Y \end{cases} \Rightarrow$$

$$\begin{cases} \mathrm{Im}\left(\mathbf{Q}_\pm^* \times \mathbf{Q}_\pm\right) = |Q_\pm|^2 \exp(-2\gamma_\pm z) \dfrac{\pm k_\pm \hat{\mathbf{x}} - \gamma_\pm \hat{\mathbf{y}}}{k_x} \Rightarrow \\ 2\mathrm{Re}\left(\mathbf{Q}_-^* \times \mathbf{Q}_+\right) \\ \quad = -Z_D \beta_+ \beta_- |\Gamma|^2 k_x^{-1} \exp\left[-(\gamma_+ + \gamma_-)z\right]\left[(k_+ - k_-)\hat{\mathbf{x}} - (\gamma_+ + \gamma_-)\hat{\mathbf{y}}\right] \end{cases}$$

$$\omega \mathbf{M} \equiv \mathrm{Im}\left(\mathbf{E}^* \times \mathbf{E}\right) = \mathrm{Im}\left(\mathbf{Q}_+^* \times \mathbf{Q}_+\right) + Z_D^2 \mathrm{Im}\left(\mathbf{Q}_-^* \times \mathbf{Q}_-\right) + 2Z_D \mathrm{Re}\left(\mathbf{Q}_-^* \times \mathbf{Q}_+\right). \quad (S7.6)$$

$$= |Q_+|^2 \exp(-2\gamma_+ z)\frac{k_+ \hat{\mathbf{x}} - \gamma_+ \hat{\mathbf{y}}}{k_x} - Z_D^2 |Q_-|^2 \exp(-2\gamma_- z)\frac{k_- \hat{\mathbf{x}} + \gamma_- \hat{\mathbf{y}}}{k_x}$$

$$- Z_D^2 \beta_+ \beta_- |\Gamma|^2 \frac{(k_+ - k_-)\hat{\mathbf{x}} - (\gamma_+ + \gamma_-)\hat{\mathbf{y}}}{k_x} \exp\left[-(\gamma_+ + \gamma_-)z\right]$$

Therefore, we come up with two (not three) nonzero components for the spin density.

$$Z_D^{-2} |\Gamma|^{-2} k_x \omega \mathbf{M} \equiv |Y|^{-2} k_x \omega \mathbf{M}$$
$$= \beta_+^2 (k_+ \hat{\mathbf{x}} - \gamma_+ \hat{\mathbf{y}})\exp(-2\gamma_+ z) - \beta_-^2 (k_- \hat{\mathbf{x}} + \gamma_- \hat{\mathbf{y}})\exp(-2\gamma_- z)$$
$$- \beta_+ \beta_- \left[(k_+ - k_-)\hat{\mathbf{x}} - (\gamma_+ + \gamma_-)\hat{\mathbf{y}}\right]\exp\left[-(\gamma_+ + \gamma_-)z\right] \qquad (S7.7)$$
$$= \begin{cases} k_+ (\beta_+)^2 \exp(-2\gamma_+ z) - k_- (\beta_-)^2 \exp(-2\gamma_- z) \\ -\beta_+ \beta_- (k_+ - k_-)\exp\left[-(\gamma_+ + \gamma_-)z\right] \end{cases} \hat{\mathbf{x}}$$
$$- \begin{cases} \gamma_+ (\beta_+)^2 \exp(-2\gamma_+ z) + \gamma_- (\beta_-)^2 \exp(-2\gamma_- z) \\ -\beta_+ \beta_- (\gamma_+ + \gamma_-)\exp\left[-(\gamma_+ + \gamma_-)z\right] \end{cases} \hat{\mathbf{y}}$$

It turns out that we do not need $\mathrm{Im}\left(\mathbf{Q}_-^* \times \mathbf{Q}_+\right)$ at all. Consequently, we end up with both longitudinal component $\left[\mathrm{Im}\left(\mathbf{E}^* \times \mathbf{E}\right)\right]_x$ and the transverse component $\left[\mathrm{Im}\left(\mathbf{E}^* \times \mathbf{E}\right)\right]_y$ [S5]. Both nonzero components are endowed with three disparate decay rates. Moreover, the depth-wise component of the spin density identically vanishes largely due to strong confinements.

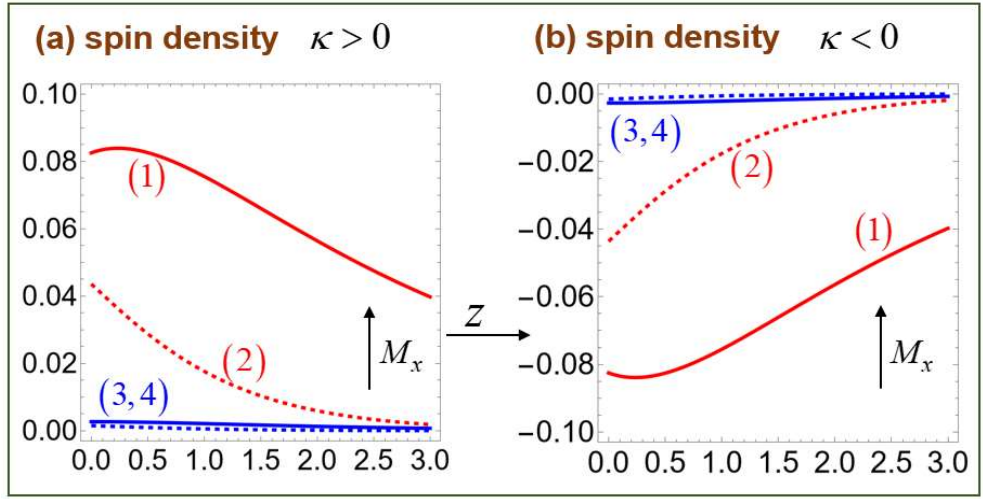

Fig. S6. The longitudinal component $Z_D^{-2}|\Gamma|^{-2}\omega M_x$ of the spin density: (a) for a positive $\kappa > 0$ and (b) for a negative $\kappa < 0$ plotted against the depth-wise $z$-coordinate. It is specified that $Z_D|\Gamma| = 1$. Four curves on panel (a) are arranged according to the orders: two red curves (1,2) for $\kappa = \frac{3}{10}$ and the two blue curves (3,4) with $\kappa = \frac{1}{10}$. For each medium chirality, the solid curves refer to lower frequencies, whereas the dotted curves refer to higher frequencies.

On Fig. S6(a), the trio of $\{\kappa, k_x, \omega\}$ is hence specified from the top to the bottom: (1) $\{\frac{3}{10}, 0.7, 0.522\}$, (2) $\{\frac{3}{10}, 0.999, 0.619\}$, (3) $\{\frac{1}{10}, 0.7, 0.537\}$, and (4) $\{\frac{1}{10}, 0.999, 0.618\}$. On Fig. S6(b), the pair of $\{\kappa, k_x, \omega\}$ is hence specified from the bottom to the top: (1) $\{-\frac{3}{10}, 0.7, 0.522\}$, (2) $\{-\frac{3}{10}, 0.999, 0.619\}$, (3) $\{-\frac{1}{10}, 0.7, 0.537\}$, and (4) $\{-\frac{1}{10}, 0.999, 0.618\}$. Given data includes $\varepsilon_D = \mu_D = \mu_M = 1$ and $\omega_p = 1$. Figure S6(b) is just the reversal of Fig. S6(a) across the horizontal line at $M_x = 0$, according to the anti-symmetric spin density such that $M_x(-\kappa) = -M_x(\kappa)$.

In addition, we find it advantageous that both nonzero components are factored out respectively into two multiplying factors.

$$
\begin{aligned}
|Y|^{-2} k_x \omega M_x &\equiv Z_D^{-2} |\Gamma|^{-2} k_x \omega M_x \\
&= \begin{bmatrix} k_+ (\beta_+)^2 \exp(-2\gamma_+ z) - k_- (\beta_-)^2 \exp(-2\gamma_- z) \\ -\beta_+ \beta_- (k_+ - k_-) \exp[-(\gamma_+ + \gamma_-)z] \end{bmatrix} \\
&= \left[ k_+ \beta_+ \exp(-\gamma_+ z) + k_- \beta_- \exp(-\gamma_- z) \right] \left[ \beta_+ \exp(-\gamma_+ z) - \beta_- \exp(-\gamma_- z) \right] \\
&\equiv \Psi_{xx}^{even}(\kappa, z) \Psi_{xy}^{odd}(\kappa, z) \\
|Y|^{-2} k_x \omega M_y &\equiv Z_D^{-2} |\Gamma|^{-2} k_x \omega M_y \\
&= -\left\{ \begin{matrix} \gamma_+ (\beta_+)^2 \exp(-2\gamma_+ z) + \gamma_- (\beta_-)^2 \exp(-2\gamma_- z) \\ -\beta_+ \beta_- (\gamma_+ + \gamma_-) \exp[-(\gamma_+ + \gamma_-)z] \end{matrix} \right\} \\
&= -\left[ \gamma_+ \beta_+ \exp(-\gamma_+ z) - \gamma_- \beta_- \exp(-\gamma_- z) \right] \left[ \beta_+ \exp(-\gamma_+ z) - \beta_- \exp(-\gamma_- z) \right] \\
&\equiv -\Psi_{yy}^{odd}(\kappa, z) \Psi_{xy}^{odd}(\kappa, z)
\end{aligned}
\tag{S7.8}
$$

Once more, we exploit the series of symmetry properties: $k_{\pm}(-\kappa) \equiv k_{\mp}(\kappa)$, $\gamma_{\pm}(-\kappa) \equiv \gamma_{\mp}(\kappa)$, and $\beta_{\pm}(-\kappa) \equiv \beta_{\mp}(\kappa)$; $\gamma_{\pm}(-k_x) \equiv \gamma_{\pm}(k_x)$ and $\beta_{\pm}(-k_x) \equiv \beta_{\pm}(k_x)$. We can thus decide even/odd properties of both nonzero components of the spin density in the following unified way.

$$
\begin{cases}
\Psi_{xx}^{even}(\kappa, z) \equiv k_+ \beta_+ \exp(-\gamma_+ z) + k_- \beta_- \exp(-\gamma_- z) \\
\Psi_{yy}^{odd}(\kappa, z) \equiv \gamma_+ \beta_+ \exp(-\gamma_+ z) - \gamma_- \beta_- \exp(-\gamma_- z) \\
\Psi_{xy}^{odd}(\kappa, z) \equiv \beta_+ \exp(-\gamma_+ z) - \beta_- \exp(-\gamma_- z)
\end{cases} \Rightarrow
$$

$$
\begin{cases} k_{\pm}(-\kappa) \equiv k_{\mp}(\kappa) \\ \gamma_{\pm}(-\kappa) \equiv \gamma_{\mp}(\kappa) \end{cases} \Rightarrow \beta_{\pm}(-\kappa) \equiv \beta_{\mp}(\kappa) \Rightarrow
\tag{S7.9}
$$

$$
\begin{cases}
\Psi_{xx}^{even}(-\kappa, z) = \Psi_{xx}^{even}(\kappa, z) \\
\Psi_{yy}^{odd}(-\kappa, z) = -\Psi_{yy}^{odd}(\kappa, z) \\
\Psi_{xy}^{odd}(-\kappa, z) = -\Psi_{xy}^{odd}(\kappa, z)
\end{cases} \Rightarrow
\begin{cases} M_x(-\kappa, z) = -M_x(\kappa, z) \\ M_y(-\kappa, z) = M_y(\kappa, z) \end{cases}
$$

Consequently, the longitudinal and transverse components of the spin density are respectively odd and even with respect to the medium chirality. Notwithstanding, the symmetry $M_y(-\kappa) \equiv M_y(\kappa)$ of the transverse spin density is physically understandable with a bit of mental resistance.

This sign flip $M_x(-\kappa, z) = -M_x(\kappa, z)$ in the longitudinal component has been known [S5] so that we can call it either a 'chirality-induced sign flip or spin flip (CISF)' or a 'chirality-induced polarization reversal (CIPR)'. Recall that we have not seen a spin flip in the third Stokes parameter $S_{D3}$ for the achiral case on surface plasmon resonances. Only chiral media offer a possibility of spin flips as the medium chirality is reversed.

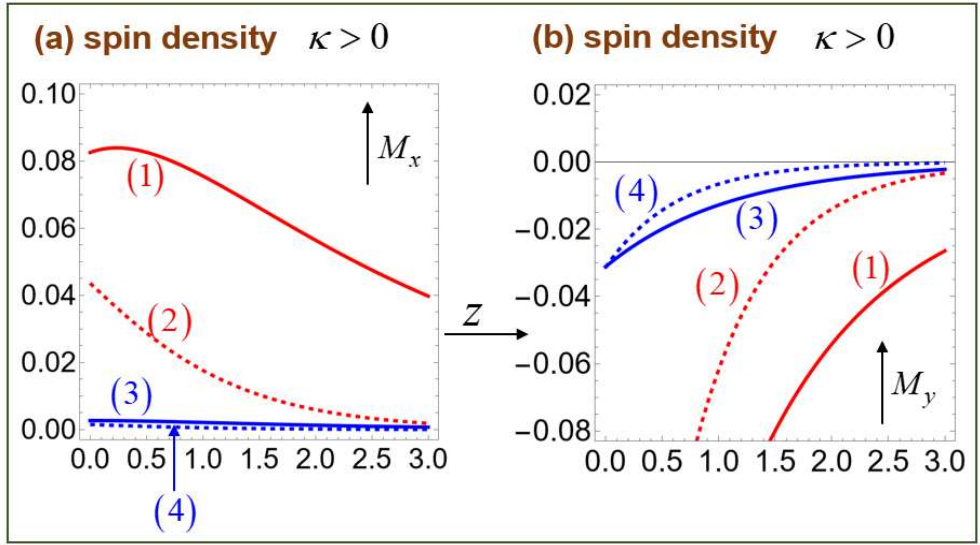

Fig. S7. (a) The longitudinal component $Z_D^{-2}|\Gamma|^{-2}\omega M_x$ and (b) the transverse component $Z_D^{-2}|\Gamma|^{-2}\omega M_y$ of the spin density plotted against the depth-wise $z$-coordinate. It is specified that $Z_D|\Gamma|=1$. Four curves on both panels are arranged according to the orders: two red curves (1,2) for $\kappa = \frac{3}{10}$ and the two blue curves (3,4) with $\kappa = \frac{1}{10}$. For each medium chirality, the solid curves refer to lower frequencies, whereas the dotted curves refer to higher frequencies.

In Fig. S7, the trio of $\{\kappa, k_x, \omega\}$ is hence specified on both panels such that (1) $\{\frac{3}{10}, 0.7, 0.522\}$, (2) $\{\frac{3}{10}, 0.999, 0.619\}$, (3) $\{\frac{1}{10}, 0.7, 0.537\}$, and (4) $\{\frac{1}{10}, 0.999, 0.618\}$. Figure S7(a) is identical to Fig. S6(a). Given data includes $\varepsilon_D = \mu_D = \mu_M = 1$ and $\omega_p = 1$. The above Fig. S7 shows that $\{|M_x|, |M_y|\}$ are of the same order of magnitudes. In addition, Fig. S7(a) shows a slight maximum feature in $|M_x|$ on curve (1) with $\{\kappa, k_x, \omega\} = \{\frac{3}{10}, 0.7, 0.522\}$.

We are now ready to take integrations of both field intensities $\{\int M_x, \int M_y\}$ over the upper half-space.

$$Z_D^{-2}|\Gamma|^{-2}k_x\omega\int M_x = \frac{k_+(\beta_+)^2}{2\gamma_+} - \frac{k_-(\beta_-)^2}{2\gamma_-} + \frac{\beta_+\beta_-(k_- - k_+)}{\gamma_+ + \gamma_-}$$

$$\neq \left(\frac{k_+\beta_+}{\sqrt{2\gamma_+}} + \frac{k_-\beta_-}{\sqrt{2\gamma_-}}\right)\left(\frac{\beta_+}{\sqrt{2\gamma_+}} - \frac{\beta_-}{\sqrt{2\gamma_-}}\right) \quad . \quad (S7.10)$$

$$Z_D^{-2}|\Gamma|^{-2}k_x\omega\int M_y = -\left[\tfrac{1}{2}(\beta_+)^2 + \tfrac{1}{2}(\beta_-)^2 - \beta_+\beta_-\right] = -\tfrac{1}{2}(\beta_+ - \beta_-)^2$$

We find that the space integral solely of the transverse component takes a perfect squared form $(\beta_+ - \beta_-)^2$. Both integrals keep their odd and even properties as their respective integrands. Since $\beta_+\beta_- < 0$ as seen from numerical results, $-\tfrac{1}{2}(\beta_+ - \beta_-)^2$ should be appreciably nonnegligible as well.

Collecting relevant formulas for the spin density,

$$Z_D^{-2}|\Gamma|^{-2}|\mathbf{E}|^2 \equiv |Y|^{-2}|\mathbf{E}|^2 = (\beta_+)^2\exp(-2\gamma_+z) + (\beta_-)^2\exp(-2\gamma_-z)$$
$$-\beta_+\beta_-\frac{\gamma_+\gamma_- - k_+k_- + k_x^2}{k_x^2}\exp\left[-(\gamma_+ + \gamma_-)z\right] \Rightarrow \quad . \quad (S7.11)$$

$$\begin{cases} Z_D^{-2}|\Gamma|^{-2}\omega M_x = \frac{1}{k_x}\begin{Bmatrix} k_+\beta_+\exp(-\gamma_+z) + k_-\beta_-\exp(-\gamma_-z) \\ [\beta_+\exp(-\gamma_+z) - \beta_-\exp(-\gamma_-z)] \end{Bmatrix} \\ Z_D^{-2}|\Gamma|^{-2}\omega M_y = -\frac{1}{k_x}\begin{Bmatrix} [\gamma_+\beta_+\exp(-\gamma_+z) + \gamma_-\beta_-\exp(-\gamma_-z)] \\ [\beta_+\exp(-\gamma_+z) + \beta_-\exp(-\gamma_-z)] \end{Bmatrix} \end{cases}$$

Consider another normalized spin density, which is component-wise identical to the degree of circular polarization (DoCP) based solely on the electric field.

$$\frac{\omega M_x}{|\mathbf{E}|^2} = \frac{Z_D^{-2}|\Gamma|^{-2}\omega M_x}{Z_D^{-2}|\Gamma|^{-2}|\mathbf{E}|^2} = \frac{1}{k_x}\frac{\begin{bmatrix} k_+\beta_+\exp(-\gamma_+z) + k_-\beta_-\exp(-\gamma_-z) \end{bmatrix}}{\begin{bmatrix} (\beta_+)^2\exp(-2\gamma_+z) + (\beta_-)^2\exp(-2\gamma_-z) \\ -\beta_+\beta_-\dfrac{\gamma_+\gamma_- - k_+k_- + k_x^2}{k_x^2}\exp\left[-(\gamma_+ + \gamma_-)z\right] \end{bmatrix}}$$

$$\frac{\omega M_y}{|\mathbf{E}|^2} = \frac{Z_D^{-2}|\Gamma|^{-2}\omega M_y}{Z_D^{-2}|\Gamma|^{-2}|\mathbf{E}|^2} = -\frac{1}{k_x}\frac{\begin{bmatrix} \gamma_+\beta_+\exp(-\gamma_+z) + \gamma_-\beta_-\exp(-\gamma_-z) \end{bmatrix}}{\begin{bmatrix} (\beta_+)^2\exp(-2\gamma_+z) + (\beta_-)^2\exp(-2\gamma_-z) \\ -\beta_+\beta_-\dfrac{\gamma_+\gamma_- - k_+k_- + k_x^2}{k_x^2}\exp\left[-(\gamma_+ + \gamma_-)z\right] \end{bmatrix}} \quad . \quad (S7.12)$$

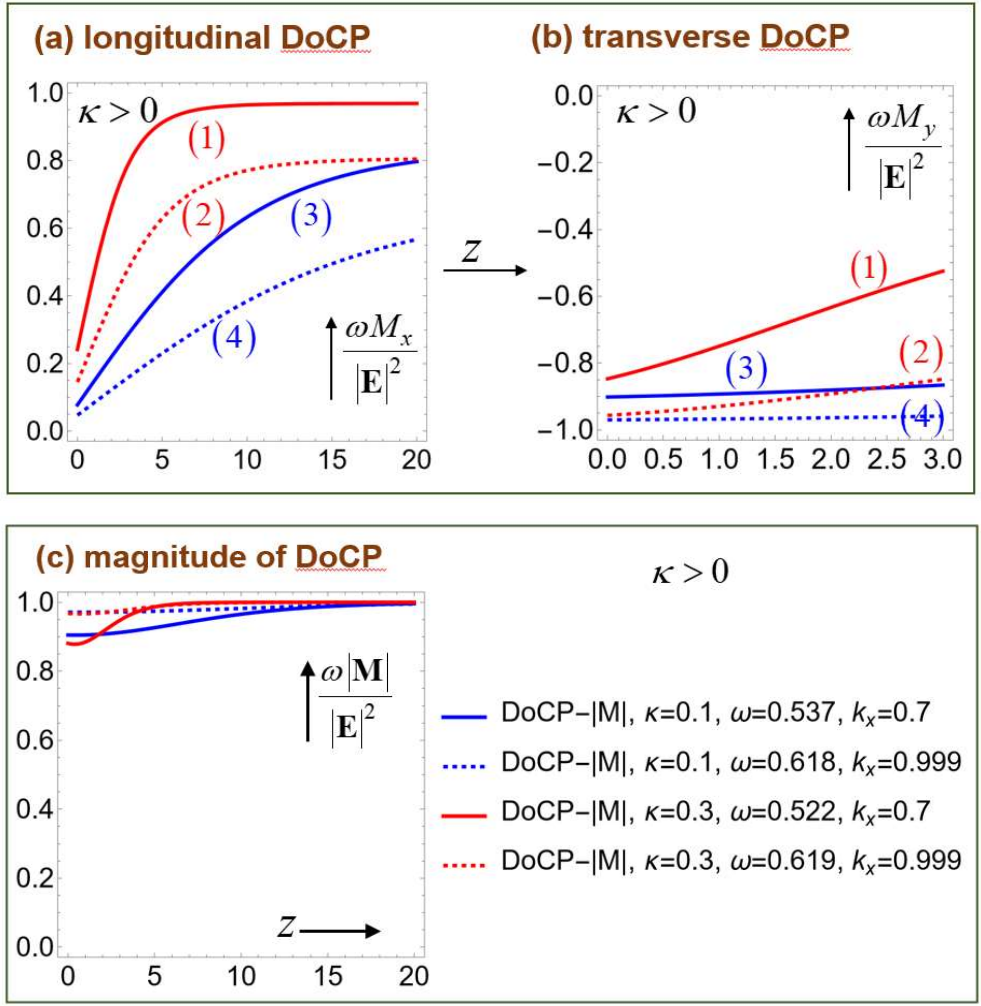

Fig. S8. The degree of circular polarization (DoCP): (a) the longitudinal component $\omega M_x/|\mathbf{E}|^2$ and (b) the transverse component $\omega M_y/|\mathbf{E}|^2$ plotted against the depth-wise $z$-coordinate. It is specified that $Z_D|\Gamma|=1$. Four curves on panel (a) are arranged according to the orders: two red curves (1,2) for $\kappa=\frac{3}{10}$ and the two blue curves (3,4) with $\kappa=\frac{1}{10}$. For each medium chirality, the solid curves refer to lower frequencies, whereas the dotted curves refer to higher frequencies.

In Fig. S8, the trio of $\{\kappa, k_x, \omega\}$ is hence specified such that (1) $\{\frac{3}{10}, 0.7, 0.522\}$, (2) $\{\frac{3}{10}, 0.999, 0.619\}$, (3) $\{\frac{1}{10}, 0.7, 0.537\}$, and (4) $\{\frac{1}{10}, 0.999, 0.618\}$. Given data includes $\varepsilon_D=\mu_D=\mu_M=1$ and $\omega_p=1$. Figure S8 shows that the DoCP on panel (a) in the longitudinal direction approaches unity as the depth-wise $z$-coordinate is increased. Such approach to unity DoCP conforms to the circular state of polarization of the chiral medium as seen from panel (c) for the magnitude $\omega|\mathbf{M}|/|\mathbf{E}|^2$ of the DoCP.

Likewise, the ratios of the spatial integrals are found below.

$$Z_D^{-2}|\Gamma|^{-2}\int|\mathbf{E}|^2 = \frac{(\beta_+)^2}{2\gamma_+} + \frac{(\beta_-)^2}{2\gamma_-} - \beta_+\beta_-\frac{\gamma_+\gamma_- - k_+k_- + k_x^2}{k_x^2(\gamma_+ + \gamma_-)}$$

$$\begin{cases} Z_D^{-2}|\Gamma|^{-2}k_x\omega\int M_x = \dfrac{k_+(\beta_+)^2}{2\gamma_+} - \dfrac{k_-(\beta_-)^2}{2\gamma_-} + \dfrac{\beta_+\beta_-(k_- - k_+)}{\gamma_+ + \gamma_-} \\ Z_D^{-2}|\Gamma|^{-2}k_x\omega\int M_y = -\tfrac{1}{2}(\beta_+ - \beta_-)^2 \end{cases} \Rightarrow$$

$$\eta_x \equiv \frac{\omega\int M_x}{\int|\mathbf{E}|^2} = \frac{Z_D^{-2}|\Gamma|^{-2}\omega\int M_x}{Z_D^{-2}|\Gamma|^{-2}\int|\mathbf{E}|^2} = \frac{1}{k_x}\cdot\frac{\dfrac{k_+(\beta_+)^2}{2\gamma_+} - \dfrac{k_-(\beta_-)^2}{2\gamma_-} + \dfrac{\beta_+\beta_-(k_- - k_+)}{\gamma_+ + \gamma_-}}{\dfrac{(\beta_+)^2}{2\gamma_+} + \dfrac{(\beta_-)^2}{2\gamma_-} - \beta_+\beta_-\dfrac{\gamma_+\gamma_- - k_+k_- + k_x^2}{k_x^2(\gamma_+ + \gamma_-)}} \quad . \quad (S7.13)$$

$$\eta_y \equiv \frac{\omega\int M_y}{\int|\mathbf{E}|^2} = \frac{Z_D^{-2}|\Gamma|^{-2}\omega\int M_y}{Z_D^{-2}|\Gamma|^{-2}\int|\mathbf{E}|^2} = \frac{1}{k_x}\cdot\frac{-\tfrac{1}{2}(\beta_+ - \beta_-)^2}{\dfrac{(\beta_+)^2}{2\gamma_+} + \dfrac{(\beta_-)^2}{2\gamma_-} - \beta_+\beta_-\dfrac{\gamma_+\gamma_- - k_+k_- + k_x^2}{k_x^2(\gamma_+ + \gamma_-)}}$$

We encounter here the appearance of a geometric mean $\dfrac{\gamma_+ + \gamma_-}{2\gamma_+\gamma_-} = \tfrac{1}{2}\left(\gamma_+^{-1} + \gamma_-^{-1}\right)$. The numerical results read the following.

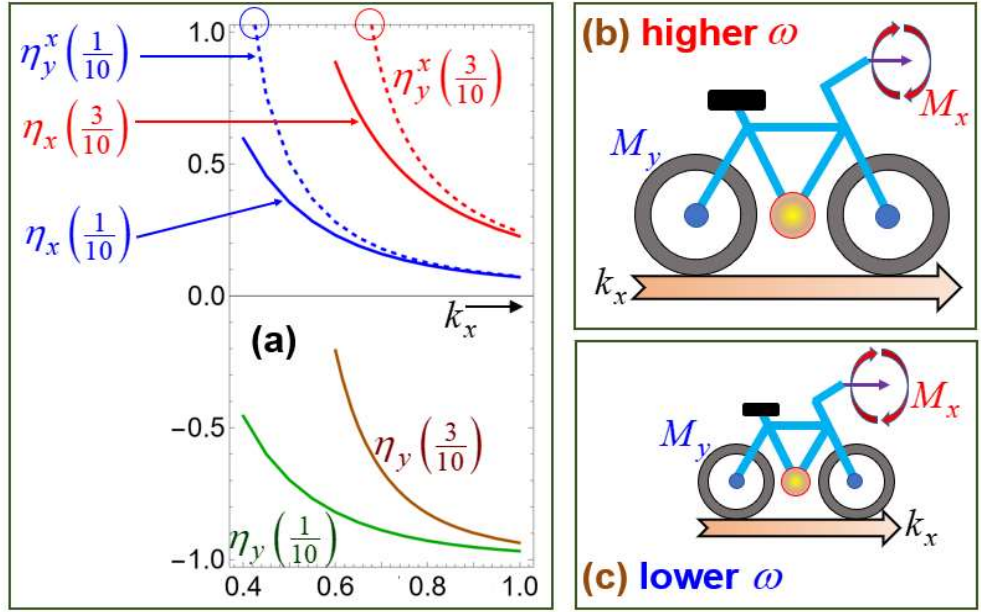

Fig. S9. (a) The degree of circular polarization based on the depth-wise integrations: the longitudinal component $\eta_x \equiv \omega \int M_x / \int |\mathbf{E}|^2$, the transverse component $\eta_y \equiv \omega \int M_y / \int |\mathbf{E}|^2$, and the ratio $\eta_y^x \equiv \int M_x / \int M_y$. They are plotted against the depth-wise $z$-coordinate. It is specified that $Z_D|\Gamma| = 1$. (b) and (c) Cartoons illustrating the distinct behaviors in two frequencies. Given data includes $\varepsilon_D = \mu_D = \mu_M = 1$ and $\omega_p = 1$.

The cartoons on Fig. S9(b) and S9(c) based on bicycles that are differentiated by the following four key parameters.

| item | parameter | relevant energy content | direction |
|---|---|---|---|
| [i] | frequency $\omega$ | electromagnetic energy | time |
| [ii] | longitudinal wave number $k_x$ | longitudinal translational energy | into the $x$-direction |
| [iii] | transverse spin density $M_y$ | transverse rotational energy | on the $zx$-plane |
| [iv] | longitudinal spin density $M_x$ | longitudinal rotational energy | on the $yz$-plane |

Four curves on Fig. S9(a) are arranged according to the orders: three curves for $\kappa = \frac{3}{10}$ and the three curves with $\kappa = \frac{1}{10}$. Notice $\omega(k_x)$ is uniformly increasing with $k_x$ as a result from the dispersion relation. Additionally, we can infer from Fig. S9(a) that $\partial|\eta_x|/\partial|\kappa| > 0$ and $\partial|\eta_y|/\partial|\kappa| < 0$.

The cartoons on Fig. S9(b) and S9(c) are self-similar to each other in sizes. An exception is the same size of the pinwheels on the right top locations of each panel. These pinwheels are intended to make rotations along the longitudinal axes pointing either into or out of the $x$-coordinate, thereby signifying the longitudinal spin density $M_x$. This rotation directly reflects the interaction between a bicycle and the environ (winds).

In comparison, the pair of rotating wheels of each bicycle is intended to represent the transverse spin density $M_y$ pointing either into or out of the $y$-coordinate. This component reflects more of the influence of a metal due to surface plasmon resonance. Since we are dealing with the product $k_x \omega \mathbf{M}$, both $\{M_x, M_y\}$ are revered in signs as the translation is made from right to left with the substitution $k_x \to -k_x$.

Because $|M_x|$ is fixed on both bicycle and $|M_y|$ is different on both bicycles, the ratio $\eta_y^x \equiv \left|\int M_x / \int M_y\right|$ can take either over-unity or under-unity value. For convenience, we call a bicycle a 'baby bicycle' if $\eta_y^x \equiv \left|\int M_x / \int M_y\right| > 1$. In contrast, we call a bicycle a 'adult bicycle' if $\eta_y^x \equiv \left|\int M_x / \int M_y\right| < 1$. Correspondingly, a 'baby-to-adult transition (BTAT)' takes place when $\eta_y^x \equiv \left|\int M_x / \int M_y\right| = 1$. From Fig. S9(a), such a BTAT takes place around $k_x = 0.42$ and $k_x = 0.68$ respectively for $\kappa = \frac{1}{10}$ and $\kappa = \frac{3}{10}$ as marked by the colored solid circles near the top boundary.

**Section S8. Helical DNA (Deoxyribo Nucleic Acid) and epilog**

From microscopic perspectives, rotational features discussed so far have something to with the molecular structures of dispersed molecules in the embedding bulk dielectric media. Suppose that a medium chirality $\kappa$ signifies a curvature of a certain helically curved material. We can then assume the energy necessary to sustain that curvature to be greater than a certain energy cut-off, namely, a frequency cut-off. The stored energy would be capable of unleashing a rotational or vortical motion in the surrounding medium. A curvature is most likely to decrease (in fact, pitch being increased and matters getting slackened) with increasing temperature above a certain critical temperature as seen from a solution consisting of chiral lyotropic chromonic liquid crystals (LCLCs) [S6].

From the viewpoints of electromagnetic-wave dynamics, a closer analogy can also be found with the surface plasmon waves along a metallic wire immersed in air, being an achiral medium. Here, propagating waves along the wire are supported only for wire radii larger than certain cut-off radius [S7]. Such cut-offs are thought to be caused by coherent interferences between two waves. This cut-off (or threshold) radius increases with increasing rotational speed for the achiral case. The 'wire analogy' is intended to show a geometrical similarity between an EM wave along a wire and our evanescent wave.

The existence of all three field components for the chiral case signifies indicates the fact that both TM and TE waves are interwoven. We can solve in principle for the surface plasmon resonances in case of a cylindrical metal-chiral-medium interface, which has not been solved as far as we know [S4], [S7], [S8].

There is a large body of literature regarding the helical structures of DNA (deoxyribo nucleic acid) [S1], [S9]. We are concerned with the 'chirality-induced spin selection (CISS)' among various interesting phenomena. We can draw several analogies between our chiral case and many features of the CISS taking place with double-helix (DH) DNAs.

One disappointing feature of a DH DNA is that the double helices of a DNA are of the same handedness. In some sense, a DH DNA is enantiopure. In contrast, our chiral medium consists of a left wave and right wave. A natural question arises as to what happens to the double helices of a DNA if the double helices are of opposite handedness.

The existence of metallic leads considered by [S1] and [S10] in modeling the CISS is analogous in the sense that our chiral medium shares a common boundary with a metal so that both left and right wave communicate with each other through a metal as well. The dephasing effect considered in [S1] is analogous to a metallic loss that would enable an energy transfer between a metal and a chiral medium [S11]. There are however many differences between our chiral case a DH DNA.

Especially, we are concerned with modeling a DNA with a helical arrangement of electric dipoles along helix axis [S10]. As with the surface plasmon resonance along a metal wire [S4], a field employed in [S10] is evanescently decaying in the radial direction of a helical configuration.